\documentclass[12pt]{article}

\usepackage{amsmath,amssymb,amsthm,appendix,bm,booktabs,caption,graphicx,multirow,natbib,physics}
\usepackage[margin=1in]{geometry}
\usepackage[singlespacing]{setspace}
\usepackage[dvipsnames]{xcolor}
\usepackage[colorlinks,citecolor=Blue,linkcolor=Blue,urlcolor=Maroon]{hyperref}

\DeclareMathOperator{\corr}{corr}
\DeclareMathOperator{\cov}{cov}
\DeclareMathOperator{\diag}{diag}
\DeclareMathOperator{\sd}{sd}
\DeclareMathOperator{\sgn}{sgn}

\newtheorem{assumption}{Assumption}
\newtheorem{definition}{Definition}
\newtheorem{lemma}{Lemma}
\newtheorem{proposition}{Proposition}

\title{Financial constraints, risk sharing, and optimal monetary policy\thanks{For helpful comments and suggestions at different stages of this research, I would like to thank Gianluca Benigno, Toni Braun, Kaiji Chen, Tasos Karantounias, Federico Mandelman, Ricardo Nunes, Juan Rubio-Ram\'{i}rez, Javier Suarez, Carlos Thomas, Tong Xu, Vivian Yue, and seminar and conference participants at Emory University, 2019 EEA-ESEM Manchester, Spring 2020 I-85 Workshop, SEA 2020 Annual Meeting, and the University of Surrey.}}
\author{Aliaksandr Zaretski\footnote{School of Social Sciences, University of Surrey, e-mail: \url{a.zaretski@surrey.ac.uk}.}}
\date{January 27, 2025}

\begin{document}

\maketitle

\begin{abstract}
I characterize optimal government policy in a sticky-price economy with different types of consumers and endogenous financial constraints in the banking and entrepreneurial sectors. The competitive equilibrium allocation is constrained inefficient due to a pecuniary externality implicit in the collateral constraint and other externalities arising from consumer type heterogeneity. These externalities can be corrected with appropriate fiscal instruments. Independently of the availability of such instruments, optimal monetary policy aims to achieve price stability in the long run and approximate price stability in the short run, as in the conventional New Keynesian environment. Compared to the competitive equilibrium, the constrained efficient allocation significantly improves between-agent risk sharing, approaching the unconstrained Pareto optimum and leading to sizable welfare gains. Such an allocation has lower leverage in the banking and entrepreneurial sectors and is less prone to the boom-bust financial crises and zero-lower-bound episodes observed occasionally in the decentralized economy.
\end{abstract}


\newpage

\section{Introduction}
In the past decade, there has been a surge in research on externalities stemming from financial constraints.\footnote{\citet{davila18} present a unifying treatment of such externalities.} This paper studies the implications of such externalities for optimal monetary policy in an economy with a banking sector and different types of consumers. This economy has both a conventional pecuniary externality working through the collateral asset price and other externalities arising from consumer type heterogeneity. To identify the externalities, I characterize a constrained efficient allocation (CEA) chosen by a benevolent social planner who faces the same constraints as private agents but internalizes the determination of market prices. The wedges between the competitive equilibrium (CE) and CEA arise in both the real and the financial sectors of the economy. The real wedges represent the inefficient demand for labor and capital. The financial wedges reflect the inefficient supply of deposits by the banking sector and demand for loans by the entrepreneurial sector, typically resulting in both overborrowing and overlending by banks. These wedges can be addressed with the appropriate fiscal instruments. A key finding of the paper is that the ability to correct the wedges with fiscal instruments does not impact the fundamental nature of Ramsey-optimal monetary policy. The latter prescribes price stability in the long run and approximate price stability in the short run, as in the basic New Keynesian environment.

The object of the analysis is a New Keynesian economy with different types of consu\-mers---workers, bankers, and entrepreneurs---and a financial sector. Workers are savers who are not directly subject to financial frictions. Bankers manage banks that issue deposits to workers and extend loans to wholesale firms subject to a leverage constraint. Entrepreneurs are the managers of wholesale firms and raise external financing subject to a collateral constraint. The entrepreneur's capital stock serves as collateral and is produced by competitive firms with a nonlinear technology. A monopolistically competitive retail sector is subject to nominal rigidities: the opportunity to adjust prices arrives stochastically according to the Calvo-pricing mechanism.

The normative analysis proceeds step-by-step, starting from a special case of a flexible-price economy with perfectly competitive markets. In this setting, I define a flexible-price competitive equilibrium (FCE) and characterize the flexible-price constrained efficient allocation (FCEA). Due to consumer type heterogeneity, the price externalities are not limited to a conventional pecuniary externality working through the collateral asset price. The social planner is subject to a consolidated budget constraint of bankers and entrepreneurs, which depends on the asset price and the wage rate. Moreover, the planner must respect the private complementary slackness conditions associated with the bank leverage constraint. As a result, the FCE has multiple wedges relative to the FCEA that arise in both the real and the financial sectors of the economy. The real wedges are in the entrepreneur's demand for labor and capital---the factors of production. The labor wedge constitutes the only intratemporal distortion, arising from consumer type heterogeneity, particularly the wage externality. The capital wedge stems from an externality due to the entrepreneur's impatience, both first-order and second-order externalities arising through the capital good production technology, and a pecuniary externality in the collateral constraint. The financial wedges are in the banker's supply of deposits and the entrepreneur's demand for loans, and they mainly result from the differences in patience, reflecting consumer type heterogeneity.

In a special case where the worker's preferences are separable in consumption and leisure and logarithmic in consumption, and the technology is such that capital good producers earn zero profits in the steady state, the FCEA has quantitatively perfect consumption risk sharing between all types of consumers, approaching the unconstrained first-best allocation. The FCEA can be decentralized in a regulated FCE with state-contingent linear taxes on the banker's supply of deposits and the entrepreneur's demand for loans, labor, and capital. I also consider a situation when the complete set of taxes is not available to the policymaker but the leverage limits---bank capital requirement and loan-to-value (LTV) ratio---can be set optimally. The resulting Ramsey allocation has the potential to enhance risk sharing but is typically inferior to the FCEA.

The analysis then moves to the benchmark sticky-price economy. Under an assumption that the social planner takes monetary policy as given, the set of wedges between the CE and CEA is similar to the flexible-price case. The financial wedges remain unchanged, while the real wedges are now affected by the presence of monopoly power and nominal rigidities. The latter reduce the extent of between-agent risk sharing in the CEA compared to the FCEA, although it remains strong quantitatively. The fact that financial wedges are not affected by nominal rigidities has two important implications. First, the fundamental nature of Ramsey-optimal monetary policy is not affected by the availability of the complete set of fiscal instruments needed to decentralize the CEA. Second, the implications of optimal monetary policy are similar to the basic New Keynesian environment: price stability is optimal in the long run, even if there is an effective lower bound (ELB) on the policy rate that does not exceed the steady-state real interest rate. In the short run, the optimal inflation rate is characterized by an Euler equation with different compensating mechanisms: the inflation rate is not necessarily zero but remains close to zero quantitatively. In the presence of an ELB, the Ramsey allocation under optimal monetary policy highlights an additional aggregate demand externality not internalized by the private agents in the CE.

Using a social-welfare consumption-equivalent measure, conditional on choosing a worker-biased vector of Pareto weights, the FCEA constitutes 98.9\% of the first best, compared to 86.2\% in the FCE; the sticky-price CEA provides 94\% of the first best, compared to 78.1\% in the CE. The flexible-price Ramsey allocation with optimal leverage limits and labor taxation---but not other fiscal instruments---gives 94.1\% of the first best, while an analogous sticky-price Ramsey allocation with optimal monetary policy stands at 90.7\%. The FCEA and CEA have perfect consumption risk sharing between bankers and entrepreneurs; the risk sharing with workers is not exactly perfect, but the correlation between the marginal utilities is close to unity. Most of the magnitude and variance of the wedges is explained by the components that arise from consumer type heterogeneity; therefore, the ability to improve between-agent risk sharing is the main source of welfare gains from the FCEA and CEA. Nominal rigidities do have a notable impact on the real wedges. In the FCEA and CEA, bank leverage is suboptimal from the planner's perspective, and the entrepreneur's leverage is lower than in the market allocations. Consequently, the FCE and CE have both overborrowing and overlending by the banking sector.

Finally, I compare the dynamics in the decentralized FCE and CE economies with the dynamics in the centralized FCEA, CEA, and Ramsey allocations around financial crises in the flexible-price settings and the episodes of hitting the zero lower bound (ZLB) on the policy rate in the sticky-price environments. A financial crisis is defined as an event that satisfies two conditions: the collateral constraint is slack for at least four quarters before the start of the crisis and is binding for at least five quarters since the start of the crisis. An event defined this way is observed in the FCE with a relative frequency of 3.2 crises per century, consistent with the data. In the FCE, such crises follow a boom-bust pattern: output, credit, and collateral asset price are increasing ahead of the crisis, followed by a sharp and persistent fall when the collateral constraint binds. In the FCEA, the collateral constraint remains slack during the whole crisis window, and the dynamics of real and financial variables resemble usual business cycle fluctuations. When the intertemporal distortions cannot be addressed but leverage limits are set optimally, the dynamics are more similar to the FCE, although the amplitude of the fluctuations is reduced.

The ZLB crises are identified similarly as events where the ZLB is slack during the year before the start of a crisis and is binding for at least three quarters, which implies a simulated frequency of 2.5 crises per century in the CE. Compared to financial crises, ZLB crises have a different pattern: before the ZLB binds, the economy is already in a recession or stagnation, and inflation is below the target. When the ZLB binds, the recession deepens, and inflation decreases further, followed by an increase due to the rise in the marginal cost. When the ZLB becomes slack, the recovery in investment and the asset price is faster than after financial crises, but the recovery in output and credit is slow. As with financial crises, the CEA dynamics are much smoother, and the ZLB is not hit. The dynamics in the Ramsey allocation with optimal labor taxation, leverage limits, and monetary policy are somewhere in between the CE and CEA, and the planner typically just avoids the ZLB. The optimal bank capital ratio and LTV ratio have countercyclical dynamics around both financial crises and ZLB episodes.

This paper is related to different sets of the literature. The theoretical model is in the class of New Keynesian economies with consumer type heterogeneity \citep{iacoviello05,andres13}. The banking sector is based on \citet{iacoviello15}, while the entrepreneurial and retail sectors have features of \citet{kiyotaki97}, \citet{bernanke99}, and \citet{iacoviello05}. The focus on the CEA in the normative analysis follows \citet{lorenzoni08}. Similar to \citet{lorenzoni08}, \citet{benigno16}, \citet{bianchi18}, \citet{davila18}, and \citet{jeanne19}, the competitive equilibrium is inefficient due to a pecuniary externality present in the collateral constraint. Unlike in most of these papers, the pecuniary externality is associated with borrowing in the domestic banking sector at an endogenous interest rate in the current paper. Moreover, the pecuniary externality is not the only externality that leads to constrained inefficiency. Due to consumer type heterogeneity, multiple wedges stem from multiple price externalities. \citet{farhi16}, \citet{korinek16}, and \citet{schmitt-grohe16} emphasize aggregate demand externalities that arise in the presence of constraints on monetary policy, fixed exchange rates, or downward sticky wages. The definition of the CEA used in this paper specifies that the social planner faces the same constraints as private agents. Hence, the CEA social planner does not internalize any monetary policy constraints. On the other hand, the Ramsey planner that determines the optimal monetary policy is generally subject to an ELB constraint. If such a constraint is present, the CE allocation has an aggregate demand externality compared to the Ramsey allocation.

By characterizing optimal monetary policy in the presence of financial frictions, this paper is related to \citet{bean10}, \citet{andres13}, \citet{curdia16}, \citet{farhi16}, \citet{collard17}, \citet{depaoli17}, \citet{ferrero18}, \citet{leduc18}, and \citet{vanderghote21}. The closest set-ups to the current paper are in \citet{andres13} and \citet{ferrero18}, who also allow for consumer type heterogeneity, collateral constraints, and financial intermediation. Both these papers have a housing market with an inelastic supply that provides collateral for entrepreneurs, while this paper considers capital stock as collateral, and the supply side is endogenous. Moreover, as in \citet{iacoviello15}, this paper considers bankers as generally risk-averse consumers, allowing for an additional degree of heterogeneity. In terms of the normative analysis, \citet{andres13} and \citet{ferrero18} adopt a linear-quadratic approach accurate in the neighborhood of the steady state. At the same time, this paper characterizes globally optimal constrained efficient and Ramsey allocations, respecting occasionally binding constraints in the theoretical derivations, as in \citet{bianchi18}. Consistent with \citet{andres13} and \citet{ferrero18}, this paper finds that optimal monetary policy does not entail perfect consumption insurance between consumers. However, this paper provides conditions under which quantitatively perfect consumption insurance is observed in the CEA. The analysis in \citet{andres13} is limited to separable preferences logarithmic in consumption, while \citet{ferrero18} restrict attention to exponential preferences. In contrast, this paper conducts normative analysis with general preferences and technology.

By proving that the optimal long-run inflation rate in the absence of uncertainty is zero even in the presence of financial frictions, this paper is consistent with \citet{curdia16}, who came to an identical conclusion in the case of a credit spread friction. In this paper, an endogenous credit spread arises from the bank leverage constraint. \citet{andres13} and \citet{collard17} have also argued that zero steady-state inflation is optimal, albeit quantitatively.

The rest of the paper is organized as follows. Section \ref{sec: model} describes the model and defines and characterizes the competitive equilibrium. Section \ref{sec: normative analysis} conducts a normative analysis in the flexible-price and sticky-price economies. Section \ref{sec: quantitative results} presents quantitative results. Section \ref{sec: conclusion} concludes. \hyperlink{appendices}{Appendices} provide additional details, including proofs of theoretical results.

\section{Model}\label{sec: model}
Consider an infinite-horizon discrete-time economy populated by consumers---workers ($w$), bankers ($b$), and entrepreneurs ($e$)---and producers of capital, retail, and final goods. Conditional on the type $i\in\mathcal{I}\equiv\{b,e,w\}$, there is a unit measure of identical risk-averse consumers. Workers are infinitely lived with certainty, but each period, a constant share of bankers and entrepreneurs exit the economy, being replaced by new consumers of the same measure who inherit the assets and liabilities of the former. As noted by \citet{andres13}, a trivial life-cycle structure of this sort facilitates a tractable normative analysis. The differences in survival rates result in the differences in effective patience: workers apply a discount factor $\beta\in(0,1)$, while bankers and entrepreneurs use $\beta_b\le\beta$ and $\beta_e\le\beta$, respectively.

Workers solve a standard consumption-saving problem and are owners of firms that produce capital, retail, and final goods. Bankers manage banks that issue deposits to workers and supply loans to entrepreneurs. Entrepreneurs manage firms that supply wholesale goods to the retail sector that operates subject to nominal rigidities, similar to \citet{bernanke99} and \citet{iacoviello05}. Capital goods are produced using a nonlinear technology as in \citet{lucas71}.

Following \citet{gertler11}, we will assume that financial assets---deposits and loans---are contracted in real terms. This assumption allows increasing the tractability of the normative analysis, since our baseline economy will have a well-defined special case of a flexible-price economy with perfectly competitive markets. Consequently, it will be easier to decipher the roles of financial frictions, consumer type heterogeneity, and nominal rigidities for the efficiency of a competitive equilibrium allocation.

Corresponding to our economy, for each $t\ge{}0$, there is a set $Z^{t}$ of histories of states of nature $z^{t}\in{}Z^{t}$. To save on notation, the dependence on histories will be hidden, but one should be aware that a variable $x_{t}$ will typically correspond to a number $x_{t}(z^{t})$, $\{x_{t}\}$ will denote a sequence $\{x_{t}\}_{t=0}^\infty$ of Borel measurable functions $x_{t}:Z^{t}\to\mathbb{R}$ for all $t\ge{}0$, and $\{x_{1,t},\dots,x_{n,t}\}$ will denote a list of $n$ such sequences.

\subsection{Workers}\label{sec: workers}
A worker's decision problem involves choosing consumption $C^w_{t}$, savings in one-period bank deposits $D_{t}$ at a risk-free gross real interest rate $R_{t}$, and labor supply $N_{t}$ given a wage rate $W_{t}$. The worker's income is augmented by the aggregate profits $\Xi_{t}$ from the ownership of retail and capital good producing firms. The final good is the numeraire, so the budget constraint is
\begin{equation*}
    C^w_{t}+D_{t}\le{}W_{t}N_{t}+R_{t-1}D_{t-1}+\Xi_{t}.
\end{equation*}

The worker's preferences are represented by $\mathbb{E}_0[\sum_{t=0}^{\infty}\beta^{t}U^w(C^w_{t},N_{t})]$, where $U^w:\mathbb{R}^2_+\to\mathbb{R}$ is twice continuously differentiable and strictly concave with $U^w_{C}(C,N)>0$ and $U^w_{N}(C,N)<0$ for all $(C,N)\in\mathbb{R}^2_{++}$, and $\lim_{C\to{0}}U^w_{C}(C,N)=\infty$ for all $N\ge{0}$. Define a stochastic discount factor $\Lambda_{t,s}\equiv\beta^{s-t}\frac{{U^w_{C,s}}}{{U^w_{C,t}}}$, where $s\ge{}t\ge{}0$. The necessary conditions for optimality include the budget constraint holding as equality, the labor supply condition \eqref{eq: workers labor supply} postulating the equality between the wage and the marginal rate of substitution of consumption for leisure, and the Euler equation \eqref{eq: workers Euler} that prices bank deposits:
\begin{align}
    W_{t}&=-\frac{U^w_{N,t}}{U^w_{C,t}},\label{eq: workers labor supply}\\
    1&=\mathbb{E}_{t}(\Lambda_{t,t+1})R_{t}.\label{eq: workers Euler}
\end{align}

\subsection{Bankers}\label{sec: bankers}
Following \citet{iacoviello15}, consider a simple banking sector where banks issue deposits to workers and use their own net worth to extend one-period loans $L_{t}$ to entrepreneurs at a state-contingent gross real loan rate $R^L_{t}$. The bank's net worth is the difference between the ex-post loan repayments from entrepreneurs and deposit repayments to workers, that is, $R^L_{t}L_{t-1}-R_{t-1}D_{t-1}$. Bankers are specialists in managing the banks and their only owners. The banking business provides a dividend $C^b_{t}$, so the banker's budget constraint is
\begin{equation}
    C^b_{t}+L_{t}\le{}R^L_{t}L_{t-1}-R_{t-1}D_{t-1}+D_{t}.\label{eq: bankers budget constraint}
\end{equation}
Furthermore, the banker's budget set is limited by a leverage constraint
\begin{equation}
    D_{t}\le(1-\kappa_{t})L_{t},\label{eq: bankers leverage constraint}
\end{equation}
where $\kappa_{t}\in[0,1]$ can be interpreted as a bank capital requirement. The leverage constraint \eqref{eq: bankers leverage constraint} may reflect agency frictions between workers and bankers or prudential regulation. We will consider $\kappa_{t}$ as a policy instrument set by a policymaker.

The banker's preferences are represented by $\mathbb{E}_0[\sum_{t=0}^{\infty}{\beta_b}^t U^b(C^b_{t})]$, where $U^b:\mathbb{R}_+\to\mathbb{R}$ is twice continuously differentiable with $U^b_{C}>0$ and $U^b_{CC}\le{}0$. Denoting the normalized Lagrange multiplier on \eqref{eq: bankers leverage constraint} as $\lambda^b_{t}$, the Karush---Kuhn---Tucker (KKT) conditions associated with the banker's problem include \eqref{eq: bankers budget constraint} as equality, \eqref{eq: bankers leverage constraint}, the Euler equations for deposits \eqref{eq: bankers Euler deposits} and loans \eqref{eq: bankers Euler loans}, and the complementary slackness conditions \eqref{eq: bankers complementary slackness}:
\begin{align}
    U^b_{C,t}&=\beta_b{}R_{t}\mathbb{E}_{t}(U^b_{C,t+1})+\lambda^b_{t},\label{eq: bankers Euler deposits}\\
    U^b_{C,t}&=\beta_b\mathbb{E}_{t}(U^b_{C,t+1}R^L_{t+1})+\lambda^b_{t}(1-\kappa_{t}),\label{eq: bankers Euler loans}\\
    0&=\lambda^b_{t}[(1-\kappa_{t})L_{t}-D_{t}],\qquad\lambda^b_{t}\ge{0}.\label{eq: bankers complementary slackness}    
\end{align}

Whenever the leverage constraint is binding, the marginal benefit of issuing deposits and borrowing from workers to consume more at $t$ exceeds the marginal cost of deposit repayments and lower consumption at $t+1$ by the shadow value $\lambda^b_{t}\ge{}0$. If the leverage constraint is slack at $t$, but there is a positive probability that it will bind at any contingency in the future, the marginal cost of issuing deposits at $t$ is higher than in the absence of the leverage constraint, which can be seen by iterating \eqref{eq: bankers Euler deposits} forward. Consequently, bankers would like to decrease borrowing to insure themselves against the future instances of a binding leverage constraint.

Both risk aversion and the leverage constraint lead to a spread between the required expected return on loans and deposits:
\begin{equation*}
    \mathbb{E}_{t}(R^L_{t+1})-R_{t}=-\cov_{t}\left[\frac{U^b_{C,t+1}}{\mathbb{E}_t(U^b_{C,t+1})},R^L_{t+1}\right]+\frac{\kappa_{t}\lambda^b_{t}}{\beta_b\mathbb{E}_t(U^b_{C,t+1})},
\end{equation*}
which follows from \eqref{eq: bankers Euler deposits} and \eqref{eq: bankers Euler loans}. The first component of the spread is a risk premium for holding an asset with procyclical payoffs, present only if bankers are risk averse. The second component arises from the leverage constraint and is positive if and only if $\kappa_{t}\lambda^b_{t}>0$. This component becomes larger when bankers are more constrained: either directly due to a higher capital requirement $\kappa_{t}$ or indirectly due to a higher value of the Lagrange multiplier $\lambda^b_{t}$.

\subsection{Entrepreneurs}\label{sec: entrepreneurs}
Similar to \citet{bernanke99} and \citet{iacoviello05}, entrepreneurs manage firms that produce wholesale goods supplied to retailers. The production process requires capital $K_{t}$ and labor $N_{t}$ and is affected by two types of exogenous stochastic disturbances: a total factor productivity (TFP) process $A_{t}$ and a capital quality process $\xi_{t}$. As in \citet{gertler10} and \citet{gertler11}, the capital stock $K_{t-1}$ purchased yesterday has an effective productive value $\xi_{t}K_{t-1}$ today. The capital quality process serves as an exogenous source of variation in the asset price and the return on capital. The effective factors of production are combined using a Cobb---Douglas technology $F:\mathbb{R}^2_{+}\to\mathbb{R}_+$; therefore, the output of the wholesale good is $Y^w_{t}\equiv{}A_{t}F(\xi_{t}K_{t-1},N_{t})$.

The entrepreneur consumes $C^e_{t}$, buys new capital goods at a relative price $Q_{t}$, demands labor from workers, sells the produced wholesale good at a price $P^w_{t}$, and obtains external financing from the banking sector. Hence, the entrepreneur's budget constraint is
\begin{equation}
    C^e_{t}+Q_{t}K_{t}+W_{t}N_{t}+R^L_{t}L_{t-1}\le{}P^w_{t}A_{t}F(\xi_{t}K_{t-1},N_{t})+Q_{t}(1-\delta)\xi_{t}K_{t-1}+L_{t}.\label{eq: entrepreneurs budget constraint}
\end{equation}
Following \citet{kiyotaki97}, external financing requires collateral. Bankers consider the possibility that entrepreneurs may default, in which case the former could recover a fraction of the value of the entrepreneur's effective capital stock $Q_{t+1}\xi_{t+1}K_{t}$. Since both the value of collateral and the value of repayment are contingent on the state, bankers will be willing to extend loans to entrepreneurs if
\begin{equation}
    \mathbb{E}_{t}(R^L_{t+1})L_{t}\le{}m_{t}\mathbb{E}_{t}(Q_{t+1}\xi_{t+1})K_{t},\label{eq: entrepreneurs collateral constraint}
\end{equation}
where $m_{t}\in[0,1]$ reflects recovery costs as perceived by the banker or a policymaker. We will use the latter interpretation and assume that $m_{t}$ is a policy instrument. Moreover, we will restrict attention to equilibria where in all contingencies, the loan rate $R^L_{t}$ is such that both bankers and entrepreneurs get strictly positive consumption, and no defaults occur ex-post.

Note how capital quality affects the entrepreneur's budget set. An expected decrease in $\xi_{t+1}$ tomorrow directly tightens the collateral constraint today, leading to a decrease in external financing. An income effect causes a decrease in the entrepreneur's spending, including the purchasing of new capital goods, which depresses $Q_{t}$ and $K_{t}$. The latter further tightens the collateral constraint, and the logic just described repeats, producing a multiplicative effect of the original shock. Moreover, if the capital quality process is persistent, a decrease in $\xi_{t}$ today would also trigger the described sequence of events due to a decrease in the anticipated capital quality tomorrow. Another source of financial amplification comes from the forward-looking nature of the asset price $Q_{t}$, as demonstrated below.

The entrepreneur's preferences are represented by $\mathbb{E}_0[\sum_{t=0}^{\infty}{\beta_e}^t U^e(C^e_{t})]$, where $U^e:\mathbb{R}_+\to\mathbb{R}$ is twice continuously differentiable with $U^e_{C}>0$ and $U^e_{CC}\le{}0$. Denoting the normalized Lagrange multiplier on \eqref{eq: entrepreneurs collateral constraint} as $\lambda^e_{t}$, the KKT conditions include \eqref{eq: entrepreneurs budget constraint} as equality, \eqref{eq: entrepreneurs collateral constraint}, the labor demand condition \eqref{eq: entrepreneurs labor demand}, the Euler equations for loans \eqref{eq: entrepreneurs Euler loans} and capital \eqref{eq: entrepreneurs Euler capital}, and the complementary slackness conditions \eqref{eq: entrepreneurs complementary slackness}:
\begin{align}
    W_{t}&=P^w_{t}A_{t}F_{N,t},\label{eq: entrepreneurs labor demand}\\
    U^e_{C,t}&=\beta_e\mathbb{E}_{t}(U^e_{C,t+1}R^L_{t+1})+\lambda^e_{t}\mathbb{E}_{t}(R^L_{t+1}),\label{eq: entrepreneurs Euler loans}\\
    U^e_{C,t}Q_{t}&=\beta_e\mathbb{E}_{t}\{U^e_{C,t+1}[P^w_{t+1}A_{t+1}F_{K,t+1}+Q_{t+1}(1-\delta)]\xi_{t+1}\}+\lambda^e_{t}m_{t}\mathbb{E}_{t}(Q_{t+1}\xi_{t+1}),\label{eq: entrepreneurs Euler capital}\\
    0&=\lambda^e_{t}[m_{t}\mathbb{E}_{t}(Q_{t+1}\xi_{t+1})K_{t}-\mathbb{E}_{t}(R^L_{t+1})L_{t}],\qquad\lambda^e_{t}\ge{0}.\label{eq: entrepreneurs complementary slackness}
\end{align}

The collateral constraint affects the entrepreneur's Euler equations \eqref{eq: entrepreneurs Euler loans} and \eqref{eq: entrepreneurs Euler capital} similar to the way the leverage constraint affects the banker's Euler equations \eqref{eq: bankers Euler deposits} and \eqref{eq: bankers Euler loans}. When the collateral constraint is binding, the marginal benefit of borrowing is greater than the marginal cost by $\lambda^e_{t}\mathbb{E}_{t}(R^L_{t+1})$. Moreover, there is self-insurance against the future states when the collateral constraint binds, as reflected by the greater marginal cost of borrowing compared to the economy without the collateral constraint. The capital Euler equation demonstrates that the asset price $Q_{t}$ is determined by the expected future payoff from capital and the marginal value of capital used as collateral, both of which depend on $Q_{t+1}$, making the asset price forward looking. Through the future asset prices, the asset price today also reflects the collateral benefits at all future states when the collateral constraint is binding.

Define the gross return on capital
\begin{equation*}
    R^K_{t}\equiv\frac{P^w_{t}A_{t}F_{K,t}+Q_{t}(1-\delta)}{Q_{t-1}}\xi_{t}.
\end{equation*}
Inspecting \eqref{eq: entrepreneurs Euler loans}--\eqref{eq: entrepreneurs complementary slackness}, we can derive a premium between the required expected returns on capital and loans:
\begin{equation*}
    \mathbb{E}_{t}(R^K_{t+1}-R^L_{t+1})=-\cov_{t}\left[\frac{U^e_{C,t+1}}{\mathbb{E}_{t}(U^e_{C,t+1})},R^K_{t+1}-R^L_{t+1}\right]+\frac{\lambda^e_{t}\mathbb{E}_{t}(R^L_{t+1})}{\beta_e\mathbb{E}_{t}(U^e_{C,t+1})}\left(1-\frac{L_{t}}{Q_{t}K_{t}}\right).
\end{equation*}
When entrepreneurs have enough internal financing to support their business so that the collateral constraint is slack, the premium is determined by the covariance between the future marginal utility and the difference in ex-post returns. The latter is numerically small, and thus in expectation, bankers recover approximately the gross return on capital, similar to \citet{gertler10}. When the collateral constraint is binding, entrepreneurs require a higher expected return on capital so that internal financing could compensate for the lack of available external financing. In this case, bankers can expect to get only a share of the return on capital, and this share is more significant when entrepreneurs fund a greater share of their capital purchases using the banking system. When the amount of external financing is enough to fund the purchase of the new capital goods fully, the expected returns on loans and capital are approximately equal independently of whether the collateral constraint is slack or binding.

\subsection{Capital, retail, and final good production}\label{sec: producers}
Producers of capital goods combine the input of final goods $I_{t}$ and the aggregate capital stock available at the beginning of the period $K_{t-1}$ to build new capital goods $\Phi\left(\frac{I_{t}}{K_{t-1}}\right)K_{t-1}$, where $\Phi:\mathbb{R}_+\to\mathbb{R}$, $\Phi'>0$, $\Phi''\le{0}$, $\lim_{x\to{0}}\Phi'(x)=\infty$, and $\lim_{x\to\infty}\Phi'(x)=0$, similar to \citet{lucas71}. A capital good producer maximizes the expected discounted profits $\mathbb{E}_0\left\{\sum_{t=0}^{\infty}\Lambda_{0,t}\left[Q_{t}\Phi\left(\frac{I_{t}}{K_{t-1}}\right)K_{t-1}-I_{t}\right]\right\}$ under perfect competition; therefore, the supply of new capital goods is described by
\begin{equation}
    Q_{t}=\left[\Phi'\left(\frac{I_{t}}{K_{t-1}}\right)\right]^{-1}.\label{eq: capital good supply}
\end{equation}

There is a unit measure of retail varieties produced by retailers. Each retailer has monopolistic power, internalizing the demand curve of the final good produces. The latter, acting under perfect competition, combine retail varieties into the final good according to a production technology with a constant elasticity of substitution $\epsilon>1$. The retail sector is subject to the pricing mechanism of \citet{calvo83} and \citet{yun96}: at any point in time and any contingency, a retailer cannot reset a price with a probability $\theta\in[0,1]$. Standard derivations (found in appendix \ref{sec: retailer's problem}) imply that retailers that can update their prices choose the same new price, and the following equations hold:
\begin{align}
    \widetilde{P}_{t}&=\frac{\epsilon}{\epsilon-1}\frac{\Omega_{1,t}}{\Omega_{2,t}},\label{eq: retailers optimal relative price}\\
    \Omega_{1,t}&=P^w_{t}Y_{t}+\theta\mathbb{E}_{t}(\Lambda_{t,t+1}\Pi_{t+1}^{\epsilon}\Omega_{1,t+1}),\label{eq: retailers marginal cost}\\
    \Omega_{2,t}&=Y_{t}+\theta\mathbb{E}_{t}(\Lambda_{t,t+1}\Pi_{t+1}^{\epsilon-1}\Omega_{2,t+1}),\label{eq: retailers marginal benefit}\\
    \Pi_{t}^{1-\epsilon}&=\theta+(1-\theta)(\Pi_{t}\widetilde{P}_{t})^{1-\epsilon},\label{eq: retailers inflation}\\
    \Delta_{t}&=\theta\Pi_{t}^{\epsilon}\Delta_{t-1}+(1-\theta){\widetilde{P}_{t}}^{-\epsilon},\label{eq: retailers price dispersion}
\end{align}
where $\widetilde{P}_{t}$ is the optimal new relative price, $\Omega_{1,t}$ defined by \eqref{eq: retailers marginal cost} reflects the retailer's expected marginal cost, $\Omega_{2,t}$ defined by \eqref{eq: retailers marginal benefit} represents the retailer's expected marginal benefit, $Y_{t}$ is the aggregate output of the final good, $\Pi_{t}$ is the gross inflation rate, and $\Delta_{t}$ is a measure of price dispersion. \eqref{eq: retailers optimal relative price} shows that the optimal relative price is set with a time-varying markup over the marginal cost, and \eqref{eq: retailers inflation} demonstrates that the optimal relative price is an increasing function of the inflation rate. According to \eqref{eq: retailers price dispersion}, price dispersion evolves recursively based on the new optimal price and the aggregate inflation rate, and these two forces affect the price dispersion in the opposite directions, implying a stationary relationship.

\subsection{Market clearing}
The capital \eqref{eq: market clearing capital good}, wholesale \eqref{eq: market clearing wholesale good}, and final \eqref{eq: market clearing final good} good market-clearing conditions are
\begin{align}
    K_{t}&=(1-\delta)\xi_{t}K_{t-1}+\Phi\left(\frac{I_{t}}{K_{t-1}}\right)K_{t-1},\label{eq: market clearing capital good}\\
    A_{t}F(\xi_{t}K_{t-1},N_{t})&=\Delta_{t}Y_{t},\label{eq: market clearing wholesale good}\\
    Y_{t}&=C^b_{t}+C^e_{t}+C^w_{t}+I_{t},\label{eq: market clearing final good}
\end{align}
where the derivation of \eqref{eq: market clearing wholesale good} is provided in appendix \ref{sec: retailer's problem}.

\subsection{Competitive equilibrium}\label{sec: competitive_equilibrium}
We are now ready to define a competitive equilibrium.
\begin{definition}\label{def: CE}
    Given exogenous stochastic processes $\{A_{t},\xi_{t}\}$ and boundary conditions, a sequential competitive equilibrium (CE) is a list of allocations $\{C^b_{t},C^e_{t},C^w_{t},D_{t},I_{t},K_{t},L_{t},\linebreak{}N_{t},Y_{t}\}$, prices $\{\widetilde{P}_{t},P^w_{t},Q_{t},R_{t},R^L_{t},W_{t}\}$, Lagrange multipliers $\{\lambda^b_{t},\lambda^e_{t}\}$, auxiliary objects\linebreak $\{\Delta_{t},\Omega_{1,t},\Omega_{2,t}\}$, and policies $\{\kappa_{t},m_{t},\Pi_{t}\}$, such that:
    \begin{enumerate}
        \item Given policies and prices, all agents solve their problems, that is, \eqref{eq: workers labor supply}--\eqref{eq: retailers price dispersion} hold. (Retailers set the prices of individual retail varieties optimally, generating $\widetilde{P}_{t}$.) 
        \item Prices are such that market-clearing conditions \eqref{eq: market clearing capital good}--\eqref{eq: market clearing final good} are satisfied.
    \end{enumerate}
\end{definition}

At this point, we have not specified the nature of the policies $\{\kappa_{t},m_{t},\Pi_{t}\}$. The normative analysis will explore how to set the policies optimally. To compute the CE, we will assume that the leverage limits $\kappa_{t}$ and $m_{t}$ are constants, and there is a central bank that targets inflation according to a Taylor rule with an effective lower bound (ELB) $\underline{R}>0$ on the gross nominal interest rate $R^N_{t}\equiv{}R_{t}\mathbb{E}_{t}(\Pi_{t+1})$. Let $R^*_{t}$ denote the nominal rate when the lower bound is slack. The policy rule can be described as follows:
\begin{equation}
    R^N_{t}=\max(R^*_{t},\underline{R}),\qquad
    R^*_{t}=(R^*_{t-1})^{\rho_R}\left[\frac{\Bar{\Pi}}{\beta}\left(\frac{\Pi_{t}}{\Bar{\Pi}}\right)^{\eta_\pi}\left(\frac{P^w_{t}}{P^w}\right)^{\eta_y}\right]^{1-\rho_R},\label{eq: Taylor rule}
\end{equation}
where $\rho_R\in[0,1)$, $\Bar{\Pi}\ge{}1$ is the central bank's gross inflation target, and $(\eta_\pi,\eta_y)\in\mathbb{R}^2_{+}$ are the response parameters. The deviation of the retailer's marginal cost from the steady state is a proxy for the output gap. (The exact relationship holds in the basic New Keynesian model.) Note that if we use \eqref{eq: Taylor rule} to determine $\{\Pi_{t}\}$, the latter is endogenous to our economy. In the context of definition \ref{def: CE}, it means that there is an implicit consistency condition that requires $\{\Pi_{t}\}$ to satisfy \eqref{eq: Taylor rule}. Although an ELB in \eqref{eq: Taylor rule} necessarily generates a multiplicity of equilibria \citep{benhabib01}, we will restrict attention to the conventional targeted-inflation regime, since it appears to be consistent with the US data \citep{aruoba18}. Using specifications similar to \eqref{eq: Taylor rule}, \citet{braun11} and \citet{fernandez-villaverde15} have also argued for selecting a conventional equilibrium. Moreover, we will show in the normative analysis that optimal monetary policy is uniquely determined even in the presence of an ELB.

Let us complete the description of the CE with two lemmas that characterize the deterministic steady state and give more insight into the optimal decisions of bankers and entrepreneurs. Define
\begin{equation*}
    \widetilde{\beta}_e\equiv\frac{\beta}{1+\kappa\left(\frac{\beta}{\beta_b}-1\right)}.
\end{equation*}
\begin{lemma}\label{lemma: steady state uniqueness}
    Conditional on $\Pi=\Bar{\Pi}$, there exists a unique steady state with positive financial flows if and only if $\beta_b<\beta$ and $\beta_e<\widetilde{\beta}_e$. In this steady state, \eqref{eq: bankers leverage constraint} and \eqref{eq: entrepreneurs collateral constraint} are binding.
\end{lemma}
The intuition for lemma \ref{lemma: steady state uniqueness} is clearer after we rewrite the inequalities $\beta_b<\beta$ and $\beta_e<\widetilde{\beta}_e$ as $\beta_b{R}<1$ and $\beta_e{R}^L<1$, which follows from \eqref{eq: workers Euler}, \eqref{eq: bankers Euler deposits}, \eqref{eq: bankers Euler loans}, and \eqref{eq: entrepreneurs Euler loans}. The latter conditions mean that bankers and entrepreneurs would like to borrow in a steady-state equilibrium because the effective rate of time preference exceeds the interest rate. This condition is consistent with the analysis of the income fluctuations problem of \citet{schechtman77}. If $\beta_b=\beta$, any amount of deposits that satisfies the leverage constraint is associated with an unstable steady state. Note that $\beta_b>\beta$ is ruled out by construction. Similarly, if $\beta_e=\widetilde{\beta}_e$, the quantity of loans is indeterminate. If $\beta_e\in(\widetilde{\beta}_e,\beta]$, then entrepreneurs would choose $L\le{}0$. To make the analysis interesting, we will assume strict inequalities in both cases.
\begin{assumption}\label{ass: discount factors}
    $\beta_b<\beta$ and $\beta_e<\widetilde{\beta}_e$.
\end{assumption}

The following lemma shows that net assets equal the lifetime stream of consumption discounted at the agent-specific stochastic discount factor for both bankers and entrepreneurs.
\begin{lemma}\label{lemma: net assets}
    At the optimum, bank capital satisfies $L_{t}-D_{t}=\frac{1}{U^b_{C,t}}\sum_{s=1}^\infty\beta_b^{s}\mathbb{E}_{t}(U^b_{C,t+s}C^b_{t+s})$. Similarly, the entrepreneur's net assets satisfy $Q_{t}K_{t}-L_{t}=\frac{1}{U^e_{C,t}}\sum_{s=1}^\infty\beta_e^{s}\mathbb{E}_{t}(U^e_{C,t+s}C^e_{t+s})$.
\end{lemma}
Note that we have simple contemporaneous relationships with logarithmic preferences: $L_{t}-D_{t}=\frac{\beta_b}{1-\beta_b}C^b_{t}$ and $Q_{t}K_{t}-L_{t}=\frac{\beta_e}{1-\beta_e}C^e_{t}$. Since the banker's net worth is $R^L_{t}L_{t-1}-R_{t-1}D_{t-1}$, using \eqref{eq: bankers budget constraint}, we see that the banker allocates the majority of her net worth---a share $\beta_b$---for bank capital, while the remaining share $1-\beta_b$ is allocated for consumption (dividends). The more patient the banker is, the greater is the share of net worth reinvested back into the banking business. Due to the Inada condition, consumption is guaranteed to be positive, which implies that bankers would optimally like to hold a positive amount of bank capital independently of the capital requirement, that is, even if $\kappa_{t}=0$.

Similarly, if we define the entrepreneur's net worth as $R^K_{t}Q_{t-1}K_{t-1}-R^L_{t}L_{t-1}$ and consider logarithmic preferences, then \eqref{eq: entrepreneurs budget constraint}, \eqref{eq: entrepreneurs labor demand}, and lemma \ref{lemma: net assets} imply that the entrepreneur's net assets take a share $\beta_e$ of net worth, while consumption takes the remaining share $1-\beta_e$. Since $C^e_{t}>0$ due to the Inada condition, and thus $L_{t}<Q_{t}K_{t}$, entrepreneurs fund the purchases of new capital goods with a nontrivial combination of internal and external financing. Consequently, when the collateral constraint is binding, entrepreneurs will require a strictly higher expected return on capital $\mathbb{E}_{t}(R^K_{t+1})$ than the loan rate $\mathbb{E}_{t}(R^L_{t+1})$, as follows from the premium derived at the end of section \ref{sec: entrepreneurs}.

Furthermore, when both \eqref{eq: bankers leverage constraint} and \eqref{eq: entrepreneurs collateral constraint} are binding, lemma \ref{lemma: net assets} implies that with logarithmic preferences, the consumption ratio of constrained bankers and entrepreneurs can be expressed as a function of policies and prices only:
\begin{equation*}
    \frac{C^e_{t}}{C^b_{t}}=\frac{1-\beta_e}{\beta_e}\frac{\beta_b}{1-\beta_b}\frac{1}{\kappa_{t}}\left[\frac{Q_{t}\mathbb{E}_{t}(R^L_{t+1})}{m_{t}\mathbb{E}_{t}(Q_{t+1}\xi_{t+1})}-1\right].
\end{equation*}
Other things equal, the more impatient agent tends to consume more. A higher bank capital requirement causes bankers to accumulate more net worth, positively affecting consumption. A greater value of collateral per unit of capital stock makes entrepreneurs use relatively more external financing, leading to lower net assets and consumption. Conversely, a higher expected loan rate decreases the available quantity of bank loans for a given value of collateral, increasing the entrepreneur's share of internal financing, net assets, and consumption. A greater price of capital at $t$ also has a positive partial effect on net assets and consumption. Note that the consumption ratio's dependence on the leverage limits anticipates the latter's ability to enhance risk sharing between constrained bankers and entrepreneurs.

\section{Normative analysis}\label{sec: normative analysis}
The purpose of this section is: first, to demonstrate how endogenous financial constraints, nominal rigidities, and consumer type heterogeneity make the CE allocation inefficient; second, to show how to decentralize the constrained efficient allocation with the appropriate fiscal instruments; and third, to characterize Ramsey-optimal leverage limits and monetary policy both when the above-mentioned fiscal instruments are available to the policymaker and when they are not. To understand the differential role of financial frictions and nominal rigidities, we will start by characterizing efficiency and Ramsey-optimal leverage limits in a flexible-price economy with a perfectly competitive retail sector. We will then study constrained efficiency in the benchmark sticky-price economy and will characterize jointly Ramsey-optimal monetary policy and leverage limits under alternative sets of available fiscal instruments.

To begin with, we must define a welfare objective. Since we have ex-ante heterogeneous consumers---workers, bankers, and entrepreneurs---a benevolent social planner should care about all of them. Due to lemma \ref{lemma: steady state uniqueness}, our economy has well-defined local dynamics only when bankers and entrepreneurs are sufficiently impatient relative to workers. Suppose we take as a welfare objective a weighted average of the agents' lifetime utility functions. Due to the differences in patience, a relatively more impatient consumer could get a socially optimal consumption plan that asymptotically converges to zero. Following \citet{andres13}, a way to achieve stationarity is to add the lifetime utilities of all future newborn impatient consumers to the welfare objective.
\begin{definition}\label{def: social welfare}
    Let $V^i_{t}\equiv\mathbb{E}_{t}(\sum_{s=0}^\infty\beta_i^{s}U^i_{t+s})$ denote the lifetime utility of a representative consumer of type $i\in\mathcal{I}$ living at $t\ge{}0$. The social welfare objective at $t\ge{}0$ is $\mathcal{W}_{t}\equiv\sum_{i\in\mathcal{I}}\omega_i\mathcal{W}^i_{t}$, where $\omega_i\ge{}0$ for all $i\in\mathcal{I}$, and $\mathcal{W}^i_{t}\equiv{}V^i_{t}+\frac{\beta-\beta_i}{\beta}\mathbb{E}_{t}(\sum_{s=1}^\infty\beta^s{}V^i_{t+s})$, with $\beta_w\equiv\beta$.
\end{definition}
Consider the aggregate welfare of type $i$ consumers $\mathcal{W}^i_{t}$: it is a sum of the lifetime utility of the representative consumer living at $t\ge{}0$ and the discounted expected lifetime utilities of all future newborns. By definition, $\beta_i$ equals $\beta$ adjusted for the survival probability. Therefore, the exit probability is $\frac{\beta-\beta_i}{\beta}$, and it equals the measure of newborns. It turns out that $\mathcal{W}^i_{t}$ has an equivalent representation independent of the type-specific survival probability.
\begin{lemma}\label{lemma: stationary welfare}
    The aggregate welfare of type $i$ consumers satisfies $\mathcal{W}^i_{t}=\mathbb{E}_{t}(\sum_{s=0}^\infty\beta^s{}U^i_{t+s})$.
\end{lemma}
The intuition for lemma \ref{lemma: stationary welfare} is that by adding the welfare of future newborns to the welfare objective, we can exactly compensate for the uncertain survival of the currently living impatient consumers.

\subsection{Flexible-price economy}\label{sec: efficiency in a flexible-price economy}
In this section, we will consider the flexible-price economy. We will, first, characterize the unconstrained Pareto-optimal allocation that will serve as a reference for welfare comparisons. Second, we will study the constrained efficient allocation and show how to decentralize it in a regulated competitive equilibrium with taxes. Finally, we will explore Ramsey-optimal leverage limits under alternative sets of fiscal instruments available to the Ramsey planner.

The flexible-price economy is a special case of the economy studied in section \ref{sec: model} after setting $\theta=0$ and $\epsilon\to\infty$. In this case, \eqref{eq: retailers optimal relative price}--\eqref{eq: retailers price dispersion} imply $\widetilde{P}_{t}=P^w_{t}=\Delta_{t}=1$, $\Omega_{1,t}=\Omega_{2,t}=Y_{t}$, and $\Pi_{t}$ becomes immaterial. Accordingly, we can revise definition \ref{def: CE} to define a competitive equilibrium in such a setting.
\begin{definition}\label{def: FCE}
    Given exogenous stochastic processes $\{A_{t},\xi_{t}\}$ and boundary conditions, a flexible-price competitive equilibrium (FCE) is a list of allocations $\{C^b_{t},C^e_{t},C^w_{t},D_{t},I_{t},\linebreak{}K_{t},L_{t},N_{t},Y_{t}\}$, prices $\{Q_{t},R_{t},R^L_{t},W_{t}\}$, Lagrange multipliers $\{\lambda^b_{t},\lambda^e_{t}\}$, and policies $\{\kappa_{t},m_{t}\}$, such that:
    \begin{enumerate}
        \item Given policies and prices, all agents solve their problems, that is, \eqref{eq: workers labor supply}--\eqref{eq: capital good supply} hold with $P^w_{t}=1$.
        \item Prices are such that market-clearing conditions \eqref{eq: market clearing capital good}--\eqref{eq: market clearing final good} are satisfied with $\Delta_{t}=1$.
    \end{enumerate}
\end{definition}

\subsubsection{First best}\label{sec: first best}
As a benchmark for welfare comparisons, consider an unconstrained Pareto-optimal alloca\-tion---``first best''---associated with the flexible-price economy. This allocation is an outcome of a planning problem where a benevolent social planner directly allocates consumption and factors of production subject to resource constraints. Conditional on Pareto weights $(\omega_b,\omega_e,\omega_w)\in\mathbb{R}^3_+$, the first-best allocation is a solution to
\begin{equation*}
    \max_{\{C^b_{t},C^e_{t},C^w_{t},I_{t},K_{t},N_{t}\}}\mathbb{E}_{0}\left(\sum_{t=0}^\infty\beta^t\sum_{i\in\mathcal{I}}\omega_i{}U^i_{t}\right)
\end{equation*}
subject to
\begin{align*}
    \lambda^K_{t}:\quad{}0&\le{}(1-\delta)\xi_{t}K_{t-1}+\Phi\left(\frac{I_{t}}{K_{t-1}}\right)K_{t-1}-K_{t},\\
    \lambda^Y_{t}:\quad{}0&\le{}A_{t}F(\xi_{t}K_{t-1},N_{t})-\sum_{i\in\mathcal{I}}C^i_{t}-I_{t}.
\end{align*}
The first-order conditions (FOCs) for $C^i_{t}$, $N_{t}$, $I_{t}$, and $K_{t}$ can be written as
\begin{align*}
    \lambda^Y_{t}&=\omega_i{}U^i_{C,t},\\
    -\frac{U^w_{N,t}}{U^w_{C,t}}&=A_{t}F_{N,t},\\
    \frac{\lambda^K_{t}}{\lambda^Y_{t}}&=\left[\Phi'\left(\frac{I_{t}}{K_{t-1}}\right)\right]^{-1},\\
    \frac{\lambda^K_{t}}{\lambda^Y_{t}}U^e_{C,t}&=\beta\mathbb{E}_{t}\biggl[U^e_{C,t+1}\biggl\{\left[A_{t+1}F_{K,t+1}+\frac{\lambda^K_{t+1}}{\lambda^Y_{t+1}}(1-\delta)\right]\xi_{t+1}\\
    &\quad\qquad\qquad\qquad+\frac{\lambda^K_{t+1}}{\lambda^Y_{t+1}}\left[\Phi\left(\frac{I_{t+1}}{K_{t}}\right)-\Phi'\left(\frac{I_{t+1}}{K_{t}}\right)\frac{I_{t+1}}{K_{t}}\right]\biggr\}\biggr].
\end{align*}

At the unconstrained Pareto optimum, we have perfect consumption risk sharing between workers, bankers, and entrepreneurs. By construction, the first-best problem ignores the occupational differences reflected in the individual budget constraints, and bankers and entrepreneurs face no financial constraints. As can be shown numerically, the marginal utility gaps in the FCE are quite significant. If all consumers have separable preferences logarithmic in consumption, workers tend to consume by an order of magnitude more than bankers and entrepreneurs, despite being more patient. Thus, we can anticipate that one of the objectives of a constrained planner in our economy is to improve between-agent consumption insurance.

The labor market equilibrium in the FCE is consistent with the first best, as follows from combining \eqref{eq: workers labor supply} and \eqref{eq: entrepreneurs labor demand} and setting $P^w_{t}=1$. By defining $Q_{t}\equiv\frac{\lambda^K_{t}}{\lambda^Y_{t}}$, we see that the competitive supply of new capital goods is efficient. On the contrary, the competitive demand for capital is inefficient, as follows from comparing the FOC for $K_{t}$ to the capital Euler equation \eqref{eq: entrepreneurs Euler capital} with $P^w_{t}=1$. On the one hand, due to uncertain survival, individual entrepreneurs underestimate the social marginal benefit of capital due to its usefulness for future newborns. On the other hand, entrepreneurs find a marginal benefit in capital stock due to its value as collateral---a motive absent in the planner's problem. Moreover, entrepreneurs do not internalize the impact of their private decisions on the productive capacity of capital good producers. This latter effect is present if and only if the technology $\Phi$ is nonlinear.

To summarize, the FCE is generally first-best inefficient, manifested in the lack of between-agent consumption risk sharing and the inefficient demand for capital.

\subsubsection{Constrained efficient allocation}\label{sec: FCEA}
Now let us turn to the second-best efficiency. Following \citet{lorenzoni08}, consider a constrained efficient allocation chosen by a benevolent planner who faces the same constraints as private agents but internalizes the impact of allocations on market prices. In our flexible-price economy, we have four market prices: $Q_{t}$, $R_{t}$, $R^L_{t}$, and $W_{t}$. In the corresponding markets for factors of production and financial assets, both the market demand and supply are endogenously determined, which implies that there are multiple concepts of constrained optimality in our framework, with potentially different implications for the welfare and efficiency of the FCE. Since the worker's problem has no financial frictions, while bankers and entrepreneurs face endogenous financial constraints, we will focus on how the planner can improve over the competitive market allocation by making decisions on behalf of bankers and entrepreneurs. We will allow the planner to intervene in all the markets mentioned above, considering the most general set-up. Since our economy features consumer type heterogeneity, the sources of constrained inefficiency may not be limited to pecuniary externalities due to prices affecting the collateral constraint.

On the banker's side, the planner chooses deposits, internalizing the demand curve $R_{t}=R(U^w_{C}(C^w_{t},N_{t}),\mathbb{E}_{t}[U^w_{C}(C^w_{t+1},N_{t+1})])$ implied by the worker's Euler equation \eqref{eq: workers Euler}. Bankers still choose consumption and loans, taking the deposit allocation as given. Hence, the implementability conditions include the banker's (binding) budget constraint \eqref{eq: bankers budget constraint}, the leverage constraint \eqref{eq: bankers leverage constraint}, the Euler equation for loans \eqref{eq: bankers Euler loans}, and the complementary slackness conditions \eqref{eq: bankers complementary slackness}. These conditions can be simplified as follows. Using the budget constraint, we can solve for the loan repayment $B_{t}\equiv{}R^L_{t}L_{t-1}=C^b_{t}+L_{t}-D_{t}+R_{t-1}D_{t-1}$. The Euler equation for loans then implies $\lambda^b_{t}(1-\kappa_{t})L_{t}=U^b_{C,t}L_{t}-\beta_b\mathbb{E}_{t}(U^b_{C,t+1}B_{t+1})$. If $\kappa_{t}<1$ and $D_{t}>0$, the leverage constraint implies $L_{t}>0$, and thus the complementary slackness conditions are equivalent to $0=\lambda^b_{t}(1-\kappa_{t})L_{t}[(1-\kappa_{t})L_{t}-D_{t}]$ and $\lambda^b_{t}(1-\kappa_{t})L_{t}\ge{0}$. If $\kappa_{t}<1$ and $D_{t}=0$, the leverage constraint is equivalent to $L_{t}\ge{}0$, which is independently implied by the nonnegativity of consumption; therefore, in this case, $\lambda^b_{t}=0$, and the complementary slackness conditions are satisfied. If $\kappa_{t}=1$, the leverage constraint leaves $D_{t}=0$ as the only choice, again implying $\lambda^b_{t}=0$. Hence, if $D_{t}=0$, we have $U^b_{C,t}L_{t}=\beta_b\mathbb{E}_{t}(U^b_{C,t+1}B_{t+1})$.

On the entrepreneur's side, the planner chooses capital stock, labor, and loans, internalizing the corresponding prices. The worker's labor supply curve \eqref{eq: workers labor supply} determines the wage rate $W_{t}=W(C^w_{t},N_{t})$. The capital good producer's supply curve \eqref{eq: capital good supply} defines the price of capital $Q_{t}=Q(K_{t-1},K_{t},\xi_{t})$ after using the capital good market-clearing condition \eqref{eq: market clearing capital good} to solve for $I_{t}=I(K_{t-1},K_{t},\xi_{t})$. The return on loans must be consistent with the banker's Euler equation, one of the implementability conditions on the banker's side. Entrepreneurs themselves only make consumption decisions, which implies that the budget constraint \eqref{eq: entrepreneurs budget constraint} is binding, and entrepreneurs consume the ``endowment'' determined by the planner's choices. Apart from the binding budget constraint, the planner faces the same collateral constraint \eqref{eq: entrepreneurs collateral constraint} as the individual entrepreneur.

Based on definition \ref{def: FCE}, the only remaining implementability constraints are the market-clearing conditions \eqref{eq: market clearing wholesale good}---with $\Delta_{t}=1$---and \eqref{eq: market clearing final good}, which can be combined in one resource constraint for the final good. The constrained efficient allocation is thus defined as follows.
\begin{definition}\label{def: FCEA}
    A flexible-price constrained efficient allocation (FCEA) is a solution to
    \begin{equation*}
        \max_{\{C^b_{t},C^e_{t},C^w_{t},D_{t},K_{t},L_{t},N_{t}\}}\mathbb{E}_{0}\left(\sum_{t=0}^\infty\beta^t\sum_{i\in\mathcal{I}}\omega_i{}U^i_{t}\right)
    \end{equation*}
    subject to
    \begin{align*}
        \lambda^b_{t}:\quad{}0&\le(1-\kappa_{t})L_{t}-D_{t},\\
        \lambda^L_{1,t}:\quad{}0&\le{}U^b_{C}(C^b_{t})L_{t}-\beta_b\mathbb{E}_{t}[U^b_{C}(C^b_{t+1})(C^b_{t+1}+L_{t+1}-D_{t+1}+R_{t}D_{t})],\quad\text{equality if }D_{t}=0,\\
        \lambda^L_{2,t}:\quad{}0&=\{U^b_{C}(C^b_{t})L_{t}-\beta_b\mathbb{E}_{t}[U^b_{C}(C^b_{t+1})(C^b_{t+1}+L_{t+1}-D_{t+1}+R_{t}D_{t})]\}[(1-\kappa_{t})L_{t}-D_{t}],\\
        \lambda^C_{t}:\quad{}0&=A_{t}F(\xi_{t}K_{t-1},N_{t})-Q(K_{t-1},K_{t},\xi_{t})[K_{t}-(1-\delta)\xi_{t}K_{t-1}]-W(C^w_{t},N_{t})N_{t}+D_{t}\\
        &\quad-R_{t-1}D_{t-1}-C^b_{t}-C^e_{t},\\
        \lambda^e_{t}:\quad{}0&\le{}m_{t}\mathbb{E}_{t}(Q(K_{t},K_{t+1},\xi_{t+1})\xi_{t+1})K_{t}-\mathbb{E}_{t}(C^b_{t+1}+L_{t+1}-D_{t+1}+R_{t}D_{t}),\\
        \lambda^Y_{t}:\quad{}0&=A_{t}F(\xi_{t}K_{t-1},N_{t})-\sum_{i\in\mathcal{I}}C^i_{t}-I(K_{t-1},K_{t},\xi_{t}),
    \end{align*}
    where $R_{t}=R(U^w_{C}(C^w_{t},N_{t}),\mathbb{E}_{t}[U^w_{C}(C^w_{t+1},N_{t+1})])$, and the functions $W$, $R$, $Q$, and $I$ are defined by \eqref{eq: workers labor supply}, \eqref{eq: workers Euler}, \eqref{eq: capital good supply}, and \eqref{eq: market clearing capital good}, respectively.
\end{definition}

Definition \ref{def: FCEA} implies that the FCE is generally constrained inefficient. The collateral constraint has a conventional pecuniary externality due to the price of capital that affects the value of collateral and an externality working through the expected loan rate affected by the banker's loan supply decisions. Moreover, since we have heterogeneous consumers, only one of the budget constraints is redundant, which we chose to be the worker's. The combined budget constraint of bankers and entrepreneurs depends on market prices, resulting in additional externalities that arise even if the collateral constraint is slack with probability one. The bank leverage constraint is not a source of inefficiency, since it is independent of prices; however, the associated market complementary slackness conditions combined with the banker's loan supply Euler equation may affect the efficiency of loan demand. If the worker's preferences are not separable in consumption and leisure, the market deposit rate depends on the labor allocation, potentially creating another externality.

Let $\lambda^L_{t}\equiv\lambda^L_{1,t}+\lambda^L_{2,t}[(1-\kappa_{t})L_{t}-D_{t}]$. The following proposition formalizes the intuitive discussion above and presents other findings.
\begin{proposition}\label{prop: FCEA}
    The FCE allocation is constrained inefficient: the right-hand sides of the planner's analogs of \eqref{eq: bankers Euler deposits} and \eqref{eq: entrepreneurs labor demand}--\eqref{eq: entrepreneurs Euler capital} have additional terms $\Psi^D_{t}$, $\Psi^L_{t}$, $\Psi^N_{t}$, and $\Psi^K_{t}$. Moreover, the FCEA has the following properties.
    \begin{enumerate}
        \item There is generally imperfect consumption insurance. The risk sharing between bankers and entrepreneurs is perfect across the contingencies where $\lambda^L_{t}=\lambda^L_{t-1}=\lambda^e_{t-1}=0$.
        \item Suppose $U^w(C^w,N)=u(C^w)-v(N)$, the steady-state profits of capital good producers are zero, and $\lambda^e=0$. The steady-state Pareto-weighted marginal utilities of all consumers are equal---risk sharing is ``approximately perfect''---if and only if $u(\cdot)=\ln(\cdot)$.
        \item There exists $\Bar{D}>0$, such that any $D\in[0,\Bar{D}]$ defines an unstable steady state. The optimal constant plan in the absence of uncertainty---optimal steady state---features $D=0$, provided that $\lambda^C>0$ if $D>0$.
    \end{enumerate}
\end{proposition}

\paragraph{Wedges}
Proposition \ref{prop: FCEA} states that the FCE is constrained inefficient due to the additional terms present in the planner's optimality conditions that reflect the wedges between the FCE and FCEA. The derivation of the wedges is provided in the proof, and here we will explore their structure.

The wedge associated with deposit supply \eqref{eq: bankers Euler deposits} is
\begin{equation*}
    \Psi^D_{t}\equiv(\beta-\beta_b)R_{t}\mathbb{E}_{t}(U^b_{C,t+1})+\frac{\lambda^Y_{t}-\beta{}R_{t}\mathbb{E}_{t}(\lambda^Y_{t+1})}{\omega_b}+\Gamma^D_{t},
\end{equation*}
where $\Gamma^D_{t}$ represents all the terms that arise from the market loan supply and complementary slackness conditions, vanishing in the neighborhood of the steady state under the baseline calibration. The term $(\beta-\beta_b)R_{t}\mathbb{E}_{t}(U^b_{C,t+1})>0$ arises from the uncertain survival of bankers: the social marginal cost of deposit issuance is greater than the private marginal cost, since future newborn bankers will have to honor the liabilities of the exiting ones. The term $\lambda^Y_{t}-\beta{}R_{t}\mathbb{E}_{t}(\lambda^Y_{t+1})$ reflects the planner's risk-sharing goals and appears because, with heterogeneous consumers, both the consolidated budget constraint of bankers and entrepreneurs and the resource constraint matter to the planner. When resources are scarce, e.g., $A_{t}$ or $\xi_{t}$ is low, then $\lambda^Y_{t}$ is higher, and the resource constraint is ``more binding,'' so the planner may need to decrease the consumption of all consumers. In such states, it is more costly for bankers to borrow from the planner's perspective because the leverage constraint would require expanding assets, bank capital, and consumption. In the steady state, $\beta{}R=1$ from \eqref{eq: workers Euler}, so the risk-sharing component is zero.

If $D_{t}>0$, the wedge corresponding to the loan demand condition \eqref{eq: entrepreneurs Euler loans} is
\begin{multline*}
    \Psi^L_{t}=(\beta-\beta_e)\mathbb{E}_{t}(U^e_{C,t+1}R^L_{t+1})+\frac{\lambda^Y_{t}-\beta{}R_{t}\mathbb{E}_{t}(\lambda^Y_{t+1})}{\omega_e}-\mathbb{E}_{t}\left[\left(\beta{}U^e_{C,t+1}+\frac{\lambda^e_{t}}{\omega_e}\right)(R^L_{t+1}-R_{t})\right]\\
    +\frac{\bm{1}_\mathbb{N}(t)}{\beta}\frac{\kappa_{t}}{1-\kappa_{t}}\frac{\lambda^e_{t-1}}{\omega_e}+\Gamma^L_{t},
\end{multline*}
where $\bm{1}_\mathbb{N}(t)$ equals $1$ if $t>0$, and $\Gamma^L_{t}$ reflects the marginal effect of $L_{t}$ on the bank loan supply and private complementary slackness conditions, vanishing in the neighborhood of the steady state. The first two components of the loan wedge are symmetric to the deposit wedge. The term $-\mathbb{E}_{t}\left[\left(\beta{}U^e_{C,t+1}+\frac{\lambda^e_{t}}{\omega_e}\right)(R^L_{t+1}-R_{t})\right]\le{}0$ demonstrates that the private marginal cost of borrowing is inefficiently high when there is a positive credit spread. This component arises because the planner borrows from workers on behalf of entrepreneurs effectively at the deposit interest rate $R_{t}$, which is a consequence of aggregating the budget constraints of bankers and entrepreneurs. The term $\frac{\kappa_{t}}{1-\kappa_{t}}\frac{\lambda^e_{t-1}}{\omega_e}\ge{}0$ demonstrates that if the collateral constraint is binding at $t-1$, the planner would like to increase the marginal cost of borrowing at the continuation histories at $t$. The lower expected borrowing at $t$ decreases the expected loan rate and relaxes the collateral constraint at $t-1$. The higher the bank capital requirement, the stronger this effect, since the bank balance sheet implies a positive relationship between the return on loans and bank capital.

If $D_{t}=0$, the loan demand wedge is
\begin{equation*}
    \Psi^L_{t}=(\beta_b-\beta_e)\mathbb{E}_{t}(U^e_{C,t+1}R^L_{t+1})-\frac{\beta-\beta_b}{\beta}\frac{\lambda^e_{t}}{\omega_e}\mathbb{E}_{t}(R^L_{t+1})+\Gamma^L_{t},
\end{equation*}
where $\Gamma^L_{t}$ is generally not identical to the term present when $D_{t}>0$ but has a similar interpretation. The component $(\beta_b-\beta_e)\mathbb{E}_{t}(U^e_{C,t+1}R^L_{t+1})$ reflects potential differences in the survival rates of bankers and entrepreneurs. If entrepreneurs are relatively more patient, the planner wants to decrease the marginal cost of borrowing and allow more external financing, leading to lower net assets and consumption. Due to $-\frac{\beta-\beta_b}{\beta}\frac{\lambda^e_{t}}{\omega_e}\mathbb{E}_{t}(R^L_{t+1})\le{}0$, the marginal cost of borrowing is lower if the banker's survival is more uncertain, and the collateral constraint is binding at $t$: higher loan demand at $t$ increases the expected loan rate and net worth of newborn bankers.

The wedge relative to the planner's analog of the labor demand condition \eqref{eq: entrepreneurs labor demand} is
\begin{equation*}
    \Psi^N_{t}=\frac{(\omega_e{}U^e_{C,t}-\omega_w{}U^w_{C,t}-\lambda^C_{t})A_{t}F_{N,t}-\lambda^C_{t}W_{N,t}N_{t}}{\omega_w{}U^w_{C,t}+\lambda^C_{t}}+\Gamma^N_{t},
\end{equation*}
where $\Gamma^N_{t}$ reflects the marginal effect of the choice of labor on the interest rate $R_{t}$ and vanishes if $U^w$ is separable in consumption and leisure. There are two sources of the labor wedge: imperfect consumption risk sharing ($\omega_e{}U^e_{C,t}\neq\omega_w{}U^w_{C,t}$) and the positive shadow value of wealth ($\lambda^C_{t}>0$). (These two sources also determine the $\Gamma^N_{t}$ term as is clear from the proof of proposition \ref{prop: FCEA}.) In the first-best allocation, risk sharing is perfect, and only the resource constraint is relevant, that is, $\lambda^C_{t}=0$; therefore, the labor wedge is zero, consistent with section \ref{sec: first best}. The term $(\omega_e{}U^e_{C,t}-\omega_w{}U^w_{C,t}-\lambda^C_{t})A_{t}F_{N,t}$ reflects the differences in the marginal utility valuation of the marginal product of labor by workers and entrepreneurs. If risk sharing is ``approximately perfect,'' then $(\omega_e{}U^e_{C,t}-\omega_w{}U^w_{C,t}-\lambda^C_{t})A_{t}F_{N,t}\approx-\lambda^C_{t}A_{t}F_{N,t}<0$. Since $W$ represents the market supply curve, $W_{N,t}>0$ and $-\lambda^C_{t}W_{N,t}N_{t}<0$, reflecting that individual entrepreneurs do not internalize how their labor demand affects the equilibrium wage. Since $\omega_w{}U^w_{C,t}+\lambda^C_{t}>0$, we have $\Psi^N_{t}<0$. For a given wage, the latter means that the planner would like to decrease labor demand. At the same time, the planner would like to decrease the wage to redistribute part of the worker's labor income to entrepreneurs and achieve some convergence in Pareto-weighted marginal utilities. In turn, by lowering the wage, the planner could support a greater labor demand. Numerically, the FCE wage is inefficiently high, and the quantity of labor is inefficiently low but to a smaller extent; therefore, the FCEA entails an increase in labor supply and a slight decrease in labor demand.

The wedge relative to the planner's analog of the market demand for capital \eqref{eq: entrepreneurs Euler capital} satisfies
\begin{multline*}
    \omega_e\Psi^K_{t}=(\beta-\beta_e)\mathbb{E}_{t}(\omega_e{}U^e_{C,t+1}R^K_{t+1})Q_{t}+\beta\mathbb{E}_{t}\left\{\lambda^Y_{t+1}\left[Q_{t+1}\Phi\left(\frac{I_{t+1}}{K_{t}}\right)-\frac{I_{t+1}}{K_{t}}\right]\right\}\\
    -\lambda^C_{t}Q_{2,t}[K_{t}-(1-\delta)\xi_{t}K_{t-1}]-\beta\mathbb{E}_{t}\{\lambda^C_{t+1}Q_{1,t+1}[K_{t+1}-(1-\delta)\xi_{t+1}K_{t}]\}\\
    +\lambda^e_{t}m_{t}\mathbb{E}_{t}(Q_{1,t+1}\xi_{t+1})K_{t}+\frac{\bm{1}_\mathbb{N}(t)}{\beta}\lambda^e_{t-1}m_{t-1}Q_{2,t}\xi_{t}K_{t-1}.
\end{multline*}
Similar to other Euler equation wedges, $(\beta-\beta_e)\mathbb{E}_{t}(\omega_e{}U^e_{C,t+1}R^K_{t+1})Q_{t}>0$ reflects uncertain survival. The component $+\beta\mathbb{E}_{t}\left\{\lambda^Y_{t+1}\left[Q_{t+1}\Phi\left(\frac{I_{t+1}}{K_{t}}\right)-\frac{I_{t+1}}{K_{t}}\right]\right\}$ demonstrates that entrepreneurs do not internalize how the choice of capital affects the future profits of capital good producers through a nonlinear technology $\Phi$, which, in turn, affects the amount of resources available for all consumers and is valued at the shadow value of output. If the steady-state profits are zero, so is this wedge component, but in the neighborhood of the steady state, it is not generally zero. The sign of $\lambda^Y_{t+1}$ is generally ambiguous but typically positive. The next two terms $-\lambda^C_{t}Q_{2,t}[K_{t}-(1-\delta)\xi_{t}K_{t-1}]\le{}0$ and $-\beta\mathbb{E}_{t}\{\lambda^C_{t+1}Q_{1,t+1}[K_{t+1}-(1-\delta)\xi_{t+1}K_{t}]\}\ge{}0$ reflect the marginal effect of an increase in capital stock at $t$ on the aggregate wealth of bankers and entrepreneurs at $t$ and $t+1$, respectively, transmitted through the price of capital. Finally, there are pecuniary externalities present in the collateral constraint. First, $\lambda^e_{t}m_{t}\mathbb{E}_{t}(Q_{1,t+1}\xi_{t+1})K_{t}\le{}0$ reflects a lower social marginal benefit of capital at $t$ due to a lower collateral asset price at $t+1$, stemming from the concave capital good technology $\Phi$. On the contrary, $\lambda^e_{t-1}m_{t-1}Q_{2,t}\xi_{t}K_{t-1}\ge{}0$ represents an additional marginal benefit of capital at $t$ due to a higher asset price and the value of collateral expected at $t-1$.

\paragraph{Risk sharing and the optimal steady state}
Consider the remaining parts of proposition \ref{prop: FCEA}. First, we do not generally have perfect consumption insurance between all types of consumers at the second best. Across the contingencies where the market loan supply and the banker's private complementary slackness conditions are slack at $t$ and $t-1$, and the collateral constraint is slack at $t-1$, insurance between bankers and entrepreneurs is perfect. The reason is that in this case, $C^b_{t}$ and $C^e_{t}$ affect the planner's budget set in an identical linear way through the consolidated budget constraint of bankers and entrepreneurs and the final good resource constraint.

Second, there is a special case when we do have approximately perfect between-agent insurance in the neighborhood of the steady state. The latter holds when workers have separable preferences over consumption and leisure with a unit constant relative risk aversion, capital good producers earn zero profits in the steady state, and the steady-state collateral constraint is slack. A sufficient condition for zero steady-state profits is $Q=\xi=1$ and $\frac{I}{K}=\delta$---standard normalizations or calibration targets. Although insurance between workers and constrained consumers is only approximately perfect, the correlation between marginal utilities is quantitatively close to one.

Third, the FCEA is locally indeterminate: any $D\in[0,\Bar{D}]$ defines a steady state, where $\Bar{D}$ corresponds to the case when either the collateral constraint or the bank leverage constraint is binding. The multiplicity is resolved if we consider the optimal constant plan in the absence of uncertainty. Any FCEA steady state satisfies the planner's constraints in the absence of uncertainty, being a feasible constant plan. It turns out that $D=0$ is part of the optimal plan if the consolidated budget constraint is relevant to the planner, that is, if $\lambda^C>0$. Intuitively, by decreasing the quantity of deposits, the planner can allocate more consumption to bankers and entrepreneurs because $-(R-1)D\le{}0$. Since $L$ must decrease to satisfy the private complementary slackness conditions, the collateral constraint is relaxed. To satisfy the resource constraint, the planner can increase both labor and the worker's consumption to achieve a Pareto improvement relative to any constant plan with $D>0$. Thus, we can restrict attention to the steady state corresponding to $D=0$---the optimal steady state.

\paragraph{Decentralization}
Consider now how to decentralize the FCEA in a regulated FCE. A natural way to address the wedges is through a proportional taxation rebated lump sum, as described in the following proposition.
\begin{proposition}\label{prop: FCEA decentralization}
    The FCEA can be decentralized in a regulated FCE with linear taxes rebated lump sum. Compared to the FCE, the banker's budget constraint is modified as
    \begin{equation*}
        C^b_{t}+L_{t}\le{}R^L_{t}L_{t-1}-R_{t-1}D_{t-1}+(1-\tau^D_{t})D_{t}+T^b_{t},
    \end{equation*}
    where $\tau^D_{t}$ and $T^b_{t}\equiv\tau^D_{t}D_{t}$ are taken as given by the individual banker. The entrepreneur's budget constraint is modified as
    \begin{multline*}
        C^e_{t}+(1+\tau^K_{t})Q_{t}K_{t}+(1+\tau^N_{t})W_{t}N_{t}+R^L_{t}L_{t-1}\\
        \le{}A_{t}F(\xi_{t}K_{t-1},N_{t})+Q_{t}(1-\delta)\xi_{t}K_{t-1}+(1-\tau^L_{t})L_{t}+T^e_{t},
    \end{multline*}
    where $(\tau^K_{t},\tau^N_{t},\tau^L_{t})$ and $T^e_{t}\equiv\tau^K_{t}Q_{t}K_{t}+\tau^N_{t}W_{t}N_{t}+\tau^L_{t}L_{t}$ are taken as given by the individual entrepreneur. The taxes defined in terms of the FCEA are
    \begin{gather*}
        \tau^D_{t}=\frac{1}{U^b_{C,t}}\left[\frac{\lambda^b_{t}}{\omega_b}-\frac{U^b_{C,t}-\beta_b\mathbb{E}_{t}(U^b_{C,t+1}R^L_{t+1})}{1-\kappa_{t}}+\Psi^D_{t}\right],\\
        \tau^N_{t}=\frac{-\Psi^N_{t}}{W_{t}},\qquad
        \tau^L_{t}=\frac{\Psi^L_{t}}{U^e_{C,t}},\qquad
        \tau^K_{t}=\frac{-\Psi^K_{t}}{U^e_{C,t}Q_{t}}.
    \end{gather*}
    Furthermore, the FCEA and $\{\tau^D_{t},\tau^N_{t},\tau^L_{t},\tau^K_{t},T^d_{t},T^e_{t}\}$ defined above constitute the allocation-policy pair chosen by the Ramsey planner that selects the best regulated FCE.
\end{proposition}
The taxes applied to entrepreneurs are simple functions of the wedges. The deposit supply tax $\tau^D_{t}$ reflects potential differences between the normalized social Lagrange multiplier on the bank leverage constraint $\frac{\lambda^b_{t}}{\omega_b}$ and the private Lagrange multiplier $\frac{U^b_{C,t}-\beta_b\mathbb{E}_{t}(U^b_{C,t+1}R^L_{t+1})}{1-\kappa_{t}}$ expressed based on \eqref{eq: bankers Euler loans}. Since $\Psi^N_{t}<0$, it must be that $\tau^N_{t}>0$: it is optimal to tax the entrepreneur's labor demand. The signs of the other wedges and taxes are generally ambiguous, necessitating quantitative analysis. The policy that decentralizes the FCEA is Ramsey optimal. Moreover, any additional taxation instruments cannot improve over the second-best optimum unless the planner can directly set prices instead of internalizing the price functions arising in the competitive markets.

\subsubsection{Optimal leverage limits}\label{sec: optimal leverage limits}
Since $\{\kappa_{t},m_{t}\}$ are exogenous to the FCE, we have so far considered them as given. Let us now study how to set these policies optimally. We will focus on two cases based on whether the Ramsey planner can address all distortions with the complete set of taxes $\{\tau^D_{t},\tau^N_{t},\tau^L_{t},\tau^K_{t},T^d_{t},T^e_{t}\}$ or the planner can only account for the intratemporal labor wedge with $\{\tau^N_{t},T^e_{t}\}$. Loosely speaking, the first case corresponds to finding the best FCEA by setting the leverage limits optimally. In the second case, the regulated FCE is constrained inefficient, and we can explore the merits of state-contingent leverage limits in mitigating the Euler equation distortions.

By proposition \ref{prop: FCEA decentralization}, conditional on $\{\kappa_{t},m_{t}\}$, setting $\{\tau^D_{t},\tau^N_{t},\tau^L_{t},\tau^K_{t},T^d_{t},T^e_{t}\}$ optimally\linebreak amounts to solving for the FCEA. Suppose the Ramsey planner can also optimize with respect to $\{\kappa_{t},m_{t}\}$. Since the leverage limits determine the strictness of inequality constraints, the optimal $\{\kappa_{t},m_{t}\}$ are generally not unique: if a leverage constraint is slack at a specific leverage limit, it is also slack at any other feasible looser limit. The associated Ramsey allocation, however, is typically unique and can be characterized using the primal approach as stated in the following lemma.
\begin{lemma}\label{lemma: OLL relaxed with unrestricted taxation}
    An allocation $\{C^b_{t},C^e_{t},C^w_{t},D_{t},K_{t},L_{t},N_{t}\}$ and policy $\{\kappa_{t},m_{t},\tau^D_{t},\tau^N_{t},\tau^L_{t},\tau^K_{t},\linebreak{}T^d_{t},T^e_{t}\}$ are part of a Ramsey equilibrium associated with the regulated FCE of proposition \ref{prop: FCEA decentralization} if and only if the allocation $\{C^b_{t},C^e_{t},C^w_{t},D_{t},K_{t},L_{t},N_{t}\}$ is a solution to a relaxed problem based on definition \ref{def: FCEA} but with $\kappa_{t}=0$, $m_{t}=1$, and no constraint corresponding to $\lambda^L_{2,t}$. Conditional on the allocation, the policy is defined as follows. Set $\kappa_{t}\equiv{}1-\frac{D_{t}}{L_{t}}$ if $U^b_{C,t}>\beta_b\mathbb{E}_{t}(U^b_{C,t+1}R^L_{t+1})$; otherwise, choose any $\kappa_{t}\in\left[0,1-\frac{D_{t}}{L_{t}}\right]$; choose any $m_{t}\in\left[\frac{\mathbb{E}_{t}(R^L_{t+1})L_{t}}{\mathbb{E}_{t}(Q_{t+1}\xi_{t+1})K_{t}},1\right]$; set the taxes to satisfy the regulated analogs of \eqref{eq: bankers Euler deposits} and \eqref{eq: entrepreneurs labor demand}--\eqref{eq: entrepreneurs Euler capital}, rebating them lump sum.
\end{lemma}

According to lemma \ref{lemma: OLL relaxed with unrestricted taxation}, without loss of generality, we can focus on the leverage limits that make the market leverage constraints binding: we can always set $\kappa_{t}\equiv{}1-\frac{D_{t}}{L_{t}}$ and $m_{t}\equiv\frac{\mathbb{E}_{t}(R^L_{t+1})L_{t}}{\mathbb{E}_{t}(Q_{t+1}\xi_{t+1})K_{t}}$. The construction of $\{\kappa_{t}\}$ ensures that the $\lambda^L_{2,t}$ constraint of definition \ref{def: FCEA} is satisfied. The relaxed problem of lemma \ref{lemma: OLL relaxed with unrestricted taxation} has a larger feasible set than the problem of definition \ref{def: FCEA}; therefore, the Ramsey allocation with optimal leverage limits weakly dominates any FCEA associated with a given policy $\{\kappa_{t},m_{t}\}$, unless the leverage constraints under $\{\kappa_{t},m_{t}\}$ are slack with probability one. It is straightforward to show that the Ramsey allocation of lemma \ref{lemma: OLL relaxed with unrestricted taxation} has the risk-sharing and steady-state properties described in proposition \ref{prop: FCEA}.

Consider the second case when only $\{\kappa_{t},m_{t},\tau^N_{t},T^e_{t}\}$ are available. Now we cannot dispense with the Euler equations \eqref{eq: bankers Euler deposits}, \eqref{eq: entrepreneurs Euler loans}, and \eqref{eq: entrepreneurs Euler capital}. To simplify the problem, we can use \eqref{eq: bankers Euler deposits} and \eqref{eq: entrepreneurs Euler loans} to solve for the private Lagrange multipliers and then use the private complementary slackness conditions to rearrange the Euler equations \eqref{eq: bankers Euler loans} and \eqref{eq: entrepreneurs Euler capital}, expressing them in terms of allocations and price functions. As in lemma \ref{lemma: OLL relaxed with unrestricted taxation}, we can characterize the Ramsey problem entirely in terms of choosing allocations.
\begin{lemma}\label{lemma: OLL relaxed with labor taxation}
    An allocation $\{C^b_{t},C^e_{t},C^w_{t},D_{t},K_{t},L_{t},N_{t}\}$ and policy $\{\kappa_{t},m_{t},\tau^N_{t},T^e_{t}\}$ are part of a Ramsey equilibrium associated with the regulated FCE of proposition \ref{prop: FCEA decentralization}---after imposing $\tau^D_{t}=\tau^L_{t}=\tau^K_{t}=0$---if and only if the allocation $\{C^b_{t},C^e_{t},C^w_{t},D_{t},K_{t},L_{t},N_{t}\}$ is a solution to a corresponding relaxed problem. Conditional on the allocation, the policy is defined as follows. If $U^b_{C,t}>\beta_b{}R_{t}\mathbb{E}_{t}(U^b_{C,t+1})$, set $\kappa_{t}\equiv{}1-\frac{D_{t}}{L_{t}}$; otherwise, choose any $\kappa_{t}\in\left[0,1-\frac{D_{t}}{L_{t}}\right]$. If $U^e_{C,t}>\beta_e\mathbb{E}_{t}(U^e_{C,t+1}R^L_{t+1})$, set $m_{t}\equiv\frac{\mathbb{E}_{t}(R^L_{t+1})L_{t}}{\mathbb{E}_{t}(Q_{t+1}\xi_{t+1})K_{t}}$; otherwise, choose any $m_{t}\in\left[\frac{\mathbb{E}_{t}(R^L_{t+1})L_{t}}{\mathbb{E}_{t}(Q_{t+1}\xi_{t+1})K_{t}},1\right]$. Set $\tau^N_{t}\equiv\frac{A_{t}F_{N,t}}{W_{t}}-1$ and $T^e_{t}\equiv\tau^N_{t}W_{t}N_{t}$.
\end{lemma}

As shown in the proof of lemma \ref{lemma: OLL relaxed with labor taxation}, the constraints corresponding to the banker's and entrepreneur's problems have a certain symmetry: in both cases, we have a leverage constraint, an asset Euler equation expressed in terms of allocations, and a constraint that requires the private Lagrange multiplier on the leverage constraint to be nonnegative; finally, we have a consolidated budget constraint. The symmetry is imperfect: while the banker's Euler equation implies that bank capital is equal to the expected discounted value of the stream of consumption, as in lemma \ref{lemma: net assets}, the entrepreneur's Euler equation does not produce a similar relationship, provided there is a nontrivial labor wedge addressed by the tax $\tau^N_{t}$. Compared to lemma \ref{lemma: OLL relaxed with unrestricted taxation}, the construction of the LTV ratio in lemma \ref{lemma: OLL relaxed with labor taxation} must be consistent with the entrepreneur's private complementary slackness conditions. The following proposition summarizes some implications of the Ramsey problem in lemma \ref{lemma: OLL relaxed with labor taxation}.
\begin{proposition}\label{prop: optimal leverage limits with labor taxation}
    An optimal allocation-policy pair in the Ramsey problem of lemma \ref{lemma: OLL relaxed with labor taxation} generally has imperfect consumption insurance. There is approximately perfect risk sharing between bankers and entrepreneurs if the relaxed collateral constraint is slack in the steady state. If, moreover, $U^w(C^w,N)=\ln(C^w)-v(N)$ and the steady-state profits of capital good producers are zero, there is approximate insurance across all consumers. A steady state is generally unique.
\end{proposition}

Similar to the FCEA and the Ramsey allocation of lemma \ref{lemma: OLL relaxed with unrestricted taxation}, there is generally imperfect consumption insurance, but it is approximately perfect under the same conditions. A difference from the former allocations is that even if the relaxed collateral constraint is slack in the neighborhood of the steady state, risk sharing between bankers and entrepreneurs is only approximate. At the same time, the relaxed leverage constraints generate a larger feasible set of leverage ratios for bankers and entrepreneurs, potentially enhancing risk sharing relative to the FCE. In contrast to the FCEA and the Ramsey allocation of lemma \ref{lemma: OLL relaxed with unrestricted taxation}, the allocation of lemma \ref{lemma: OLL relaxed with labor taxation} generally has a unique steady state with $D>0$, which is a consequence of respecting the intertemporal Euler equations and the arguments related to the proof of lemma \ref{lemma: steady state uniqueness}.

\subsection{Sticky-price economy}
Consider now the general environment with nominal rigidities. Given the analysis in section \ref{sec: efficiency in a flexible-price economy}, the exposition can be significantly simplified. Apart from exploring the implications of nominal rigidities for constrained efficiency, the main objective of this subsection is to characterize jointly optimal leverage limits and monetary policy.

\subsubsection{Constrained efficient allocation}\label{sec: CEA}
Compared to the flexible-price economy, we have two additional markets: wholesale goods and retail varieties. Retailers act as monopolists, internalizing the demand curve of the final good producers; hence, the only additional way to achieve an improvement over the CE allocation is to intervene in the competitive market for wholesale goods.

A social planner, making decisions on behalf of the entrepreneur, internalizes the determination of the wholesale good price $P^w_{t}$ from the retailer's optimality conditions. Like the individual agents, the planner takes policies $\{\kappa_{t},m_{t},\Pi_{t}\}$ as given. If inflation is pinned down by a Taylor rule \eqref{eq: Taylor rule} in the CE, then it must be so in the centralized allocation. In this case, \eqref{eq: Taylor rule} must not be part of the planner's implementability conditions: instead, it augments the planner's optimality conditions. Note that \eqref{eq: retailers inflation} yields a conditional solution for the retailer's optimal relative price $\widetilde{P}_{t}=\widetilde{P}(\Pi_{t})$, which allows constructing the price dispersion sequence $\{\Delta_{t}\}$ recursively based on $\{\Pi_{t},\widetilde{P}_{t}\}$ and an initial condition $\Delta_{-1}$, using \eqref{eq: retailers price dispersion}. Hence, effectively, the planner takes as given $\{\Delta_{t},\Pi_{t},\widetilde{P}_{t}\}$. Using \eqref{eq: retailers optimal relative price} and \eqref{eq: market clearing wholesale good}, we can solve for the measure of the retailer's marginal benefit $\Omega_{2,t}=\frac{\epsilon}{\epsilon-1}\frac{\Omega_{1,t}}{\widetilde{P}_{t}}$ and final good output $Y_{t}=\frac{A_{t}}{\Delta_{t}}F(\xi_{t}K_{t-1},N_{t})$. Then \eqref{eq: retailers marginal cost} defines the retailer's demand curve for wholesale goods:
\begin{equation*}
    P^w_{t}=\frac{\Delta_{t}}{A_{t}F(\xi_{t}K_{t-1},N_{t})}\left\{\Omega_{1,t}-\frac{\beta\theta\mathbb{E}_{t}[U^w_{C}(C^w_{t+1},N_{t+1})\Pi_{t+1}^{\epsilon}\Omega_{1,t+1}]}{U^w_{C}(C^w_{t},N_{t})}\right\}.
\end{equation*}

Relative to the FCEA problem, we have one additional control variable---a measure of the retailer's marginal cost $\Omega_{1,t}$. Similarly, there is one additional implementability condition---the recursive definition of the retailer's marginal benefit \eqref{eq: retailers marginal benefit}. Since $\Omega_{1,t}$ is an auxiliary variable, the set of potential wedges does not change. The constrained efficient allocation can then be defined as follows.
\begin{definition}\label{def: CEA}
    A constrained efficient allocation (CEA) is a solution to
    \begin{equation*}
        \max_{\{C^b_{t},C^e_{t},C^w_{t},D_{t},K_{t},L_{t},N_{t}\textcolor{BrickRed}{,\Omega_{1,t}}\}}\mathbb{E}_{0}\left(\sum_{t=0}^\infty\beta^t\sum_{i\in\mathcal{I}}\omega_i{}U^i_{t}\right)
    \end{equation*}
    subject to the same constraints as in definition \ref{def: FCEA}---with the consolidated budget constraint of bankers and entrepreneurs and the resource constraint modified as shown below---and \eqref{eq: retailers marginal benefit}. The modified and additional constraints are
    \begin{align*}
        \lambda^C_{t}:\quad{}0&=\textcolor{BrickRed}{\Delta_{t}\left\{\Omega_{1,t}-\frac{\beta\theta\mathbb{E}_{t}[U^w_{C}(C^w_{t+1},N_{t+1})\Pi_{t+1}^{\epsilon}\Omega_{1,t+1}]}{U^w_{C}(C^w_{t},N_{t})}\right\}}-Q(K_{t-1},K_{t},\xi_{t})[K_{t}\\
        &\quad-(1-\delta)\xi_{t}K_{t-1}]-W(C^w_{t},N_{t})N_{t}+D_{t}-R_{t-1}D_{t-1}-C^b_{t}-C^e_{t},\\
        \lambda^Y_{t}:\quad{}0&=\frac{A_{t}}{\textcolor{BrickRed}{\Delta_{t}}}F(\xi_{t}K_{t-1},N_{t})-\sum_{i\in\mathcal{I}}C^i_{t}-I(K_{t-1},K_{t},\xi_{t}),\\
        \textcolor{BrickRed}{\lambda^\Omega_{t}}:\quad{}0&=\frac{\epsilon-1}{\epsilon}\widetilde{P}_{t}\frac{A_{t}}{\Delta_{t}}F(\xi_{t}K_{t-1},N_{t})-\Omega_{1,t}+\frac{\beta\theta\mathbb{E}_{t}\left[U^w_{C}(C^w_{t+1},N_{t+1})\Pi_{t+1}^{\epsilon-1}\frac{\widetilde{P}_{t}}{\widetilde{P}_{t+1}}\Omega_{1,t+1}\right]}{U^w_{C}(C^w_{t},N_{t})},
    \end{align*}
    where $\{\Delta_{t},\Pi_{t},\widetilde{P}_{t}\}$ are taken as given.
\end{definition}

For the convenience of the reader using the electronic version of this paper, the modifications relative to the flexible-price economy are in color. The implementability conditions on the banker's side and the collateral constraint on the entrepreneur's side are identical to the flexible-price case. Consequently, the financial wedges corresponding to deposit supply and loan demand will be identical to those in the FCEA. The real wedges corresponding to the entrepreneur's demand for factors of production do change in the sticky-price economy. First, the planner internalizes how the optimal retail pricing affects the relative price of wholesale goods $P^w_{t}$, which directly affects the entrepreneur's revenue and the combined income of constrained consumers. Second, there is an output loss due to price dispersion $\Delta_{t}\ge{}1$, which limits the consumption of all consumers. (In the flexible-price economy, we have $P^w_{t}=\Delta_{t}=1$ for all $t\ge{}0$.) The following proposition formalizes this discussion and compares the CEA and FCEA.
\begin{proposition}\label{prop: CEA}
    The CE allocation is constrained inefficient, reflected in additional terms $\Psi^D_{t}$, $\Psi^L_{t}$, $\Psi^N_{t}$, and $\Psi^K_{t}$, as in proposition \ref{prop: FCEA}. The financial wedges $\Psi^D_{t}$ and $\Psi^L_{t}$ are identical to those in the flexible-price economy. There is perfect consumption insurance between bankers and entrepreneurs when the collateral constraint is slack. The CEA is locally indeterminate, and the optimal steady state features $D=0$. The CEA can be decentralized in a regulated CE with linear taxes $\{\tau^D_{t},\tau^N_{t},\tau^L_{t},\tau^K_{t},T^d_{t},T^e_{t}\}$ identically to proposition \ref{prop: FCEA decentralization}, and the policy is Ramsey optimal.
\end{proposition}

Unlike in the FCEA, we do not have a special case of approximate full risk sharing in the CEA because we would need to have zero steady-state profits of retailers. Since retailers act as monopolists, setting a time-varying markup over the marginal cost, their steady-state profits are positive. The other CEA properties are identical to those of the FCEA, except for the differences in real wedges.

The labor wedge is now
\begin{equation*}
    \Psi^N_{t}=\frac{[(\omega_e{}U^e_{C,t}\textcolor{BrickRed}{+\lambda^\Omega_{t}\frac{\epsilon-1}{\epsilon}\widetilde{P}_{t}-\lambda^C_{t}})\textcolor{BrickRed}{(P^w_{t}\Delta_{t})^{-1}}-\omega_w{}U^w_{C,t}-\lambda^C_{t}]\textcolor{BrickRed}{P^w_{t}}A_{t}F_{N,t}-\lambda^C_{t}W_{N,t}N_{t}}{\omega_w{}U^w_{C,t}+\lambda^C_{t}}+\Gamma^N_{t},
\end{equation*}
where $\Gamma^N_{t}$ is identical to the one in the FCEA. The marginal product of labor is now priced at $P^w_{t}<1$, and the marginal utility gap between workers and entrepreneurs is affected by nominal rigidities. The term $\lambda^\Omega_{t}\frac{\epsilon-1}{\epsilon}\widetilde{P}_{t}-\lambda^C_{t}$ arises because the consolidated budget constraint does not directly contain the entrepreneur's output---it is now present in the constraint that reflects the retailer's marginal benefit. In the steady state, $\lambda^\Omega\widetilde{P}=\lambda^C$, and thus typically $\lambda^\Omega_{t}\frac{\epsilon-1}{\epsilon}\widetilde{P}_{t}-\lambda^C_{t}<0$. Moreover, there is a multiplicative factor $(P^w_{t}\Delta_{t})^{-1}$, greater than unity under a reasonable calibration. Quantitatively, the second effect dominates and $(\omega_e{}U^e_{C,t}+\lambda^\Omega_{t}\frac{\epsilon-1}{\epsilon}\widetilde{P}_{t}-\lambda^C_{t})(P^w_{t}\Delta_{t})^{-1}>\omega_e{}U^e_{C,t}$, which implies that the difference between the marginal utility gap and the shadow value of wealth tends to decrease, and so does the magnitude of the labor wedge.

The capital wedge now satisfies
\begin{multline*}
    \omega_e\Psi^K_{t}=(\beta-\beta_e)\mathbb{E}_{t}(\omega_e{}U^e_{C,t+1}\textcolor{BrickRed}{R^K_{t+1}})Q_{t}+\beta\mathbb{E}_{t}\left\{\lambda^Y_{t+1}\left[Q_{t+1}\Phi\left(\frac{I_{t+1}}{K_{t}}\right)-\frac{I_{t+1}}{K_{t}}\right]\right\}\\
    -\lambda^C_{t}Q_{2,t}[K_{t}-(1-\delta)\xi_{t}K_{t-1}]-\beta\mathbb{E}_{t}\{\lambda^C_{t+1}Q_{1,t+1}[K_{t+1}-(1-\delta)\xi_{t+1}K_{t}]\}\\
    +\lambda^e_{t}m_{t}\mathbb{E}_{t}(Q_{1,t+1}\xi_{t+1})K_{t}+\frac{\bm{1}_\mathbb{N}(t)}{\beta}\lambda^e_{t-1}m_{t-1}Q_{2,t}\xi_{t}K_{t-1}\\
    \textcolor{BrickRed}{+\beta\mathbb{E}_{t}\left\{\left[\omega_e{}U^e_{C,t+1}(1-P^w_{t+1}\Delta_{t+1})+\lambda^\Omega_{t+1}\frac{\epsilon-1}{\epsilon}\widetilde{P}_{t+1}-\lambda^C_{t+1}\right]\frac{A_{t+1}}{\Delta_{t+1}}F_{K,t+1}\xi_{t+1}\right\}}.
\end{multline*}
The effects of nominal rigidities parallel those for the labor wedge but applied to the effective marginal product of capital. First, $P^w_{t+1}<1$ affects the future return on capital. Second, $\lambda^\Omega_{t+1}\frac{\epsilon-1}{\epsilon}\widetilde{P}_{t+1}-\lambda^C_{t+1}<0$. Third, $\omega_e{}U^e_{C,t+1}(1-P^w_{t+1}\Delta_{t+1})>0$. Quantitatively, the magnitude of the wedge tends to increase.

\subsubsection{Optimal monetary policy and leverage limits}
Consider now how to set $\{\kappa_{t},m_{t},\Pi_{t}\}$ optimally. In section \ref{sec: optimal leverage limits}, we argued that the optimal leverage limits are not unique, but the corresponding allocation is uniquely determined as a solution to a relaxed planning problem. We can use a similar approach here, except inflation $\{\Pi_{t}\}$ will be a control variable. Although \eqref{eq: retailers price dispersion} allows us to construct $\{\Delta_{t}\}$ conditional on $\{\Pi_{t}\}$ and an initial condition, any $\Delta_{t}$ will be history dependent, complicating the optimization with respect to inflation. It is more tractable to add $\{\Delta_{t}\}$ to the set of controls and \eqref{eq: retailers price dispersion} to the implementability conditions. Furthermore, we will allow for an ELB. Our relaxed problems will thus feature the additional constraints\footnote{Note that the ELB constraint highlights an aggregate demand externality that was absent in the comparison of the CE and CEA allocations. This externality is similar to that emphasized in \citet{farhi16} and \citet{korinek16}. Unlike the Ramsey planner, the individual agents do not internalize that their consumption-saving choice affects the strictness of the ELB constraint through the worker's Euler equation \eqref{eq: workers Euler}. A thorough analysis of this externality in our environment is left for future research.}
\begin{align*}
    \textcolor{BrickRed}{\lambda^\Delta_{t}}:\quad{}0&=\theta\Pi_{t}^{\epsilon}\Delta_{t-1}+(1-\theta)(\widetilde{P}(\Pi_{t}))^{-\epsilon}-\Delta_{t},\\
    \textcolor{BrickRed}{\lambda^R_{t}}:\quad{}0&\le{}R_{t}\mathbb{E}_{t}(\Pi_{t+1})-\underline{R}.
\end{align*}

As in section \ref{sec: optimal leverage limits}, consider two alternative Ramsey problems. The first problem---case 1---allows the Ramsey planner to set $\{\kappa_{t},m_{t},\Pi_{t},\tau^D_{t},\tau^N_{t},\tau^L_{t},\tau^K_{t},T^d_{t},T^e_{t}\}$ optimally. By proposition \ref{prop: CEA}, conditional on $\{\kappa_{t},m_{t},\Pi_{t}\}$, we get a CEA if the ELB constraint is slack with probability one. Therefore, the case 1 Ramsey allocation approximately corresponds to the best CEA. The constraint set of the relaxed problem is formed by taking the constraints from lemma \ref{lemma: OLL relaxed with unrestricted taxation}, modifying the $\lambda^C_{t}$ and $\lambda^Y_{t}$ constraints and adding the $\lambda^\Omega_{t}$ constraint as in definition \ref{def: CEA}, and adding the $\lambda^\Delta_{t}$ and $\lambda^R_{t}$ constraints above. The ELB affects the worker's consumption through the real interest rate $R_{t}$, but there is no direct effect on bankers and entrepreneurs; therefore, the case 1 Ramsey allocation has the same risk-sharing and steady-state properties as the CEA. Let us postpone the characterization of optimal monetary policy until after we have described our alternative problem.

The second problem---case 2---has only $\{\kappa_{t},m_{t},\Pi_{t},\tau^N_{t},T^e_{t}\}$ as policy instruments. The case 2 Ramsey allocation is thus constrained inefficient. The constraint set in the relaxed problem is formed by taking the constraints from lemma \ref{lemma: OLL relaxed with labor taxation}, modifying or adding the $\lambda^C_{t}$, $\lambda^Y_{t}$, $\lambda^\Omega_{t}$, $\lambda^\Delta_{t}$, and $\lambda^R_{t}$ constraints identically to case 1, and modifying the $\lambda^K_{t}$ constraint as follows:
\begin{multline*}
    \lambda^K_{t}:\quad{}0=\beta_e\mathbb{E}_{t}\biggl\{U^e_{C}(C^e_{t+1})\biggl[\textcolor{BrickRed}{\alpha\Delta_{t+1}\left\{\Omega_{1,t+1}-\frac{\beta\theta\mathbb{E}_{t+1}[U^w_{C}(C^w_{t+2},N_{t+2})\Pi_{t+2}^{\epsilon}\Omega_{1,t+2}]}{U^w_{C}(C^w_{t+1},N_{t+1})}\right\}}\\
    +Q(K_{t},K_{t+1},\xi_{t+1})(1-\delta)\xi_{t+1}K_{t}-C^b_{t+1}-L_{t+1}+D_{t+1}-R_{t}D_{t}\biggr]\biggr\}\\
    -U^e_{C}(C^e_{t})(Q(K_{t-1},K_{t},\xi_{t})K_{t}-L_{t}),
\end{multline*}
where $\alpha\equiv\frac{F_{K}(\xi_{t}K_{t-1},N_{t})\xi_{t}K_{t-1}}{F(\xi_{t}K_{t-1},N_{t})}$ is the capital share. (It is constant because $F$ is Cobb---Douglas.) The modified term corresponds to $P^w_{t+1}A_{t+1}F_{K,t+1}\xi_{t+1}K_{t}$, reflecting the determination of $P^w_{t}$ from the retailer's problem. As in proposition \ref{prop: optimal leverage limits with labor taxation}, the case 2 Ramsey allocation has partial risk sharing between bankers and entrepreneurs in the neighborhood of the steady state if the relaxed collateral constraint is slack, and there typically exists a unique steady state with $D>0$.

Note that inflation $\{\Pi_{t}\}$ affects the planner's constraints in both problems identically with one exception: in case 2, future inflation affects the future return on capital in the $\lambda^K_{t}$ constraint through $P^w_{t+1}$. It turns out that we can define an auxiliary variable that captures the combined shadow value of the effect through $P^w_{t+1}$ with the effect through $P^w_{t}$, where the latter is common to both problems. Conditional on this auxiliary variable, the optimal monetary policy has identical long-run and short-run characteristics. The following proposition summarizes and formalizes our discussion.
\begin{proposition}\label{prop: optimal leverage limits and monetary policy}
    The case 1 and 2 Ramsey allocations have the risk-sharing and steady-state properties of Propositions \ref{prop: CEA} and \ref{prop: optimal leverage limits with labor taxation}, respectively, except for the special case of approximate full insurance.
    
    In both cases, the optimal monetary policy is characterized as follows. The long-run gross inflation rate in the absence of uncertainty is uniquely determined as
    \begin{equation*}
        \Pi=
        \begin{cases}
            1 & \text{if }\underline{R}\le\frac{1}{\beta},\\
            \beta\underline{R} & \text{if }\underline{R}>\frac{1}{\beta}.
        \end{cases}
    \end{equation*}
    The short-run inflation behavior is represented by the Euler equation
    \begin{multline*}
        0=\lambda^\Omega_{t}\widetilde{P}'(\Pi_{t})\left[\frac{\epsilon-1}{\epsilon}Y_{t}+\frac{\beta\theta\mathbb{E}_{t}\left(U^w_{C,t+1}\Pi_{t+1}^{\epsilon-1}\frac{\Omega_{1,t+1}}{\widetilde{P}_{t+1}}\right)}{U^w_{C,t}}\right]+\lambda^\Delta_{t}\epsilon\left[\theta\Pi_{t}^{\epsilon-1}\Delta_{t-1}-(1-\theta)\frac{\widetilde{P}'(\Pi_{t})}{\widetilde{P}_{t}^{\epsilon+1}}\right]\\
        -\bm{1}_\mathbb{N}(t)\theta\Pi_{t}^{\epsilon-1}\Omega_{1,t}\frac{U^w_{C,t}}{U^w_{C,t-1}}\left[\widetilde{\lambda}^C_{t-1}\Delta_{t-1}\epsilon-\lambda^\Omega_{t-1}\frac{\widetilde{P}_{t-1}}{\widetilde{P}_{t}}\left(\frac{\epsilon-1}{\Pi_{t}}-\frac{\widetilde{P}'(\Pi_{t})}{\widetilde{P}_{t}}\right)\right]+\frac{\bm{1}_\mathbb{N}(t)}{\beta}\lambda^R_{t-1}R_{t-1},
    \end{multline*}
    where $\widetilde{\lambda}^C_{t}\equiv\lambda^C_{t}$ in case 1, and $\widetilde{\lambda}^C_{t}\equiv\lambda^C_{t}+\frac{\bm{1}_\mathbb{N}(t)}{\beta}\lambda^K_{t-1}\beta_e{}U^e_{C,t}\alpha$ in case 2.
\end{proposition}

Since an ELB typically satisfies $\underline{R}\le{}1<\frac{1}{\beta}$, by proposition \ref{prop: optimal leverage limits and monetary policy}, the long-run price stability is optimal independently of whether the planner can address the intertemporal distortions. One can demonstrate that steady-state price stability is optimal even if inflation is the only policy instrument. Stable prices eliminate output losses due to price dispersion $\Delta\ge{}1$, and the optimality of price stability is consistent with the normative analyses of the basic New Keynesian economies \citep{schmitt-grohe10,woodford10}. Moreover, \citet{curdia16} obtained a similar result in a model with a credit spread friction that may be viewed as an approximation of the endogenous credit spread that arises in our model due to the leverage constraint. \citet{coibion12} found that a slightly positive steady-state inflation could be optimal in the context of Taylor rules with an ELB in the basic New Keynesian model, but they showed quantitatively that inflation is close to zero under the optimal policy with commitment. In our economy, depending on $\underline{R}$, positive inflation might arise in a stochastic steady state due to the planner's precautionary motive to insure against the binding ELB.

It is worth emphasizing that the steady-state inflation rate under the optimal monetary policy is uniquely determined, unlike in the case of an ad hoc Taylor rule with an ELB, where multiple equilibria are an inherent property of the functional form \citep{benhabib01}. The Ramsey planner is not subject to any functional form restrictions and chooses a state-contingent plan subject to an ELB inequality constraint. At the same time, \citet{armenter18} has shown that multiple Markov equilibria might arise under an optimal discretionary policy with an ELB.

In the short run, stabilizing prices state-by-state is not generally optimal, and the inflation dynamics are characterized by an Euler equation that balances different forces. Note that if $\Pi_{t}\approx{}1$, then $\widetilde{P}_{t}=\frac{1}{\Pi_{t}}\left(\frac{\Pi_{t}^{1-\epsilon}-\theta}{1-\theta}\right)^\frac{1}{1-\epsilon}\approx{}1$ and $\widetilde{P}'(\Pi_{t})=\frac{\widetilde{P}_{t}}{\Pi_{t}}\left(\frac{\widetilde{P}_{t}^{\epsilon-1}}{1-\theta}-1\right)>0$. First, a greater inflation rate positively affects welfare by raising the retailer's marginal benefit through the higher optimal relative price, reflected in the term $\lambda^\Omega_{t}\widetilde{P}'(\Pi_{t})>0$. Second, inflation affects price dispersion: both positively, by expanding the price dispersion inherited from the previous period, $\theta\Pi_{t}^{\epsilon-1}\Delta_{t-1}>0$, and negatively, by raising the retailer's optimal price, reflected in $-(1-\theta)\frac{\widetilde{P}'(\Pi_{t})}{\widetilde{P}_{t}^{\epsilon+1}}<0$. In the steady state, the net effect is zero. Third, by raising inflation at $t$, the planner affects the expectation of retailers at $t-1$ regarding the marginal cost at $t$, having a negative effect $-\theta\Pi_{t}^{\epsilon-1}\Omega_{1,t}\frac{U^w_{C,t}}{U^w_{C,t-1}}\widetilde{\lambda}^C_{t-1}\Delta_{t-1}\epsilon<0$. Fourth, higher inflation at $t$ also affects the retailer's marginal benefit expected at $t-1$, which has a positive effect $\theta\Pi_{t}^{\epsilon-1}\Omega_{1,t}\frac{U^w_{C,t}}{U^w_{C,t-1}}\lambda^\Omega_{t-1}\frac{\widetilde{P}_{t-1}}{\widetilde{P}_{t}}\left(\frac{\epsilon-1}{\Pi_{t}}-\frac{\widetilde{P}'(\Pi_{t})}{\widetilde{P}_{t}}\right)>0$ since $\frac{\epsilon-1}{\Pi_{t}}-\frac{\widetilde{P}'(\Pi_{t})}{\widetilde{P}_{t}}=\frac{1}{\Pi_{t}}\left(\epsilon-\frac{\widetilde{P}_{t}^{\epsilon-1}}{1-\theta}\right)>0$, provided that the elasticity of substitution $\epsilon$ is sufficiently greater than the price duration $\frac{1}{1-\theta}$. Fifth, higher inflation at $t$ raises the expected inflation at $t-1$ and relaxes the ELB at $t-1$.

None of the components in the inflation Euler equation are directly related to financial constraints. In our economy, the Tinbergen separation principle applies: the implications of the Ramsey allocations with optimal leverage limits are similar to those in the flexible-price environment, and the approximate price stability is optimal as in the basic New Keynesian model. \citet{collard17} also found support for such independence of policy goals exploring jointly optimal bank capital requirements and monetary policy in the absence of collateral constraints and consumer type heterogeneity but with nominal contracts. The latter indicates that our choice to proceed with real contracts is mostly without loss of generality. Indeed, one can show that the optimal monetary policy would still have steady-state price stability and similar short-run dynamics.

\section{Quantitative results}\label{sec: quantitative results}
This section describes the model calibration and computation, quantifies the welfare losses due to the constrained inefficiency and welfare benefits of optimal policies, compares the extent of consumption insurance observed in the different types of decentralized and centralized allocations, and studies the economic dynamics around financial crises and binding ELB events.

\subsection{Calibration}
To simplify the interpretation of quantitative results, we will assume that all consumers have logarithmic preferences over period consumption. With $U^b(C^b)=\ln(C^b)$ and $U^e(C^e)=\ln(C^e)$, lemma \ref{lemma: net assets} implies that the banker's and entrepreneur's net assets are proportional to consumption. The worker's preferences are separable in consumption and leisure, taking the form $U^w(C^w,N)=\ln(C^w)-\chi\frac{N^{1+\phi}}{1+\phi}$, where $\chi>0$ is the labor disutility scale, and $\phi\ge{}0$ is the inverse of the Frisch elasticity of labor supply. Thus, the worker's preferences are consistent with the special case of full insurance of Propositions \ref{prop: FCEA} and \ref{prop: optimal leverage limits with labor taxation}. As for technology, the entrepreneur's output is produced according to $F(\xi{}K,N)=(\xi{}K)^\alpha{}N^{1-\alpha}$ with $\alpha\in(0,1)$, and capital goods are built using $\Phi(x)=\zeta+\kappa_1{}x^\psi$ with $\zeta\in\mathbb{R}$, $\kappa_1>0$, and $\psi\in(0,1]$. The logarithms of the exogenous stochastic processes $\{A_{t}\}$ and $\{\xi_{t}\}$ are independent Gaussian AR(1) with autocorrelations $(\rho_a,\rho_\xi)$ and shock standard deviations $(\sigma_a,\sigma_\xi)$, respectively, implying a steady-state normalization $A=\xi=1$.

Table \ref{tab: parameters} reports the model parameter values.
\begin{table}[ht!]
    \caption{Parameter values}\label{tab: parameters}
    \centering
    \begin{tabular*}{\textwidth}{l@{\extracolsep{\fill}}ll}
        \toprule
        Parameter & Value & Target\\
        \midrule
        \multicolumn{3}{l}{Baseline policy}\\
        \midrule
        $\Bar{\kappa}$ & 0.105 & Basel III total capital requirement + conservation buffer \\
        $\Bar{m}$ & 0.7 & FDIC LTV limits for raw land (65\%) and land development (75\%) \\
        $\Bar{\Pi}$ & 1.005 & annual inflation $=2\%$ \\
        $\underline{R}$ & 1 & zero lower bound \\
        \midrule
        \multicolumn{3}{l}{Preferences and technology}\\
        \midrule
        $\alpha$ & 0.404 & average nonfarm labor share $\approx{}59.6\%$ \\
        $\beta$ & 0.995 & annualized real interest rate $=2\%$ \\
        $\beta_b$ & 0.972 & annual NAICS 52 establishment exit rate $\approx{}9.1\%$ \\
        $\beta_e$ & 0.974 & annual NAICS 31--33 establishment exit rate $\approx{}8.2\%$ \\
        $\delta$ & 0.02 & annual depreciation rate $\approx{}7.6\%$ \\
        $\epsilon$ & 9.093 & average retail markup $=1.125$ \\
        $\zeta$ & -0.002 & $\frac{I}{K}=\delta$ and $Q=1$ \\
        $\theta$ & 0.75 & average price duration $=4$ quarters \\
        $\kappa_1$ & 0.781 & $\frac{I}{K}=\delta$ and $Q=1$ \\
        $\phi$ & 0.625 & microfounded aggregate Frisch elasticity = 1.6 \\
        $\chi$ & 0.94 & $N=1$ in the FCE \\
        $\psi$ & 0.75 & panel data evidence \\
        \midrule
        \multicolumn{3}{l}{Exogenous stochastic processes}\\
        \midrule
        $\rho_a$ & 0.918 & \multirow{4}{*}{\begin{minipage}{0.57\textwidth}
        First-step MSM estimation based on the FCE, targeting $\corr(\widehat{Y}_{t},\widehat{Y}_{t-1})$, $\sd(\widehat{I}_{t})$, $\sd(\widehat{Y}_{t})$, and $\corr(\widehat{I}_{t},\widehat{Y}_{t})$.\end{minipage}} \\
        $\rho_\xi$ & 0.935 & \\
        $\sigma_a$ & 0.005 & \\
        $\sigma_\xi$ & 0.003 & \\
        \midrule
        \multicolumn{3}{l}{Taylor rule}\\
        \midrule
        $\rho_R$ & 0.897 & \multirow{3}{*}{\begin{minipage}{0.57\textwidth}
        Second-step MSM estimation based on the CE, targeting $\corr(\widehat{Y}_{t},\widehat{Y}_{t-1})$, $\sd(\widehat{Y}_{t})$, $\corr(\widehat{\Pi}_{t},\widehat{\Pi}_{t-1})$, $\sd(\widehat{\Pi}_{t})$, $\corr(\widehat{\Pi}_{t},\widehat{Y}_{t})$, and $\Pr(R^N_{t}=\underline{R})$.\end{minipage}} \\
        $\eta_\pi$ & 3.366 & \\
        $\eta_y$ & 3.104 & \\
        \bottomrule
    \end{tabular*}
    \begin{tabular}{@{}p{\textwidth}@{}}
        {\small Notes: $\widehat{X}_{t}$ denotes the cyclical component of $\ln(X_{t})$ extracted using the HP filter with $\lambda=1600$.}
    \end{tabular}
\end{table}
To calibrate the structural parameters, we need to determine the baseline policies taken as given by the private agents. The leverage limits are set to constant values $\kappa_{t}=\Bar{\kappa}$ and $m_{t}=\Bar{m}$ for all $t\ge{}0$. The capital requirement $\Bar{\kappa}$ corresponds to the Basel III minimum total capital requirement that includes the conservation buffer. The LTV ratio $\Bar{m}$ is set to the average of the Federal Deposit Insurance Corporation's (FDIC) recommended maximum LTV limits for raw land and land development---a proxy for commercial loans. The inflation target $\Bar{\Pi}$ corresponds to the annual target of 2\%, and the effective lower bound $\underline{R}$ is the zero lower bound (ZLB).

The structural parameters that affect the steady state are either based on micro evidence or target various long-run moments in the US quarterly---or annual if not available---data for 1990--2019 or the largest available subset. The remaining parameters are estimated using the method of simulated moments (MSM) of \citet{mcfadden89}. The procedure is described in appendix \ref{sec: MSM estimation}.

From the preference parameters, the discount factor $\beta$ corresponds to the annualized real interest rate of 2\%. The effective discount factors of bankers and entrepreneurs are based on the average annual establishment exit rates in finance and insurance (NAICS 52) and manufacturing (NAICS 31--33), respectively, using Business Dynamics Statistics data. The inverse of the Frisch elasticity of the worker's labor supply $\phi$ targets the average of the microfounded estimates of the aggregate Frisch elasticity for males \citep{erosa16} and females \citep{attanasio18}. The labor disutility scale $\chi$ is set to normalize $N=1$ in the FCE.

Turning to the technology parameters, the capital share $\alpha$ targets the average labor share in the nonfarm business sector based on US Bureau of Labor Statistics data. The depreciation rate $\delta$ is based on the average depreciation rate of the current-cost net stock of private fixed assets and consumer durables in Bureau of Economic Analysis data. The capital good technology elasticity $\psi$ is based on the panel data evidence \citep{gertler20}. Conditional on $\psi$, there is a one-to-one correspondence between the location and scale parameters $(\zeta,\kappa_1)$ and a steady-state pair $(\frac{I}{K},Q)$. Using \eqref{eq: capital good supply} and \eqref{eq: market clearing capital good}, we get $\kappa_1=\frac{1}{\psi{}Q}\left(\frac{I}{K}\right)^{1-\psi}$ and $\zeta=1-(1-\delta)\xi-\kappa_1\left(\frac{I}{K}\right)^\psi$. We have already normalized $\xi=1$. By targeting $\frac{I}{K}=\delta$ and normalizing $Q=1$, the steady-state profits of capital good producers are zero, and thus the calibration is consistent with the special case of perfect insurance in Propositions \ref{prop: FCEA} and \ref{prop: optimal leverage limits with labor taxation}. The Calvo price stickiness parameter targets the average price duration $\frac{1}{1-\theta}$, and the elasticity of substitution between retail varieties $\epsilon$ is mapped to the markup $\frac{1}{P^w}$ in retail, solving a steady-state equation $P^w=\frac{\epsilon-1}{\epsilon}\frac{1-\beta\theta\Pi^\epsilon}{1-\beta\theta\Pi^{\epsilon-1}}\left(\frac{1-\theta}{1-\theta\Pi^{\epsilon-1}}\right)^\frac{1}{\epsilon-1}$ that follows from combining \eqref{eq: retailers optimal relative price}--\eqref{eq: retailers inflation}. The targets are consistent with the micro evidence as in \citet{gali15}.

To account for multiple occasionally binding constraints in simulations of both competitive equilibria and centralized allocations, I use the piecewise linear perturbation approach of \citet{guerrieri15}, extending it to handle an arbitrary number of regime-switching constraints.\footnote{The extension is available at \url{https://github.com/azaretski/occbin-n}.} In some exercises, I use a standard second-order perturbation, taking advantage of the possibility to approximate theoretical moments when the system stays close to the steady state. To get a locally unique approximation for the FCEA, CEA, the Ramsey allocation of lemma \ref{lemma: OLL relaxed with unrestricted taxation}, and the case 1 Ramsey allocation of proposition \ref{prop: optimal leverage limits and monetary policy}, I fix the quantity of deposits at the optimal steady-state value of zero. The welfare benefits of the corresponding allocations are thus generally underestimated.

\subsection{Welfare comparison}
Starting from this subsection, we will use additional notation, referring to the Ramsey allocations of Lemmas \ref{lemma: OLL relaxed with unrestricted taxation} and \ref{lemma: OLL relaxed with labor taxation} as FCEA OLL and OLL, respectively, where ``OLL'' means ``optimal leverage limits.'' We will call the case 1 and case 2 Ramsey allocations of proposition \ref{prop: optimal leverage limits and monetary policy} CEA OLLMP and OLLMP, respectively, where ``OLLMP'' corresponds to ``optimal leverage limits and monetary policy.''

Table \ref{tab: welfare} reports the welfare ranking of alternative environments conditional on a Pareto vector $\omega=(\omega_b,\omega_e,\omega_w)'=(0.1,0.1,0.8)'$.
\begin{table}[ht!]
    \caption{Welfare in consumption equivalents, \% of first best}\label{tab: welfare}
    \centering
    \begin{tabular*}{\textwidth}{l@{\extracolsep{\fill}}llll}
        \toprule
        & bankers & entrepreneurs & workers & social welfare \\
        \midrule
        First best & 100 & 100 & 100 & 100 \\
        FCE & 28.8 & 109.7 & 95.9 & 86.2 \\
        FCEA & 94.2 & 94.2 & 100.1 & 98.9 \\
        FCEA OLL & 94.2 & 94.2 & 100.1 & 98.9 \\
        OLL & 71.8 & 88.6 & 98.1 & 94.1 \\
        CE & 21.0 & 79.8 & 91.7 & 78.1 \\
        CEA & 78.6 & 78.6 & 98.4 & 94.0 \\
        CEA OLLMP & 79.0 & 79.0 & 98.9 & 94.5 \\
        OLLMP & 77.4 & 60.3 & 97.4 & 90.7 \\
        \bottomrule
    \end{tabular*}
    \begin{tabular}{@{}p{\textwidth}@{}}
        {\small Notes: Second-order accurate theoretical moments in the neighborhood of the steady state, conditional on a Pareto vector $(\omega_b,\omega_e,\omega_w)'=(0.1,0.1,0.8)'$. The OLLMP row is based on $\beta_b\approx{}0.989$---the nearest neighbor where the Blanchard---Kahn conditions for local uniqueness hold.}
    \end{tabular}
\end{table}
There is a unit measure of all types of consumers in the model, so one might want to choose comparable Pareto weights for all agents. On the other hand, the real-world population of workers is significantly greater than that of bankers or entrepreneurs, which suggests a worker-biased weighting. The chosen Pareto vector reflects these two margins: it is worker biased, but the banker's and entrepreneur's weights are still sizable. As a welfare benchmark, we will consider the first-best allocation. The differences relative to the first best are represented in consumption equivalents. Let $\mathcal{W}^i_{FB}$ and $\mathcal{W}^i$ denote the expected welfare of type $i$ consumers at the first best and an alternative set-up, respectively. We can solve for $\lambda^i$ that satisfies $\mathcal{W}^i=\mathbb{E}\left[\sum_{t=0}^\infty\beta^{t}U^i(\lambda^i{}C^i_{FB,t})\right]$, where $\{C^i_{FB,t}\}$ is the first-best consumption plan. By construction, $\lambda^i\in(0,1)$ is the proportion of the first-best consumption plan---applied in all contingencies---that yields the same welfare for agent $i$ as the alternative consumption allocation. With logarithmic preferences, we have a closed-form solution $\lambda^i=\exp[(1-\beta)(\mathcal{W}^i-\mathcal{W}^i_{FB})]$. Similarly, we can get a social welfare ranking by computing $\lambda=\exp[(1-\beta)(\mathcal{W}-\mathcal{W}_{FB})]$, where $\mathcal{W}$ and $\mathcal{W}_{FB}$ denote the expected social welfare of the alternative and first-best allocations, respectively, and $\lambda\in(0,1)$ is the proportion of the first-best consumption plan---applied in all contingencies and for all consumers---that yields the same value of social welfare as the alternative consumption allocation.

Compared to the first best, constrained bankers and entrepreneurs are more worse off than workers in most environments, reflecting the worker-biased Pareto vector. Due to nominal rigidities, the sticky-price environments tend to be welfare dominated by their flexible-price counterparts. The welfare gains from constrained efficiency---FCEA over FCE and CEA over CE---are rather significant. The FCEA and FCEA OLL allocations have identical welfare implications because both financial constraints are locally slack in the FCEA: bank leverage is suboptimal, and the optimal entrepreneur's LTV ratio is lower than the calibrated limit. Relaxing the leverage constraints might impact precautionary savings, but we cannot account for this effect using our computation method. The OLL allocation is Pareto dominated by the FCEA, since bankers have positive leverage, and the relaxed collateral constraint is binding. At the same time, the OLL allocation constitutes a significant social welfare gain over the FCE.

Although leverage constraints are locally slack in the CEA, with nominal rigidities, there is a distinction between the CEA and CEA OLLMP allocations, since the latter has optimal monetary policy, compared to an ad hoc Taylor rule in the CEA. Similar to the flexible-price case, the OLLMP allocation is between the CE and CEA in social welfare terms, although optimal monetary policy reduces the relative distance to the CEA.

\subsection{Risk sharing}
Table \ref{tab: risk sharing} reports the correlations between the HP-filtered logged marginal utilities of consumption across consumers in the alternative allocations.
\begin{table}[ht!]
    \caption{Consumption risk sharing}\label{tab: risk sharing}
    \centering
    \begin{tabular*}{\textwidth}{l@{\extracolsep{\fill}}lll}
        \toprule
        & $\corr(\widehat{U}^b_{C,t},\widehat{U}^e_{C,t})$ & $\corr(\widehat{U}^b_{C,t},\widehat{U}^w_{C,t})$ & $\corr(\widehat{U}^e_{C,t},\widehat{U}^w_{C,t})$ \\
        \midrule
        First best  & 1 & 1 & 1 \\
        FCE & 0.07 & 0.57 & -0.51 \\
        FCEA & 1 & 1.0 & 1.0 \\
        FCEA OLL & 1 & 1.0 & 1.0 \\
        OLL & 0.92 & -0.55 & -0.71 \\
        CE & -0.1 & 0.59 & -0.57 \\
        CEA & 1 & 0.99 & 0.99 \\
        CEA OLLMP & 1 & 0.99 & 0.99 \\
        OLLMP & 0.8 & -0.85 & -0.97 \\
        \bottomrule
    \end{tabular*}
    \begin{tabular}{@{}p{\textwidth}@{}}
        {\small Notes: Second-order accurate theoretical correlations in the neighborhood of the steady state, conditional on a Pareto vector $(\omega_b,\omega_e,\omega_w)'=(0.1,0.1,0.8)'$. The decimal point in 1.0 indicates that the correlation is not exactly 1. $\widehat{X}_{t}$ denotes the cyclical component of $\ln(X_{t})$ extracted using the HP filter with $\lambda=1600$. The OLLMP row is based on $\beta_b\approx{}0.989$---the nearest neighbor where the Blanchard---Kahn conditions for local uniqueness hold.}
    \end{tabular}
\end{table}
The first-best allocation is the only one that has perfect consumption insurance. Consistent with Propositions \ref{prop: FCEA}, \ref{prop: CEA}, and \ref{prop: optimal leverage limits and monetary policy}, the FCEA, FCEA OLL, CEA, and CEA OLLMP allocations have perfect risk sharing between bankers and entrepreneurs, since the collateral constraint is locally slack. The latter is not the case in the OLL and OLLMP allocations. In the FCE and CE, financial constraints are locally binding, and consumption insurance is largely imperfect. Since our calibration is consistent with the special case of proposition \ref{prop: FCEA}, the FCEA and FCEA OLL allocations have nearly perfect insurance across all consumers: the correlation between the worker's marginal utility and the marginal utility of constrained consumers is near unity. With nominal rigidities, the correlation is only slightly lower.

Although perfect consumption risk sharing is a feature of the first best, stronger risk sharing between consumers is not a prerequisite for higher welfare, as Tables \ref{tab: welfare} and \ref{tab: risk sharing} demonstrate. For example, insurance is much stronger in the CEA than in the OLL allocation, but the latter has no nominal rigidities and has greater social welfare. Conditional on a flexible-price or a sticky-price environment, stronger risk sharing is indeed associated with higher welfare.

\subsection{Wedges and overborrowing}
Table \ref{tab: wedges} quantifies the financial and real wedges. As shown in Sections \ref{sec: FCEA} and \ref{sec: CEA}, each wedge can be decomposed into several components.
\begin{table}[ht!]
    \caption{Wedges}\label{tab: wedges}
    \centering
    \begin{tabular*}{\textwidth}{l@{\extracolsep{\fill}}lllll}
        \toprule
         & & \multicolumn{2}{c}{FCEA} & \multicolumn{2}{c}{CEA} \\
         & & mean, \% & variance, \% & mean, \% & variance, \% \\
        \midrule
        \multicolumn{2}{l}{$\Psi^D_{t}$, \% of $U^b_{C,t}$} & 2.4 & 0.1 & 2.4 & 0.1 \\
         & uncertain survival: bankers & 99.9 & 73.6 & 99.9 & 65.6 \\
         & consumer type heterogeneity & 0.1 & 19.6 & 0.1 & 29.1 \\
        \multicolumn{2}{l}{$\Psi^L_{t}$, \% of $U^e_{C,t}$} & -0.3 & 0.0 & -0.3 & 0.0 \\
         & survival rate differences: $\beta_b\neq\beta_e$ & 100 & 100 & 100 & 100 \\
         & uncertain survival: bankers & 0 & 0 & 0 & 0 \\
        \multicolumn{2}{l}{$\Psi^N_{t}$, \% of $W_{t}$} & -8.7 & 8.6 & -17.8 & 4.4 \\
         & consumer type heterogeneity & 64.0 & 42.3 & 24.0 & 61.3 \\
         & $W$-externality & 36.0 & 12.2 & 61.6 & 3.1 \\
         & nominal rigidities & 0 & 0 & 14.5 & 0.6 \\
        \multicolumn{2}{l}{$\Psi^K_{t}$, \% of $U^e_{C,t}Q_{t}$} & 2.1 & 0.1 & 2.3 & 0.1 \\
         & uncertain survival: entrepreneurs & 99.7 & 67.1 & 91.1 & 60.1 \\
         & $\Phi$-externality & 0.6 & 3.4 & 0.4 & 1.7 \\
         & $Q$-externality & -0.3 & 0.9 & -1.1 & 1.4 \\
         & nominal rigidities & 0 & 0 & 9.5 & 1.0 \\
        \bottomrule
    \end{tabular*}
    \begin{tabular}{@{}p{\textwidth}@{}}
        {\small Notes: Second-order accurate theoretical moments in the neighborhood of the steady state, conditional on a Pareto vector $(\omega_b,\omega_e,\omega_w)'=(0.1,0.1,0.8)'$. Components of wedges are in \% of the mean or variance of the corresponding wedge. ``Consumer type heterogeneity'' reflects marginal utility gaps and terms that arise because $\lambda^Y_{t}\neq\omega_w{}U^w_{C,t}$. The $W$-, $\Phi$-, and $Q$-externalities are the externalities through the wage, the capital good production technology directly, and the capital good price, respectively.}
    \end{tabular}
\end{table}
By definition, the means of components add up to 100\%. Since the components are generally correlated, the sum of the variances need not be equal to the variance of the corresponding wedge.

The expected value of the deposit wedge $\Psi^D_{t}$ is almost entirely based on the survival externality and entirely in the steady state. The survival externality is dominant in terms of the variance, but the consumer type heterogeneity component also has a nonnegligible variation. The loan wedge $\Psi^L_{t}$ is entirely determined by the difference in the survival rates of bankers and entrepreneurs because the collateral constraint is locally slack in both the FCEA and the CEA. By proposition \ref{prop: CEA}, nominal rigidities do not affect the expressions of financial wedges, which results in an identical decomposition of means.

In the FCEA, about two-thirds of the expected value of the labor wedge stems from the direct implications of consumer type heterogeneity, and the rest is explained by the wage externality---an indirect consequence of consumer type heterogeneity. In the CEA, the order is reversed, and nominal rigidities play an additional role. In both environments, consumer type heterogeneity explains a significant part of the variance, especially in the CEA. Although the additive term arising from nominal rigidities in the CEA contributes to only 14.5\% of the expected value, nominal rigidities also affect the consumer type heterogeneity component in a multiplicative way, so their impact cannot be easily decoupled. The absolute value of the consumer type heterogeneity component is significantly less in the CEA than in the FCEA, as predicted in section \ref{sec: CEA}, although the magnitude of the wedge is greater in the CEA due to the other two components.

The uncertain survival of entrepreneurs explains a major part of the expected value and variance of the capital wedge. Nominal rigidities constitute the second strongest direct source of the wedge in the CEA, and they also have an indirect multiplicative effect through the price of wholesale goods that affects the return on capital and the uncertain survival component. The role of the asset-price externality is modest. Since the collateral constraint is locally slack in the FCEA and CEA, the pecuniary externality only has precautionary savings effects. Hence, the asset-price externality works exclusively through the consolidated budget constraint of bankers and entrepreneurs. Although our calibration ensures that the steady-state profits of capital good producers are zero, the expected value is slightly positive, and so is the first-order externality that works through the capital good production technology $\Phi$ directly. 

As a result of constrained inefficiency, our economy has inefficient borrowing in the financial markets. There are two types of borrowing: banks' borrowing from workers and entrepreneurs' borrowing from banks. By Propositions \ref{prop: FCEA} and \ref{prop: CEA}, the constrained efficient bank leverage is zero in the optimal steady state, implying extreme overborrowing by banks in the competitive equilibria. The intertemporal inefficiency of the entrepreneur's borrowing is reflected in the wedge $\Psi^L_{t}$. As shown in figure \ref{tab: wedges}, the wedge is negative since $\beta_b<\beta_e$. Although the competitive demand for bank loans is inefficiently low, overborrowing by the banking sector results in an inefficiently large supply, which tends to make the competitive quantity of bank loans inefficiently large if the Pareto vector is sufficiently worker biased.

\begin{figure}[ht!]
    \caption{Histograms of bank loans}\label{fig: histograms}
    \centering
    \includegraphics[trim={2.1cm 14.7cm 2.1cm 7.3cm},clip,width=0.75\textwidth]{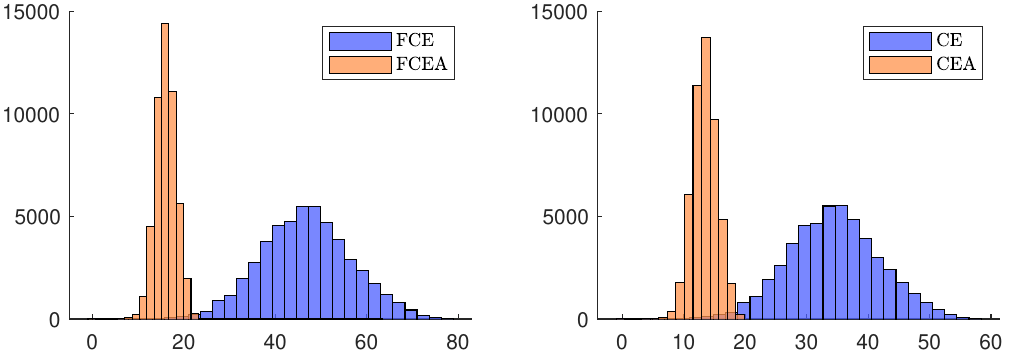}
    \begin{tabular}{@{}p{\textwidth}@{}}
        {\small Notes: 50,000-period simulation conditional on a Pareto vector $(\omega_b,\omega_e,\omega_w)'=(0.1,0.1,0.8)'$.}
    \end{tabular}
\end{figure}

Figure \ref{fig: histograms} displays the histograms of bank loans in the FCE and CE compared to the FCEA and CEA, respectively.
By construction, in the FCEA and CEA, there is no variation in the quantity of deposits that are fixed at the optimal steady-state value of zero; consequently, the variance of bank loans is smaller in the FCEA and CEA. The expected values are considerably smaller, reflecting overlending in the FCE and CE. Nominal rigidities tend to decrease the level of economic activity, shifting the distributions of bank loans to the left.

\subsection{Financial crises}
This subsection explores the economic dynamics around financial crises. The focus is on the flexible-price economy to isolate the effect of the occasionally binding collateral constraint. Financial crises are defined similarly as in \citet{mendoza10}. To be qualified as a financial crisis that starts at $t$, two conditions must be true: first, the collateral constraint is slack at $[t-4,t-1]$; second, the collateral constraint is binding at $[t,t+4]$. Such an event is observed in the FCE with a frequency of 3.2 crises per century, consistent with the data.

\begin{figure}[ht!]
    \caption{Financial crises}\label{fig: financial crises}
    \centering
    \includegraphics[trim={2.3cm 6.3cm 2.3cm 7.1cm},clip,width=0.75\textwidth]{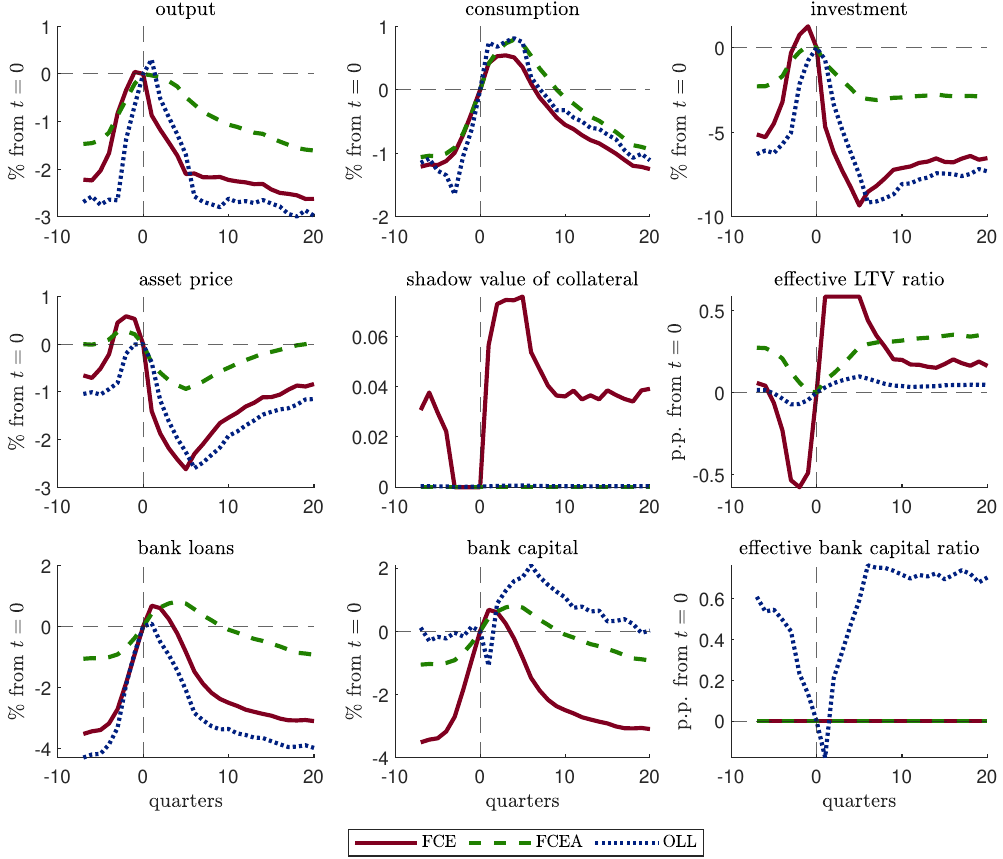}
    \begin{tabular}{@{}p{\textwidth}@{}}
        {\small Notes: Each line is based on an average of 399 crisis episodes over a 50,000-period simulation conditional on a Pareto vector $(\omega_b,\omega_e,\omega_w)'=(0.1,0.1,0.8)'$. A crisis starts at $t=1$ and lasts at least five quarters. The shadow value of collateral is in levels. The effective LTV ratio is the ratio of the expected loan repayment to the value of collateral. The effective bank capital ratio is the ratio of bank capital to bank loans. ``p.p.'' is ``percentage points.''}
    \end{tabular}
\end{figure}

Figure \ref{fig: financial crises} illustrates the dynamics around financial crises in alternative environments based on a 50,000-period simulation conditional on an identical sequence of exogenous shocks drawn randomly from the corresponding distributions. The financial crisis events are identified in the FCE simulation, and the identified dates are used to extract the corresponding paths in the FCEA and OLL simulations. The dynamics around identified crises are averaged, and each crisis is normalized to start at $t=1$, lasting at least until $t=5$.

Ahead of a typical crisis in the FCE, the economy is booming: output, consumption, and investment are increasing, so is bank lending and---for most of the period---the collateral asset price. By construction, the collateral constraint is slack during the year before the start of the crisis, so the shadow value of collateral is zero during that time. The asset price starts to fall a few quarters ahead of the crisis, leading to a decrease in the value of collateral and triggering a switch of the collateral constraint from the slack to the binding regime. Output and bank lending immediately start to drop, while investment starts to fall earlier, responding to a fall in the asset price. As the collateral constraint returns to a slack regime, which occurs at different times in each crisis, the asset price and investment start to recover, and the fall in output and bank lending slows down, plateauing gradually. There is a one percentage point increase in the entrepreneur's LTV ratio just before the crisis until it hits the LTV limit $\Bar{m}$ during the crisis. The bank leverage constraint remains binding in the FCE, so the bank capital ratio is constant at $\Bar{\kappa}$.

The FCEA and FCEA OLL allocations have identical dynamics, since both leverage constraints remain slack in the simulation; therefore, the figure shows only the dynamics in the FCEA. What happens to be a financial crisis in the FCE is reminiscent of a cyclical slowdown in the FCEA, which is a consequence of the fact that the optimal entrepreneur's leverage is smaller, and the collateral constraint is slack. The fluctuations in the LTV ratio are small. Since bank leverage is constant at zero in the FCEA, the capital ratio is constant at one.

In the OLL allocation, the dynamics are more similar to the FCE than to the FCEA. A fall in real quantities and the asset price tends to be initially smaller, but the eventual decrease is similar to that in the FCE. The amplitude of the relative changes in investment and asset price is slightly smaller than that in the FCE, while the opposite is true for output and bank lending. The variation in the entrepreneur's optimal LTV ratio is negligible. The Ramsey planner keeps bank capital at a stable level ahead of a crisis and provides additional capital during the crisis. Combined with the credit dynamics, the optimal bank capital ratio decreases ahead of the crisis and increases during the crisis, although the changes are in the range of one percentage point.

\subsection{Zero lower bound}
\begin{figure}[ht!]
    \caption{Zero-lower-bound crises}\label{fig: ZLB crises}
    \centering
    \includegraphics[trim={2.2cm 6.3cm 2.2cm 7.1cm},clip,width=0.75\textwidth]{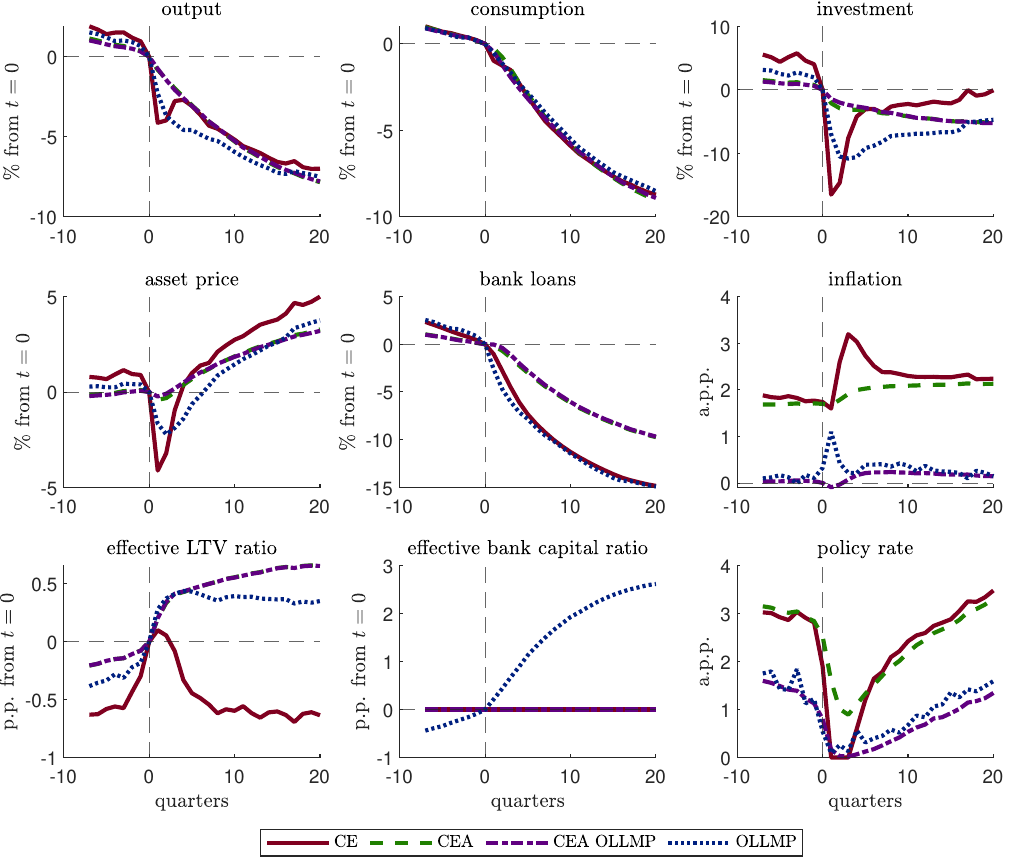}
    \begin{tabular}{@{}p{\textwidth}@{}}
        {\small Notes: Each line is based on an average of 309 crisis episodes over a 50,000-period simulation---with an exception below---conditional on a Pareto vector $(\omega_b,\omega_e,\omega_w)'=(0.1,0.1,0.8)'$. A crisis starts at $t=1$ and lasts at least three quarters. The effective LTV ratio is the ratio of the expected loan repayment to the value of collateral. The effective bank capital ratio is the ratio of bank capital to bank loans. ``p.p.'' is ``percentage points,'' and ``a.p.p.'' is ``annualized percentage points.'' The OLLMP paths are the averages over the crises observed in $0\le{}t\le{}24,108$, after which the simulation algorithm encounters numerical problems. The OLLMP simulation is based on setting $\beta_b=0.995<\beta$, which is the nearest neighbor that satisfies the Blanchard---Kahn conditions for local uniqueness and permits a relatively long simulation.}
    \end{tabular}
\end{figure}

This subsection considers a different type of crisis that occurs when the ZLB binds. A ZLB crisis that starts at $t$ is an event that satisfies two conditions: the ZLB constraint is slack at $[t-4,t-1]$ and is binding at $[t,t+2]$. An event defined this way is observed in the CE with a frequency of 2.5 crises per century. Figure \ref{fig: ZLB crises} illustrates the dynamics around a typical ZLB crisis and is constructed similar to figure \ref{fig: financial crises}.

Unlike financial crisis episodes that follow a boom-bust pattern, ZLB events occur when the economy is either already in a recession or a state of stagnation. The latter is reflected in the paths of both the real sector variables---output, consumption, and investment---and the financial sector variables, such as the collateral asset price and bank loans. Ahead of a ZLB crisis, the central bank consistently fails to match a 2\% annualized inflation target. When the ZLB binds at $t=1$, there is a further decrease in inflation, followed by a spike reflecting an increase in the retailer's marginal cost due to the drop in the entrepreneur's supply of wholesale goods. In our economy, a ZLB crisis results from a persistent decrease in the TFP and capital quality processes, leading to a sharp drop in output, investment, and asset price. The decrease in consumption and bank loans accelerates. When the ZLB becomes slack, the asset price and investment start to recover, but a decrease in output continues, and the recovery is slow. The collateral constraint is typically binding during a ZLB crisis, and there is a spike in the shadow value of collateral when the ZLB binds, reflected in the rise in the entrepreneur's LTV ratio. The bank leverage constraint remains binding during the whole crisis window, so the bank capital ratio is constant at $\Bar{\kappa}$.

Except for the paths of inflation and the policy rate, the dynamics in the CEA and CEA OLLMP allocations are similar, consistent with the flexible-price analysis. Although the economy is stagnating ahead of ZLB crises, followed by a deep recession, there are no sharp changes in output growth, no drop in the asset price, and the dynamics of investment are reminiscent of a cyclical decline. Bank loans eventually decrease by about five percentage points less than in the CE. A key reason for these differences is that in the CEA and CEA OLLMP allocations, the collateral constraint remains slack around a ZLB crisis, which allows for an increase in the entrepreneur's LTV ratio, supporting investment and asset price and smoothing out a decrease in output and credit. Since bank leverage is suboptimal in the CEA, the capital ratio is constant at one. In the CEA, monetary policy is determined by the same Taylor rule as in the CE. However, the ZLB is not hit, and inflation stays close to the target. In the CEA OLLMP allocation, there is optimal monetary policy, inflation stays close to the long-run level of zero throughout the crisis window, consistent with proposition \ref{prop: optimal leverage limits and monetary policy}, and the Ramsey planner typically just avoids the ZLB.

A long simulation of the OLLMP allocation is prone to numerical problems because it entails accounting for five regime-switching constraints: the private complementary slackness conditions of bankers and entrepreneurs, the corresponding planner's complementary slackness conditions, and the planner's effective lower bound constraint. After increasing the banker's survival rate, a relatively long simulation is possible, but the results are not directly comparable to those in the other environments. Considering this limitation, we see that the dynamics of the real sector variables and the asset price are roughly a convex combination of the CE and CEA dynamics. A drop in investment and asset price is less than in the CE, since the relaxed collateral constraint allows the Ramsey planner to increase the entrepreneur's LTV ratio. Although inflation is close to zero during most of the crisis window, there is a spike to about one percentage point after the ZLB binds in the CE. By increasing the inflation rate, the Ramsey planner just evades the ZLB, which allows the planner to smooth out fluctuations, facilitated by the planner's ability to increase the bank capital ratio.

\section{Conclusion}\label{sec: conclusion}
Financial constraints combined with consumer type heterogeneity lead to multiple sources of the inefficiency of the CE allocation. The inefficiency is reflected in both the real sector wedges in the demand for factors of production---labor and capital---and the financial sector wedges in the supply of bank deposits and the demand for bank loans. Nominal rigidities affect the real wedges but not the financial wedges. Consequently, optimal monetary policy in the presence of financial constraints and consumer type heterogeneity is reminiscent of the basic New Keynesian economy: stabilizing prices is optimal, exactly in the long run and approximately in the short run.

If a policymaker has the appropriate fiscal instruments to correct the intertemporal and intratemporal distortions in the CE allocation, the resulting CEA entails significant welfare gains. Under certain assumptions, such an allocation is close to an unconstrained Pareto optimum, having quantitatively perfect consumption insurance not only within consumer types but also between types. Furthermore, the CEA has lower leverage in both the banking and the entrepreneurial sectors. These features help eliminate or mitigate both the boom-bust financial crises and zero-lower-bound crises observed occasionally in the decentralized economy.

Correcting the Euler equation distortions might constitute an ambitious task. If that is not possible, but the leverage limits can be set optimally, the policymaker can still smooth out fluctuations by setting the leverage ratios in a state-contingent manner. Both the optimal bank capital and the LTV ratios appear to be countercyclical around financial and ZLB crises.

\bibliography{lit}

\appendix

\hypertarget{appendices}{\appendixpage}

\section{Retailer's problem}\label{sec: retailer's problem}
The final good production technology is $Y_{t}\equiv\left(\int_0^1 {Y_{i,t}}^\frac{\epsilon-1}{\epsilon}\dd{i}  \right)^\frac{\epsilon}{\epsilon-1}$, where $Y_{i,t}$ is the output of variety $i\in[0,1]$. Let $P_{i,t}$ denote the price of variety $i$. Minimizing costs $\int_0^1 P_{i,t}Y_{i,t}\dd{i}$ leads to the demand function $Y_{i,t}=\left(\frac{P_{i,t}}{P_{t}}\right)^{-\epsilon}Y_{t}$, where $P_{t}\equiv\left(\int_0^1 {P_{i,t}}^{1-\epsilon}\dd{i}  \right)^\frac{1}{1-\epsilon}$ is the aggregate price level. Define a random variable $\mathcal{U}_{i,t}$:
\begin{equation*}
    \mathcal{U}_{i,t}=
        \begin{cases}
            1 & \text{if price $P_{i,t}$ can be updated in period $t$,}\\
            0 & \text{otherwise.}
        \end{cases}
\end{equation*}
Let $P^*_{i,t}$ denote the updated price if $\mathcal{U}_{i,t}=1$. A retailer's problem is then described by the Bellman equation
\begin{equation*}
    V_{t}(P_{i,t-1},\mathcal{U}_{i,t})=\max_{P^*_{i,t}}\{\Phi_{t}(P_{i,t})+\mathbb{E}_{t}(\Lambda_{t,t+1}V_{t+1}(P_{i,t},\mathcal{U}_{i,t+1}))\}
\end{equation*}
subject to
\begin{align*}
    P_{i,t}&=(1-\mathcal{U}_{i,t})P_{i,t-1}+\mathcal{U}_{i,t}P^*_{i,t},\\
    \Phi_{t}(P_{i,t})&=\left(\frac{P_{i,t}}{P_{t}}-P^w_{t}\right)\left(\frac{P_{i,t}}{P_{t}}\right)^{-\epsilon}Y_{t}.
\end{align*}
Assuming that $V$ is differentiable with respect to the first argument and denoting this derivative as $V'$, the first-order and envelope conditions are
\begin{align*}
    0&=\mathcal{U}_{i,t}[\Phi_{t}'(P_{i,t})+\mathbb{E}_{t}(\Lambda_{t,t+1}V_{t+1}'(P_{i,t},\mathcal{U}_{i,t+1}))],\\
    V_{t}'(P_{i,t-1},\mathcal{U}_{i,t})&=(1-\mathcal{U}_{i,t})[\Phi_{t}'(P_{i,t})+\mathbb{E}_{t}(\Lambda_{t,t+1}V_{t+1}'(P_{i,t},\mathcal{U}_{i,t+1}))],
\end{align*}
where
\begin{equation*}
    \Phi_{t}'(P_{i,t})=\frac{1}{P_{t}}\left(\frac{P_{i,t}}{P_{t}}\right)^{-\epsilon}Y_{t}\left(1-\epsilon+\epsilon{}P^w_{t}\frac{P_{t}}{P_{i,t}}\right).
\end{equation*}
Observing that $\mathcal{U}_{i,t+s}=0$ with probability $\theta$ for all $i\in[0,1]$, $t\ge{0}$ and $s\in\mathbb{Z}_+$, iterating the envelope condition forward, and applying the law of iterated expectations, one could obtain
\begin{equation*}
    V_{t}'(P_{i,t-1},\mathcal{U}_{i,t})=(1-\mathcal{U}_{i,t})\mathbb{E}_{t}\left(\sum_{s=0}^\infty\theta^s\Lambda_{t,t+s}\Phi_{t+s}'(P_{i,t})+\lim_{s\rightarrow\infty}\theta^{s-1}\Lambda_{t,t+s}V_{t+s}'(P_{i,t},\mathcal{U}_{i,t+s})\right).
\end{equation*}
Since $\theta\in[0,1]$ and $\beta\in(0,1)$, under appropriate boundedness assumptions, the limit term is zero almost surely. Then the first-order condition (FOC) becomes
\begin{equation*}
    \mathcal{U}_{i,t}\mathbb{E}_{t}\left(\sum_{s=0}^\infty\theta^s\Lambda_{t,t+s}\Phi_{t+s}'(P_{i,t})\right)=0.
\end{equation*}
If $\mathcal{U}_{i,t}=0$, a retailer cannot update the price, so $P^*_{i,t}$ is irrelevant, or, in other words, any $P^*_{i,t}\in\mathbb{R}$ is a solution to the maximization problem. The case $\mathcal{U}_{i,t}=1$ and $P_{i,t}=P^*_{i,t}$ is the only interesting one. Substituting the derivative of the profit function and rearranging, one obtains
\begin{align*}
    &\mathbb{E}_{t}\left[\sum_{s=0}^\infty\theta^s\Lambda_{t,t+s}\left(\frac{P_{t+s}}{P_{t}}\right)^{\epsilon-1}Y_{t+s}\right]\frac{P^*_{i,t}}{P_{t}}=\frac{\epsilon}{\epsilon-1}\mathbb{E}_{t}\left[\sum_{s=0}^\infty\theta^s\Lambda_{t,t+s}\left(\frac{P_{t+s}}{P_{t}}\right)^{\epsilon}P^w_{t+s}Y_{t+s}\right].
\end{align*}
This equation clarifies that the optimal price is the same for all retailers---from now on, we will use the notation $P^*_{t}$ instead of $P^*_{i,t}$. For computational convenience, the infinite sums can be represented as follows.
\begin{align*}
    \Omega_{1,t}&\equiv\mathbb{E}_{t}\left[\sum_{s=0}^\infty\theta^s\Lambda_{t,t+s}\left(\frac{P_{t+s}}{P_{t}}\right)^{\epsilon}P^w_{t+s}Y_{t+s}\right]\\
    &=\mathbb{E}_{t}\left[P^w_{t}Y_{t}+\sum_{s=1}^\infty\theta^s\Lambda_{t,t+s}\left(\frac{P_{t+s}}{P_{t}}\right)^{\epsilon}P^w_{t+s}Y_{t+s}\right]\\
    &=P^w_{t}Y_{t}+\theta\mathbb{E}_{t}\left\{\Lambda_{t,t+1}\Pi_{t+1}^{\epsilon}\mathbb{E}_{t+1}\left[\sum_{s=0}^\infty\theta^{s}\Lambda_{t+1,t+1+s}\left(\frac{P_{t+1+s}}{P_{t+1}}\right)^{\epsilon}P^w_{t+1+s}Y_{t+1+s}\right]\right\}\\
    &=P^w_{t}Y_{t}+\theta\mathbb{E}_{t}(\Lambda_{t,t+1}\Pi_{t+1}^{\epsilon}\Omega_{1,t+1}),
\end{align*}
which is \eqref{eq: retailers marginal cost}. Similarly,
\begin{align*}
    \Omega_{2,t}&\equiv\mathbb{E}_{t}\left[\sum_{s=0}^\infty\theta^s\Lambda_{t,t+s}\left(\frac{P_{t+s}}{P_{t}}\right)^{\epsilon-1}Y_{t+s}\right]\\
    &=Y_{t}+\theta\mathbb{E}_{t}(\Lambda_{t,t+1}\Pi_{t+1}^{\epsilon-1}\Omega_{2,t+1}),
\end{align*}
which is \eqref{eq: retailers marginal benefit}. Hence, the optimal relative price can be expressed as $\widetilde{P}_{t}\equiv\frac{P^*_{t}}{P_{t}}=\frac{\epsilon}{\epsilon-1}\frac{\Omega_{1,t}}{\Omega_{2,t}}$, which is \eqref{eq: retailers optimal relative price}. Recognizing that $P^*_{i,t}=P^*_{t}$ and noting the independence of $\mathcal{U}_{i,t}$ both across $i$ and $t$, the aggregate version of the price evolution equation can be obtained:
\begin{align*}
    P_{i,t}&=(1-\mathcal{U}_{i,t})P_{i,t-1}+\mathcal{U}_{i,t}P^*_{t}\\
    \Longleftrightarrow {P_{i,t}}^{1-\epsilon}&=(1-\mathcal{U}_{i,t}){P_{i,t-1}}^{1-\epsilon}+\mathcal{U}_{i,t}{P^*_{t}}^{1-\epsilon}\\
    \Longrightarrow\int_0^1{P_{i,t}}^{1-\epsilon}\dd{i}&=\int_{\{i\mid\mathcal{U}_{i,t}=0\}}{P_{i,t-1}}^{1-\epsilon}\dd{i}+{P^*_{t}}^{1-\epsilon}\int_{\{i\mid\mathcal{U}_{i,t}=1\}}\dd{i}\\
    &=\theta\int_0^1{P_{i,t-1}}^{1-\epsilon}\dd{i}+(1-\theta){P^*_{t}}^{1-\epsilon}\\
    \Longleftrightarrow{P_{t}}^{1-\epsilon}&=\theta{P_{t-1}}^{1-\epsilon}+(1-\theta){P^*_{t}}^{1-\epsilon}.
\end{align*}
Normalizing, one could get $\Pi_{t}^{1-\epsilon}=\theta+(1-\theta)(\Pi_{t}\widetilde{P}_{t})^{1-\epsilon}$, which is \eqref{eq: retailers inflation}. Equivalently, $\widetilde{P}_{t}=\left(\frac{1-\theta}{1-\theta\Pi_{t}^{\epsilon-1}}\right)^\frac{1}{\epsilon-1}$. Note that the wholesale good market clears as $Y^w_{t}=\int_0^1{}Y_{i,t}\dd{i}=Y_{t}\int_0^1\left(\frac{P_{i,t}}{P_{t}}\right)^{-\epsilon}\dd{i}$. Define $\Delta_{t}\equiv\int_0^1\left(\frac{P_{i,t}}{P_{t}}\right)^{-\epsilon}\dd{i}$, which is a measure of price dispersion. Then we get $Y^w_{t}=\Delta_{t}Y_{t}$, which is \eqref{eq: market clearing wholesale good}. Note that
\begin{align*}
    \Delta_{t}&=\int_{\{i\mid\mathcal{U}_{i,t}=0\}}\left(\frac{P_{i,t-1}}{P_{t}}\right)^{-\epsilon}\dd{i}+\left(\frac{P^*_{t}}{P_{t}}\right)^{-\epsilon}\int_{\{i\mid\mathcal{U}_{i,t}=1\}}\dd{i}\\
    &={\Pi_{t}}^\epsilon\int_{\{i\mid\mathcal{U}_{i,t}=0\}}\left(\frac{P_{i,t-1}}{P_{t-1}}\right)^{-\epsilon}\dd{i}+(1-\theta){\widetilde{P}_{t}}^{-\epsilon}\\
    &=\theta\Pi_{t}^{\epsilon}\Delta_{t-1}+(1-\theta){\widetilde{P}_{t}}^{-\epsilon},
\end{align*}
which is \eqref{eq: retailers price dispersion}.

\section{Proofs}\label{sec: proofs}

\subsection{Lemma \ref{lemma: steady state uniqueness}}
Note that \eqref{eq: workers Euler} and \eqref{eq: bankers Euler deposits} imply $\lambda^b=U^b_{C}\left(1-\frac{\beta_b}{\beta}\right)$. Using the latter, \eqref{eq: bankers Euler loans}, \eqref{eq: entrepreneurs Euler loans}, and the definition of $\widetilde{\beta}_e$, we get $\lambda^e=\frac{U^e_{C}}{R^L}\left(1-\frac{\beta_e}{\widetilde{\beta}_e}\right)$.

\paragraph{If}
Suppose $\beta_b<\beta$ and $\beta_e<\widetilde{\beta}_e$. Then $\lambda^b>0$ and $\lambda^e>0$, which implies that \eqref{eq: bankers leverage constraint} and \eqref{eq: entrepreneurs collateral constraint} are binding. Section \ref{sec: SS CE} provides a closed-form sequential solution for a unique steady state, where we set $\tau^D=\tau^K=\tau^L=\tau^N=0$. As shown there, the binding collateral constraint is used to solve for $L>0$ conditional on $K>0$. The binding leverage constraint is then used to solve for $D>0$ conditional on $L>0$.

\paragraph{Only if}
Suppose there exists a unique steady state with $D>0$ and $L>0$. Since $\lambda^b\ge{}0$ and $\lambda^e\ge{}0$, we must have $\beta_b\le\beta$ and $\beta_e\le\widetilde{\beta}_e$. If $\beta_b=\beta$, then \eqref{eq: bankers Euler deposits} is equivalent to $\lambda^b=0$. Moreover, the complementary slackness conditions \eqref{eq: bankers complementary slackness} are automatically satisfied. Since $L>0$ by the premise, any $D\in(0,(1-\kappa)L]$ can be part of an unstable steady state, which contradicts uniqueness. It follows that $\beta_b<\beta$. An identical argument applied to \eqref{eq: entrepreneurs collateral constraint} and \eqref{eq: entrepreneurs Euler loans} demonstrates that we must have $\beta_e<\widetilde{\beta}_e$.$\quad\blacksquare$

\subsection{Lemma \ref{lemma: net assets}}

\paragraph{Bankers}
Multiply both sides of \eqref{eq: bankers Euler deposits} by $D_{t}$, multiply both sides of \eqref{eq: bankers Euler loans} by $L_{t}$, and subtract the former from the latter:
\begin{equation*}
    U^b_{C,t}(L_{t}-D_{t})=\beta_b\mathbb{E}_{t}[U^b_{C,t+1}(R^L_{t+1}L_{t}-R_{t}D_{t})]+\lambda^b_{t}[(1-\kappa_{t})L_{t}-D_{t}].
\end{equation*}
Using \eqref{eq: bankers budget constraint} and \eqref{eq: bankers complementary slackness},
\begin{equation*}
    U^b_{C,t}(L_{t}-D_{t})=\beta_b\mathbb{E}_{t}(U^b_{C,t+1}C^b_{t+1})+\beta_b\mathbb{E}_{t}[U^b_{C,t+1}(L_{t+1}-D_{t+1})].
\end{equation*}
Iterating this equation forward, we obtain
\begin{equation*}
    L_{t}-D_{t}=\frac{1}{U^b_{C,t}}\sum_{s=1}^\infty\beta_b^{s}\mathbb{E}_{t}(U^b_{C,t+s}C^b_{t+s}).
\end{equation*}

\paragraph{Entrepreneurs}
The argument is symmetric to the case of bankers. Multiply \eqref{eq: entrepreneurs Euler loans} by $L_{t}$ and \eqref{eq: entrepreneurs Euler capital} by $K_{t}$, subtract the former from the latter and use \eqref{eq: entrepreneurs budget constraint}, \eqref{eq: entrepreneurs labor demand}, and \eqref{eq: entrepreneurs complementary slackness}, noting that $F$ is Cobb---Douglas, to obtain
\begin{equation*}
    U^e_{C,t}(Q_{t}K_{t}-L_{t})=\beta_e\mathbb{E}_{t}(U^e_{C,t+1}C^e_{t+1})+\beta_e\mathbb{E}_{t}[U^e_{C,t+1}(Q_{t+1}K_{t+1}-L_{t+1})].
\end{equation*}
Iterating forward, we get
\begin{equation*}
    Q_{t}K_{t}-L_{t}=\frac{1}{U^e_{C,t}}\sum_{s=1}^\infty\beta_e^{s}\mathbb{E}_{t}(U^e_{C,t+s}C^e_{t+s}).\quad\blacksquare
\end{equation*}

\subsection{Lemma \ref{lemma: stationary welfare}}
The definition of $\mathcal{W}^i_{t}$ implies
\begin{align*}
    \frac{\beta}{\beta-\beta_i}(\mathcal{W}^i_{t}-V^i_{t})&=\mathbb{E}_{t}\left(\sum_{s=1}^\infty\beta^s{}V^i_{t+s}\right)\\
    &=\beta\mathbb{E}_{t}(V^i_{t+1})+\beta\mathbb{E}_{t}\left[\mathbb{E}_{t+1}\left(\sum_{s=1}^\infty\beta^{s}{}V^i_{t+1+s}\right)\right]\\
    &=\beta\mathbb{E}_{t}(V^i_{t+1})+\beta\mathbb{E}_{t}\left[\frac{\beta}{\beta-\beta_i}(\mathcal{W}^i_{t+1}-V^i_{t+1})\right].
\end{align*}
Hence,
\begin{align*}
    \mathcal{W}^i_{t}&=V^i_{t}-\beta_i\mathbb{E}_{t}(V^i_{t+1})+\beta\mathbb{E}_{t}(\mathcal{W}^i_{t+1})\\
    &=U^i_{t}+\beta\mathbb{E}_{t}(\mathcal{W}^i_{t+1})\\
    &=\mathbb{E}_{t}\left(\sum_{s=0}^\infty\beta^s U^i_{t+s}\right).\quad\blacksquare
\end{align*}

\subsection{Proposition \ref{prop: FCEA}}
Define $\lambda^L_{t}\equiv\lambda^L_{1,t}+\lambda^L_{2,t}[(1-\kappa_{t})L_{t}-D_{t}]$. The FOCs are
\begin{align*}
    C^b_{t}:\quad{}0&=\omega_b{}U^b_{C,t}-\lambda^Y_{t}-\lambda^C_{t}+\lambda^L_{t}U^b_{CC,t}L_{t}-\frac{\bm{1}_\mathbb{N}(t)}{\beta}[\lambda^L_{t-1}\beta_b(U^b_{CC,t}R^L_{t}L_{t-1}+U^b_{C,t})+\lambda^e_{t-1}],\\
    C^e_{t}:\quad{}0&=\omega_e{}U^e_{C,t}-\lambda^Y_{t}-\lambda^C_{t},\\
    C^w_{t}:\quad{}0&=\omega_w{}U^w_{C,t}-\lambda^Y_{t}-\lambda^C_{t}W_{C,t}N_{t}-[\lambda^L_{t}\beta_b\mathbb{E}_{t}(U^b_{C,t+1})+\beta\mathbb{E}_{t}(\lambda^C_{t+1})+\lambda^e_{t}]R_{1,t}U^w_{CC,t}D_{t}\\
    &\quad-\frac{\bm{1}_\mathbb{N}(t)}{\beta}[\lambda^L_{t-1}\beta_b\mathbb{E}_{t-1}(U^b_{C,t})+\beta\mathbb{E}_{t-1}(\lambda^C_{t})+\lambda^e_{t-1}]R_{2,t-1}U^w_{CC,t}D_{t-1},\\
    D_{t}:\quad{}0&\ge-\lambda^b_{t}-\lambda^L_{2,t}[U^b_{C,t}-\beta_b\mathbb{E}_{t}(U^b_{C,t+1}R^L_{t+1})]L_{t}+\lambda^C_{t}-[\lambda^L_{t}\beta_b\mathbb{E}_{t}(U^b_{C,t+1})+\beta\mathbb{E}_{t}(\lambda^C_{t+1})\\
    &\quad+\lambda^e_{t}]R_{t}+\frac{\bm{1}_\mathbb{N}(t)}{\beta}(\lambda^L_{t-1}\beta_b{}U^b_{C,t}+\lambda^e_{t-1}),\qquad\text{equality if }D_{t}>0,\\
    K_{t}:\quad{}0&=-\lambda^C_{t}\{Q_{2,t}[K_{t}-(1-\delta)\xi_{t}K_{t-1}]+Q_{t}\}+\lambda^e_{t}m_{t}\mathbb{E}_{t}[(Q_{1,t+1}K_{t}+Q_{t+1})\xi_{t+1}]-\lambda^Y_{t}I_{2,t}\\
    &\quad+\beta\mathbb{E}_{t}[(\lambda^C_{t+1}+\lambda^Y_{t+1})A_{t+1}F_{K,t+1}\xi_{t+1}+\lambda^C_{t+1}\{Q_{t+1}(1-\delta)\xi_{t+1}\\
    &\quad-Q_{1,t+1}[K_{t+1}-(1-\delta)\xi_{t+1}K_{t}]\}-\lambda^Y_{t+1}I_{1,t+1}]+\frac{\bm{1}_\mathbb{N}(t)}{\beta}\lambda^e_{t-1}m_{t-1}Q_{2,t}\xi_{t}K_{t-1},\\
    L_{t}:\quad{}0&=\{\lambda^b_{t}+\lambda^L_{2,t}[U^b_{C,t}-\beta_b\mathbb{E}_{t}(U^b_{C,t+1}R^L_{t+1})]L_{t}\}(1-\kappa_{t})+\lambda^L_{t}U^b_{C,t}\\
    &\quad-\frac{\bm{1}_\mathbb{N}(t)}{\beta}(\lambda^L_{t-1}\beta_b{}U^b_{C,t}+\lambda^e_{t-1}),\\
    N_{t}:\quad{}0&=\omega_w{}U^w_{N,t}+(\lambda^C_{t}+\lambda^Y_{t})A_{t}F_{N,t}-[\lambda^L_{t}\beta_b\mathbb{E}_{t}(U^b_{C,t+1})+\beta\mathbb{E}_{t}(\lambda^C_{t+1})+\lambda^e_{t}]R_{1,t}U^w_{CN,t}D_{t}\\
    &\quad-\lambda^C_{t}(W_{N,t}N_{t}+W_{t})\\
    &\quad-\frac{\bm{1}_\mathbb{N}(t)}{\beta}[\lambda^L_{t-1}\beta_b\mathbb{E}_{t-1}(U^b_{C,t})+\beta\mathbb{E}_{t-1}(\lambda^C_{t})+\lambda^e_{t-1}]R_{2,t-1}U^w_{CN,t}D_{t-1}.
\end{align*}
The complementary slackness conditions are
\begin{align*}
    0&=\lambda^b_{t}[(1-\kappa_{t})L_{t}-D_{t}],\qquad\lambda^b_{t}\ge{}0,\\
    0&=\lambda^L_{1,t}[U^b_{C,t}L_{t}-\beta_b\mathbb{E}_{t}(U^b_{C,t+1}B_{t+1})],\qquad{}D_{t}\lambda^L_{1,t}\ge{}0,\\
    0&=\lambda^e_{t}[m_{t}\mathbb{E}_{t}(Q_{t+1}\xi_{t+1})K_{t}-\mathbb{E}_{t}(B_{t+1})],\qquad\lambda^e_{t}\ge{}0.
\end{align*}

\subsubsection{Constrained inefficiency}\label{sec: FCEA wedges}
Follows from inspecting the planner's analogs of \eqref{eq: bankers Euler deposits} and \eqref{eq: entrepreneurs labor demand}--\eqref{eq: entrepreneurs Euler capital}. Consider them one-by-one.

\paragraph{Deposit supply}
The FOCs for $C^b_{t}$ and $D_{t}$ imply
\begin{equation*}
    U^b_{C,t}\le\beta_b{}R_{t}\mathbb{E}_{t}(U^b_{C,t+1})+\frac{\lambda^b_{t}}{\omega_b}+\Psi^D_{t},\qquad\text{equality if }D_{t}>0,
\end{equation*}
where
\begin{multline*}
    \omega_b\Psi^D_{t}\equiv(\beta-\beta_b)R_{t}\mathbb{E}_{t}(\omega_b{}U^b_{C,t+1})+\lambda^Y_{t}-\beta{}R_{t}\mathbb{E}_{t}(\lambda^Y_{t+1})+\lambda^L_{2,t}[U^b_{C,t}-\beta_b\mathbb{E}_{t}(U^b_{C,t+1}R^L_{t+1})]L_{t}\\
    -\lambda^L_{t}[U^b_{CC,t}+\beta_b{}R_{t}\mathbb{E}_{t}(U^b_{CC,t+1}R^L_{t+1})]L_{t}+\beta{}R_{t}\mathbb{E}_{t}(\lambda^L_{t+1}U^b_{CC,t+1}L_{t+1})+\frac{\bm{1}_\mathbb{N}(t)}{\beta}\lambda^L_{t-1}\beta_b{}U^b_{CC,t}R^L_{t}L_{t-1}.
\end{multline*}

\paragraph{Loan demand}
If $D_{t}>0$, the FOCs for $C^e_{t}$, $D_{t}$, and $L_{t}$ imply
\begin{equation*}
    U^e_{C,t}=\beta_e\mathbb{E}_{t}(U^e_{C,t+1}R^L_{t+1})+\frac{\lambda^e_{t}}{\omega_e}\mathbb{E}_{t}(R^L_{t+1})+\Psi^L_{t},
\end{equation*}
where
\begin{align*}
    \omega_e\Psi^L_{t}&=(\beta-\beta_e)\mathbb{E}_{t}(\omega_e{}U^e_{C,t+1}R^L_{t+1})-\mathbb{E}_{t}[(\beta\omega_e{}U^e_{C,t+1}+\lambda^e_{t})(R^L_{t+1}-R_{t})]+\lambda^Y_{t}-\beta{}R_{t}\mathbb{E}_{t}(\lambda^Y_{t+1})\\
    &\quad-\lambda^L_{t}\left[\frac{U^b_{C,t}}{1-\kappa_{t}}-\beta_b{}R_{t}\mathbb{E}_{t}(U^b_{C,t+1})\right]+\frac{\bm{1}_\mathbb{N}(t)}{\beta}\frac{\kappa_{t}}{1-\kappa_{t}}(\lambda^L_{t-1}\beta_b{}U^b_{C,t}+\lambda^e_{t-1}).
\end{align*}
If $D_{t}=0$, we still have $L_{t}>0$, so the FOC for $L_{t}$ holds. To see this, note that the leverage constraint implies $C^b_{t+1}+L_{t+1}-D_{t+1}\ge{}0$, and the inequality is strict if $C^b_{t+1}>0$. Provided that $C^b_{t+1}>0$ with positive measure, which is guaranteed if bankers are risk averse and the Inada condition holds, $D_{t}=L_{t}=0$ would contradict the constraint associated with $\lambda^L_{1,t}$. Note that the FOCs for $C^b_{t}$ and $C^e_{t}$ yield the following general relationship between the marginal utilities
\begin{equation*}
    \omega_b{}U^b_{C,t}=\omega_e{}U^e_{C,t}-\lambda^L_{t}U^b_{CC,t}L_{t}+\frac{\bm{1}_\mathbb{N}(t)}{\beta}[\lambda^L_{t-1}\beta_b(U^b_{CC,t}R^L_{t}L_{t-1}+U^b_{C,t})+\lambda^e_{t-1}].
\end{equation*}
With $D_{t}=0$, we have $U^b_{C,t}=\beta_b\mathbb{E}_{t}(U^b_{C,t+1}R^L_{t+1})$. The FOC for $L_{t}$ then implies $\lambda^L_{t}U^b_{C,t}=\frac{\bm{1}_\mathbb{N}(t)}{\beta}(\lambda^L_{t-1}\beta_b{}U^b_{C,t}+\lambda^e_{t-1})$ at $t$. Combining these results, if $D_{t}=0$, the wedge satisfies
\begin{multline*}
    \omega_e\Psi^L_{t}=(\beta_b-\beta_e)\mathbb{E}_{t}(\omega_e{}U^e_{C,t+1}R^L_{t+1})\\
    +\lambda^L_{t}\left\{U^b_{CC,t}L_{t}-U^b_{C,t}+\frac{\beta_b^2}{\beta}\mathbb{E}_{t}[(U^b_{CC,t+1}R^L_{t+1}L_{t}+U^b_{C,t+1})R^L_{t+1}]\right\}\\
    -\beta_b\mathbb{E}_{t}(\lambda^L_{t+1}U^b_{CC,t+1}L_{t+1}R^L_{t+1})-\frac{\bm{1}_\mathbb{N}(t)}{\beta}\lambda^L_{t-1}\beta_b{}U^b_{CC,t}R^L_{t}L_{t-1}-\frac{\beta-\beta_b}{\beta}\lambda^e_{t}\mathbb{E}_{t}(R^L_{t+1}).
\end{multline*}

\paragraph{Labor demand}
The FOCs for $C^e_{t}$, $C^w_{t}$, and $N_{t}$ combined with the definition of $W_{t}$ imply
\begin{equation*}
    W_{t}=A_{t}F_{N,t}+\Psi^N_{t},
\end{equation*}
where
\begin{multline*}
    \Psi^N_{t}=\frac{(\omega_e{}U^e_{C,t}-\omega_w{}U^w_{C,t}-\lambda^C_{t})A_{t}F_{N,t}-\lambda^C_{t}W_{N,t}N_{t}}{\omega_w{}U^w_{C,t}+\lambda^C_{t}}\\
    -\frac{U^w_{CN,t}}{U^w_{CC,t}}\frac{\omega_w{}U^w_{C,t}-\omega_e{}U^e_{C,t}+\lambda^C_{t}(1-W_{C,t}N_{t})}{\omega_w{}U^w_{C,t}+\lambda^C_{t}}.
\end{multline*}

\paragraph{Capital demand}
The FOCs for $C^e_{t}$ and $K_{t}$ imply
\begin{equation*}
    U^e_{C,t}Q_{t}=\beta_e\mathbb{E}_{t}\{U^e_{C,t+1}[A_{t+1}F_{K,t+1}+Q_{t+1}(1-\delta)]\xi_{t+1}\}+\frac{\lambda^e_{t}}{\omega_e}m_{t}\mathbb{E}_{t}(Q_{t+1}\xi_{t+1})+\Psi^K_{t},
\end{equation*}
where, using the form of $I$,
\begin{multline*}
    \omega_e\Psi^K_{t}=(\beta-\beta_e)\mathbb{E}_{t}(\omega_e{}U^e_{C,t+1}R^K_{t+1})Q_{t}+\beta\mathbb{E}_{t}\left\{\lambda^Y_{t+1}\left[Q_{t+1}\Phi\left(\frac{I_{t+1}}{K_{t}}\right)-\frac{I_{t+1}}{K_{t}}\right]\right\}\\
    -\lambda^C_{t}Q_{2,t}[K_{t}-(1-\delta)\xi_{t}K_{t-1}]-\beta\mathbb{E}_{t}\{\lambda^C_{t+1}Q_{1,t+1}[K_{t+1}-(1-\delta)\xi_{t+1}K_{t}]\}\\
    +\lambda^e_{t}m_{t}\mathbb{E}_{t}(Q_{1,t+1}\xi_{t+1})K_{t}+\frac{\bm{1}_\mathbb{N}(t)}{\beta}\lambda^e_{t-1}m_{t-1}Q_{2,t}\xi_{t}K_{t-1}.
\end{multline*}

\subsubsection{Risk sharing}
That consumption insurance is generally imperfect follows immediately from inspecting the FOCs with respect to $C^b_{t}$, $C^e_{t}$, and $C^w_{t}$. The same applies to partial risk sharing between bankers and entrepreneurs. Note that the FOCs for $D_{t}$ and $L_{t}$ imply a steady-state relationship $\lambda^e=\lambda^L(\beta-\beta_b)U^b_{C}(C^b)$. Hence, $\lambda^e$ and $\lambda^L$ are either both zero or both positive.

Suppose workers have separable preferences $U^w(C^w,N)=u(C^w)-v(N)$ and $\lambda^e=\lambda^L=0$. In this case, $\omega_b{}U^b_C(C^b)=\omega_e{}U^e_C(C^e)=\lambda^Y+\lambda^C$. Using the definition of functions $R$ and $W$, we have $R_{1}=\frac{R}{u'(C^w)}$, $R_{2}=-\frac{\beta{}R^2}{u'(C^w)}$, $W_{C}=-W\frac{u''(C^w)}{u'(C^w)}$, and $\beta{}R=1$. The FOC for $C^w$ then implies
\begin{align*}
    0&=\omega_w{}u'(C^w)-\lambda^Y+\lambda^C\frac{u''(C^w)}{u'(C^w)}[W{}N+(R-1)D]\\
    &=\omega_w{}u'(C^w)-\lambda^Y+\lambda^C\frac{u''(C^w)}{u'(C^w)}C^w,
\end{align*}
where the second equality is true if the steady-state profits of capital good producers are zero so that the worker's budget constraint implies $C^w=W{}N+(R-1)D$. It follows that $\omega_w{}u'(C^w)=\lambda^Y+\lambda^C$ if and only if $(-C^w)\frac{u''(C^w)}{u'(C^w)}=1$ if and only if $u(\cdot)=\ln(\cdot)$.

\subsubsection{Indeterminacy and optimal steady state}
Section \ref{sec: SS FCEA} shows that the steady state construction reduces to considering two cases, $\lambda^L=0$ and $\lambda^L>0$. If $\lambda^L=0$, $D$ must satisfy the rearranged collateral constraint:
\begin{equation*}
    C^b+(R-1)D+\max\left\{\frac{1}{1-\kappa}D,\frac{\beta_b}{1-\beta_b}[C^b+(R-1)D]\right\}\le{}m{}Q\xi{}K.
\end{equation*}
If $\lambda^L>0$, we instead have a rearranged bank leverage constraint: $D\le\frac{\beta_b(1-\kappa)}{1-\beta_b[1+(1-\kappa)(R-1)]}C^b$. In both cases, there is a generally infinite set of solutions $D\in[0,\Bar{D}]$ for some $\Bar{D}>0$. Since there is an uncountable infinity of steady states, each such steady state is unstable, and the FCEA is locally indeterminate. Numerical analysis under the baseline calibration demonstrates that each choice of $D$ yields either a unique solution to a nonlinear system or no solutions, and welfare $\mathcal{W}$ is strictly decreasing in $D$. The latter is related to the problem of finding an optimal steady state.

Consider the planner's problem with no uncertainty, restricting attention to constant plans. An optimal plan of this sort will define the optimal steady state. In the steady state, $R=\frac{1}{\beta}$, $\frac{I}{K}=\Phi^{-1}[1-(1-\delta)\xi]$, and $Q=\left[\Phi'\left(\frac{I}{K}\right)\right]^{-1}$. Moreover, the constraints associated with $\lambda^b$, $\lambda^L_{1}$, and $\lambda^L_2$ are equivalent to
\begin{equation*}
    L=\max\left\{\frac{1}{1-\kappa}D,\frac{\beta_b}{1-\beta_b}[C^b+(R-1)D]\right\},
\end{equation*}
conditional on $(C^b,D)$. The optimal steady state is then a solution to
\begin{equation*}
    \max_{(C^b,C^e,C^w,D,K,N)}\sum_{i\in\mathcal{I}}\omega_i{}U^i
\end{equation*}
subject to
\begin{align*}
    \lambda^C:\quad{}0&=A{}F(\xi{}K,N)-Q[1-(1-\delta)\xi]K-W(C^w,N)N-(R-1)D-C^b-C^e,\\
    \lambda^e:\quad{}0&\le{}m{}Q\xi{}K-[C^b+(R-1)D+L],\\
    \lambda^Y:\quad{}0&=A{}F(\xi{}K,N)-\sum_{i\in\mathcal{I}}C^i-\frac{I}{K}K.
\end{align*}
Conditional on $C^b$, $L$ is a strictly increasing function of $D$, differentiable everywhere except at the kink. We can assume without loss of generality that the derivative at the kink is an average of the left and right derivatives. Suppose $(C^b,C^e,C^w,D,K,N)$ is optimal, where $D>0$. It must satisfy the FOC for $D$:
\begin{equation*}
    0=-\lambda^C(R-1)-\lambda^e\left(R-1+\frac{\partial{}L}{\partial{}D}\right).
\end{equation*}
Note that $R>1$, $\lambda^e\ge{}0$, and $\frac{\partial{}L}{\partial{}D}>0$. If, moreover, $\lambda^C>0$, we have $-\lambda^C(R-1)-\lambda^e\left(R-1+\frac{\partial{}L}{\partial{}D}\right)<0$, which is a contradiction. Therefore, $D=0$ is optimal.

Intuitively, $\lambda^C$ must be positive since it is the shadow value of wealth associated with the aggregate budget constraint of bankers and entrepreneurs. Assume separable preferences and combine the FOCs for $C^w$ and $N$ together with the definition of $W$ to obtain
\begin{equation*}
    \lambda^C=\frac{\omega_w{}u'(C^w)(A{}F_{N}-W)}{(W_{C}N-1)A{}F_{N}+W_{N}N+W}.
\end{equation*}
By definition, $W_C>0$ and $W_N>0$. Hence, if $N$ and $C^w$ are less than in the first-best allocation, $F_N$ must be greater and $W$ less; therefore, $A{}F_N-W>0$. A sufficient---but not necessary---condition for the denominator to be positive is $W_C{}N\ge{}1$. If $u$ has constant relative risk aversion $\gamma_w>0$, as is the case in the quantitative analysis, $W_C=-W\frac{u''(C^w)}{u'(C^w)}=\frac{W}{C^w}\gamma_w$. If $\gamma_w$ is large enough, we are done. Alternatively, if $\gamma_w\approx{1}$ and $D\approx{0}$, then $C_w\approx{}W{}N$ and $W_C{}N\approx{1}$; therefore, $(W_{C}N-1)A{}F_{N}\approx{0}$. Since $W_{N}N+W>0$, we then have $\lambda^C>0$.$\quad\blacksquare$

\subsection{Proposition \ref{prop: FCEA decentralization}}

\paragraph{Bankers}
Note that the form of $T^b_{t}$ ensures that \eqref{eq: bankers budget constraint} is true in equilibrium. The Euler equation for deposits is now
\begin{equation*}
    U^b_{C,t}(1-\tau^D_{t})\le\beta_b{}R_{t}\mathbb{E}_{t}(U^b_{C,t+1})+\lambda^b_{t},\qquad\text{equality if }D_{t}>0.
\end{equation*}
Using \eqref{eq: bankers Euler loans}---which remains unchanged relative to the FCE---to solve for $\lambda^b_{t}$, the Euler equation for deposits can be rearranged as
\begin{equation*}
    U^b_{C,t}\le\beta_b{}R_{t}\mathbb{E}_{t}(U^b_{C,t+1})+\frac{U^b_{C,t}-\beta_b\mathbb{E}_{t}(U^b_{C,t+1}R^L_{t+1})}{1-\kappa_{t}}+\tau^D_{t}U^b_{C,t},\qquad\text{equality if }D_{t}>0.
\end{equation*}
As follows from section \ref{sec: FCEA wedges}, the right-hand side is equivalent to the one in the FCEA if and only if
\begin{equation*}
    \tau^D_{t}=\frac{1}{U^b_{C,t}}\left[\frac{\lambda^b_{t}}{\omega_b}-\frac{U^b_{C,t}-\beta_b\mathbb{E}_{t}(U^b_{C,t+1}R^L_{t+1})}{1-\kappa_{t}}+\Psi^D_{t}\right].
\end{equation*}

\paragraph{Entrepreneurs}
The form of $T^e_{t}$ guarantees that \eqref{eq: entrepreneurs budget constraint} holds in equilibrium. Without loss of generality, let $\frac{\lambda^e_{t}}{\omega_e}$ denote the scaled Lagrange multiplier on the collateral constraint. The modified FOCs are
\begin{align*}
    (1+\tau^N_{t})W_{t}&=A_{t}F_{N,t},\\
    U^e_{C,t}(1-\tau^L_{t})&=\beta_e\mathbb{E}_{t}(U^e_{C,t+1}R^L_{t+1})+\frac{\lambda^e_{t}}{\omega_e}\mathbb{E}_{t}(R^L_{t+1}),\\
    U^e_{C,t}(1+\tau^K_{t})Q_{t}&=\beta_e\mathbb{E}_{t}\{U^e_{C,t+1}[A_{t+1}F_{K,t+1}+Q_{t+1}(1-\delta)]\xi_{t+1}\}+\frac{\lambda^e_{t}}{\omega_e}m_{t}\mathbb{E}_{t}(Q_{t+1}\xi_{t+1}).
\end{align*}
Section \ref{sec: FCEA wedges} then immediately implies that we must set
\begin{equation*}
    \tau^N_{t}=\frac{-\Psi^N_{t}}{W_{t}},\qquad
    \tau^L_{t}=\frac{\Psi^L_{t}}{U^e_{C,t}},\qquad
    \tau^K_{t}=\frac{-\Psi^K_{t}}{U^e_{C,t}Q_{t}}.
\end{equation*}

\paragraph{Ramsey equilibrium}
On the banker's side, we can use the regulated deposit supply Euler equation to solve for $\tau^b_{t}$ in terms of allocations and prices. The remaining constraints are identical to those faced by the social planner in the definition of an FCEA. Similarly, on the entrepreneur's side, we can use the regulated demand conditions for labor, loans, and capital to back out the corresponding tax rates $\tau^N_{t}$, $\tau^L_{t}$, and $\tau^K_{t}$. Guessing that the private complementary slackness conditions associated with the collateral constraint are not binding, we are left with the entrepreneur's budget constraint and the collateral constraint---the same set of constraints as in the FCEA definition. After solving for prices and the investment function as in the FCEA, the complete set of constraints faced by the Ramsey planner is identical to the one in the FCEA definition. Therefore, the FCEA is exactly the allocation that is part of the Ramsey equilibrium. Finally, we can verify that the individual entrepreneur's complementary slackness conditions are indeed not binding because they are implied by the planner's analogous complementary slackness conditions.$\quad\blacksquare$

\subsection{Lemma \ref{lemma: OLL relaxed with unrestricted taxation}}
The relaxed problem is
\begin{equation*}
    \max_{\{C^b_{t},C^e_{t},C^w_{t},D_{t},K_{t},L_{t},N_{t}\}}\mathbb{E}_{0}\left(\sum_{t=0}^\infty\beta^t\sum_{i\in\mathcal{I}}\omega_i{}U^i_{t}\right)
\end{equation*}
subject to
\begin{align*}
    \lambda^b_{t}:\quad{}0&\le{}L_{t}-D_{t},\\
    \lambda^L_{t}:\quad{}0&\le{}U^b_{C}(C^b_{t})L_{t}-\beta_b\mathbb{E}_{t}[U^b_{C}(C^b_{t+1})(C^b_{t+1}+L_{t+1}-D_{t+1}+R_{t}D_{t})],\qquad\text{equality if }D_{t}=0,\\
    \lambda^C_{t}:\quad{}0&=A_{t}F(\xi_{t}K_{t-1},N_{t})-Q(K_{t-1},K_{t},\xi_{t})[K_{t}-(1-\delta)\xi_{t}K_{t-1}]-W(C^w_{t},N_{t})N_{t}+D_{t}\\
    &\quad-R_{t-1}D_{t-1}-C^b_{t}-C^e_{t},\\
    \lambda^e_{t}:\quad{}0&\le\mathbb{E}_{t}(Q(K_{t},K_{t+1},\xi_{t+1})\xi_{t+1})K_{t}-\mathbb{E}_{t}(C^b_{t+1}+L_{t+1}-D_{t+1}+R_{t}D_{t}),\\
    \lambda^Y_{t}:\quad{}0&=A_{t}F(\xi_{t}K_{t-1},N_{t})-\sum_{i\in\mathcal{I}}C^i_{t}-I(K_{t-1},K_{t},\xi_{t}),
\end{align*}
where $R_{t}=R(U^w_{C}(C^w_{t},N_{t}),\mathbb{E}_{t}[U^w_{C}(C^w_{t+1},N_{t+1})])$, the functions $W$, $R$, $Q$, and $I$ are the same as in definition \ref{def: FCEA}. An allocation-policy pair is part of a Ramsey equilibrium if---combined with the associated prices and Lagrange multipliers---it constitutes a regulated competitive equilibrium with the maximum level of welfare over all feasible allocation-policy pairs.

Consider a feasible policy $\{\kappa_{t},m_{t},\tau^D_{t},\tau^N_{t},\tau^L_{t},\tau^K_{t}\}\subset[0,1]^2\times\mathbb{R}^4$ and the corresponding regulated FCE allocation $\{C^b_{t},C^e_{t},C^w_{t},D_{t},K_{t},L_{t},N_{t}\}$. The policy is consistent with the construction in the lemma. If $U^b_{C,t}>\beta_b\mathbb{E}_{t}(U^b_{C,t+1}R^L_{t+1})$, then \eqref{eq: bankers Euler loans} implies that $\lambda^b_{t}>0$, and thus the leverage constraint is binding, which implies $\kappa_{t}=1-\frac{D_{t}}{L_{t}}$; otherwise, $\kappa_{t}\ge{}0$ combined with the leverage constraint is equivalent to $\kappa_{t}\in\left[0,1-\frac{D_{t}}{L_{t}}\right]$. The collateral constraint combined with $m_{t}\le{}1$ is equivalent to $m_{t}\in\left[\frac{\mathbb{E}_{t}(R^L_{t+1})L_{t}}{\mathbb{E}_{t}(Q_{t+1}\xi_{t+1})K_{t}},1\right]$. The tax rates are consistent with the regulated analogs of \eqref{eq: bankers Euler deposits} and \eqref{eq: entrepreneurs labor demand}--\eqref{eq: entrepreneurs Euler capital}. Moreover, as argued in proposition \ref{prop: FCEA decentralization}, the allocation is feasible for the FCEA problem. Since $D_{t}\le(1-\kappa_{t})L_{t}\le{}L_{t}$ and $\mathbb{E}_{t}(R^L_{t+1})L_{t}\le{}m_{t}\mathbb{E}_{t}(Q_{t+1}\xi_{t+1})K_{t}\le\mathbb{E}_{t}(Q_{t+1}\xi_{t+1})K_{t}$, the allocation is feasible for the relaxed problem.

Conversely, suppose an allocation $\{C^b_{t},C^e_{t},C^w_{t},D_{t},K_{t},L_{t},N_{t}\}$ is feasible for the relaxed problem and construct the corresponding policy as described in the lemma. The construction of $\kappa_{t}$ ensures that the FCE version of the bank leverage constraint and the private complementary slackness conditions are satisfied. The construction of $m_{t}$ guarantees that the FCE version of the collateral constraint is respected. The construction of the tax rates makes sure that the regulated analogs of \eqref{eq: bankers Euler deposits} and \eqref{eq: entrepreneurs labor demand}--\eqref{eq: entrepreneurs Euler capital} hold. The policy is feasible, that is, $\{\kappa_{t},m_{t},\tau^D_{t},\tau^N_{t},\tau^L_{t},\tau^K_{t}\}\subset[0,1]^2\times\mathbb{R}^4$. It follows that the allocation and the constructed policy---combined with the associated prices and Lagrange multipliers---constitute an FCE.

We have established that the two problems have identical feasible sets of allocation-policy pairs. Since the objective functions are equivalent, the two problems yield identical optimal allocation-policy pairs.$\quad\blacksquare$

\subsection{Lemma \ref{lemma: OLL relaxed with labor taxation}}
The relaxed problem is
\begin{equation*}
    \max_{\{C^b_{t},C^e_{t},C^w_{t},D_{t},K_{t},L_{t},N_{t}\}}\mathbb{E}_{0}\left(\sum_{t=0}^\infty\beta^t\sum_{i\in\mathcal{I}}\omega_i{}U^i_{t}\right)
\end{equation*}
subject to
\begin{align*}
    \lambda^b_{t}:\quad{}0&\le{}L_{t}-D_{t},\\
    \lambda^L_{t}:\quad{}0&=\beta_b\mathbb{E}_{t}[U^b_{C}(C^b_{t+1})(C^b_{t+1}+L_{t+1}-D_{t+1})]-U^b_{C}(C^b_{t})(L_{t}-D_{t}),\\
    \lambda^D_{t}:\quad{}0&\le{}U^b_{C}(C^b_{t})-\beta_b{}R_{t}\mathbb{E}_{t}(U^b_{C}(C^b_{t+1})),\\
    \lambda^C_{t}:\quad{}0&=A_{t}F(\xi_{t}K_{t-1},N_{t})-Q(K_{t-1},K_{t},\xi_{t})[K_{t}-(1-\delta)\xi_{t}K_{t-1}]-W(C^w_{t},N_{t})N_{t}+D_{t}\\
    &\quad-R_{t-1}D_{t-1}-C^b_{t}-C^e_{t},\\
    \lambda^e_{t}:\quad{}0&\le\mathbb{E}_{t}(Q(K_{t},K_{t+1},\xi_{t+1})\xi_{t+1})K_{t}-\mathbb{E}_{t}(C^b_{t+1}+L_{t+1}-D_{t+1}+R_{t}D_{t}),\\
    \lambda^K_{t}:\quad{}0&=\beta_e\mathbb{E}_{t}[U^e_{C}(C^e_{t+1})\{[A_{t+1}F_{K}(\xi_{t+1}K_{t},N_{t+1})+Q(K_{t},K_{t+1},\xi_{t+1})(1-\delta)]\xi_{t+1}K_{t}-C^b_{t+1}\\
    &\quad-L_{t+1}+D_{t+1}-R_{t}D_{t}\}]-U^e_{C}(C^e_{t})(Q(K_{t-1},K_{t},\xi_{t})K_{t}-L_{t}),\\
    \lambda^B_{t}:\quad{}0&\le{}U^e_{C}(C^e_{t})L_{t}-\beta_e\mathbb{E}_{t}[U^e_{C}(C^e_{t+1})(C^b_{t+1}+L_{t+1}-D_{t+1}+R_{t}D_{t})],\\
    \lambda^Y_{t}:\quad{}0&=A_{t}F(\xi_{t}K_{t-1},N_{t})-\sum_{i\in\mathcal{I}}C^i_{t}-I(K_{t-1},K_{t},\xi_{t}),
\end{align*}
where $R_{t}=R(U^w_{C}(C^w_{t},N_{t}),\mathbb{E}_{t}[U^w_{C}(C^w_{t+1},N_{t+1})])$, the functions $W$, $R$, $Q$, and $I$ are the same as in definition \ref{def: FCEA}.

In the absence of taxation on the banker's side, \eqref{eq: bankers budget constraint}--\eqref{eq: bankers complementary slackness} must be respected by the planner. As before, we can use \eqref{eq: bankers budget constraint} to solve for $B_{t}\equiv{}R^L_{t}L_{t-1}=C^b_{t}+L_{t}-D_{t}+R_{t-1}D_{t-1}$. Now we can use \eqref{eq: bankers Euler deposits} to express $\lambda^b_{t}=U^b_{C,t}-\beta_b{}R_{t}\mathbb{E}_{t}(U^b_{C,t+1})$. Multiplying \eqref{eq: bankers Euler deposits} by $D_{t}$ and \eqref{eq: bankers Euler loans} by $L_{t}$, subtracting the former from the latter, and using the complementary slackness conditions \eqref{eq: bankers complementary slackness}, \eqref{eq: bankers Euler loans} can be expressed in terms of allocations only. Hence, the implementability conditions that go to the Ramsey problem from the banker's side are
\begin{align*}
    0&\le(1-\kappa_{t})L_{t}-D_{t},\\
    0&=\beta_b\mathbb{E}_{t}[U^b_{C,t+1}(C^b_{t+1}+L_{t+1}-D_{t+1})]-U^b_{C,t}(L_{t}-D_{t}),\\
    0&\le{}U^b_{C,t}-\beta_b{}R_{t}\mathbb{E}_{t}(U^b_{C,t+1}),\\
    0&=[U^b_{C,t}-\beta_b{}R_{t}\mathbb{E}_{t}(U^b_{C,t+1})][(1-\kappa_{t})L_{t}-D_{t}].
\end{align*}

Consider the entrepreneur's problem. As before, we can use the regulated analog of \eqref{eq: entrepreneurs labor demand} to solve for $\tau^N_{t}\equiv\frac{A_{t}F_{N,t}}{W_{t}}-1$. Using \eqref{eq: entrepreneurs Euler loans}, we can express $\lambda^e_{t}\mathbb{E}_{t}(B_{t+1})=U^e_{C,t}L_{t}-\beta_e\mathbb{E}_{t}(U^e_{C,t+1}B_{t+1})$. By multiplying \eqref{eq: entrepreneurs Euler loans} by $L_{t}$ and \eqref{eq: entrepreneurs Euler capital} by $K_{t}$, subtracting the former from the latter, and using the complementary slackness conditions \eqref{eq: entrepreneurs complementary slackness}, \eqref{eq: entrepreneurs Euler capital} can be identically expressed in terms of allocations. Using the definition of $B_{t}$ based on \eqref{eq: bankers budget constraint}, the implementability conditions from the entrepreneur's side are
\begin{align*}
    0&=A_{t}F(\xi_{t}K_{t-1},N_{t})-Q_{t}[K_{t}-(1-\delta)\xi_{t}K_{t-1}]-W_{t}N_{t}+D_{t}-R_{t-1}D_{t-1}-C^b_{t}-C^e_{t},\\
    0&\le{}m_{t}\mathbb{E}_{t}(Q_{t+1}\xi_{t+1})K_{t}-\mathbb{E}_{t}(B_{t+1}),\\
    0&=\beta_e\mathbb{E}_{t}[U^e_{C,t+1}\{[A_{t+1}F_{K}(\xi_{t+1}K_{t},N_{t+1})+Q_{t+1}(1-\delta)]\xi_{t+1}K_{t}-B_{t+1}\}]-U^e_{C,t}(Q_{t}K_{t}-L_{t}),\\
    0&\le{}U^e_{C,t}L_{t}-\beta_e\mathbb{E}_{t}(U^e_{C,t+1}B_{t+1}),\\
    0&=[U^e_{C,t}L_{t}-\beta_e\mathbb{E}_{t}(U^e_{C,t+1}B_{t+1})][m_{t}\mathbb{E}_{t}(Q_{t+1}\xi_{t+1})K_{t}-\mathbb{E}_{t}(B_{t+1})].
\end{align*}

The remaining implementability conditions are constituted in the functions $W$, $R$, $Q$, and $I$, defined by \eqref{eq: workers labor supply}, \eqref{eq: workers Euler}, \eqref{eq: capital good supply}, and \eqref{eq: market clearing capital good}, as well as the resource constraint obtained by combining \eqref{eq: market clearing wholesale good} and \eqref{eq: market clearing final good}.

The equivalence between the feasible sets of allocation-policy pairs that satisfy the implementability conditions above and the constraints of the relaxed problem follows from the arguments that are identical to the proof of lemma \ref{lemma: OLL relaxed with unrestricted taxation}. Now we have only one tax rate $\tau^N_{t}$ that can be constructed from the regulated version of \eqref{eq: entrepreneurs labor demand}, and both $\kappa_{t}$ and $m_{t}$ are set such that the private complementary slackness conditions are satisfied.$\quad\blacksquare$

\subsection{Proposition \ref{prop: optimal leverage limits with labor taxation}}
The FOCs for the problem of lemma \ref{lemma: OLL relaxed with labor taxation} are
\begin{align*}
    C^b_{t}:\quad{}0&=\omega_b{}U^b_{C,t}-\lambda^Y_{t}-\lambda^C_{t}-[\lambda^L_{t}(L_{t}-D_{t})-\lambda^D_{t}]U^b_{CC,t}+\frac{\bm{1}_\mathbb{N}(t)}{\beta}\{[\lambda^L_{t-1}(C^b_{t}+L_{t}-D_{t})\\
    &\quad-\lambda^D_{t-1}R_{t-1}]\beta_b{}U^b_{CC,t}+\lambda^L_{t-1}\beta_b{}U^b_{C,t}-\lambda^e_{t-1}-(\lambda^K_{t-1}+\lambda^B_{t-1})\beta_e{}U^e_{C,t}\},\\
    C^e_{t}:\quad{}0&=\omega_e{}U^e_{C,t}-\lambda^Y_{t}-\lambda^C_{t}-[\lambda^K_{t}(Q_{t}K_{t}-L_{t})-\lambda^B_{t}L_{t}]U^e_{CC,t}\\
    &\quad+\frac{\bm{1}_\mathbb{N}(t)}{\beta}[\lambda^K_{t-1}R^K_{t}Q_{t-1}K_{t-1}-(\lambda^K_{t-1}+\lambda^B_{t-1})R^L_{t}L_{t-1}]\beta_e{}U^e_{CC,t},\\
    C^w_{t}:\quad{}0&=\omega_w{}U^w_{C,t}-\lambda^Y_{t}-\lambda^C_{t}W_{C,t}N_{t}-\{\lambda^D_{t}\beta_b\mathbb{E}_{t}(U^b_{C,t+1})+[\beta\mathbb{E}_{t}(\lambda^C_{t+1})+\lambda^e_{t}\\
    &\quad+(\lambda^K_{t}+\lambda^B_{t})\beta_e\mathbb{E}_{t}(U^e_{C,t+1})]D_{t}\}R_{1,t}U^w_{CC,t}-\frac{\bm{1}_\mathbb{N}(t)}{\beta}\{\lambda^D_{t-1}\beta_b\mathbb{E}_{t-1}(U^b_{C,t})\\
    &\quad+[\beta\mathbb{E}_{t-1}(\lambda^C_{t})+\lambda^e_{t-1}+(\lambda^K_{t-1}+\lambda^B_{t-1})\beta_e\mathbb{E}_{t-1}(U^e_{C,t})]D_{t-1}\}R_{2,t-1}U^w_{CC,t},\\
    D_{t}:\quad{}0&=-\lambda^b_{t}+\lambda^L_{t}U^b_{C,t}+\lambda^C_{t}-[\beta\mathbb{E}_{t}(\lambda^C_{t+1})+\lambda^e_{t}+(\lambda^K_{t}+\lambda^B_{t})\beta_e\mathbb{E}_{t}(U^e_{C,t+1})]R_{t}\\
    &\quad+\frac{\bm{1}_\mathbb{N}(t)}{\beta}[-\lambda^L_{t-1}\beta_b{}U^b_{C,t}+\lambda^e_{t-1}+(\lambda^K_{t-1}+\lambda^B_{t-1})\beta_e{}U^e_{C,t}],\\
    K_{t}:\quad{}0&=-\lambda^C_{t}\{Q_{2,t}[K_{t}-(1-\delta)\xi_{t}K_{t-1}]+Q_{t}\}+\lambda^e_{t}\mathbb{E}_{t}[(Q_{1,t+1}K_{t}+Q_{t+1})\xi_{t+1}]-\lambda^Y_{t}I_{2,t}\\
    &\quad+\lambda^K_{t}\{\beta_e\mathbb{E}_{t}[U^e_{C,t+1}\{[A_{t+1}F_{KK,t+1}\xi_{t+1}+Q_{1,t+1}(1-\delta)]\xi_{t+1}K_{t}+R^K_{t+1}Q_{t}\}]\\
    &\quad-U^e_{C,t}(Q_{2,t}K_{t}+Q_{t})\}+\beta\mathbb{E}_{t}[(\lambda^C_{t+1}+\lambda^Y_{t+1})A_{t+1}F_{K,t+1}\xi_{t+1}\\
    &\quad+\lambda^C_{t+1}\{Q_{t+1}(1-\delta)\xi_{t+1}-Q_{1,t+1}[K_{t+1}-(1-\delta)\xi_{t+1}K_{t}]\}-\lambda^K_{t+1}U^e_{C,t+1}Q_{1,t+1}K_{t+1}\\
    &\quad-\lambda^Y_{t+1}I_{1,t+1}]+\frac{\bm{1}_\mathbb{N}(t)}{\beta}[\lambda^e_{t-1}+\lambda^K_{t-1}\beta_e{}U^e_{C,t}(1-\delta)]Q_{2,t}\xi_{t}K_{t-1},\\
    L_{t}:\quad{}0&=\lambda^b_{t}-\lambda^L_{t}U^b_{C,t}+(\lambda^K_{t}+\lambda^B_{t})U^e_{C,t}+\frac{\bm{1}_\mathbb{N}(t)}{\beta}[\lambda^L_{t-1}\beta_b{}U^b_{C,t}-\lambda^e_{t-1}-(\lambda^K_{t-1}+\lambda^B_{t-1})\beta_e{}U^e_{C,t}],\\
    N_{t}:\quad{}0&=\omega_w{}U^w_{N,t}+(\lambda^C_{t}+\lambda^Y_{t})A_{t}F_{N,t}-\lambda^C_{t}(W_{N,t}N_{t}+W_{t})-\{\lambda^D_{t}\beta_b\mathbb{E}_{t}(U^b_{C,t+1})\\
    &\quad+[\beta\mathbb{E}_{t}(\lambda^C_{t+1})+\lambda^e_{t}+(\lambda^K_{t}+\lambda^B_{t})\beta_e\mathbb{E}_{t}(U^e_{C,t+1})]D_{t}\}R_{1,t}U^w_{CN,t}\\
    &\quad-\frac{\bm{1}_\mathbb{N}(t)}{\beta}\{\lambda^D_{t-1}\beta_b\mathbb{E}_{t-1}(U^b_{C,t})+[\beta\mathbb{E}_{t-1}(\lambda^C_{t})+\lambda^e_{t-1}\\
    &\quad+(\lambda^K_{t-1}+\lambda^B_{t-1})\beta_e\mathbb{E}_{t-1}(U^e_{C,t})]D_{t-1}\}R_{2,t-1}U^w_{CN,t}+\frac{\bm{1}_\mathbb{N}(t)}{\beta}\lambda^K_{t-1}\beta_e{}U^e_{C,t}A_{t}F_{KN,t}\xi_{t}K_{t-1}.
\end{align*}
The complementary slackness conditions are
\begin{align*}
    0&=\lambda^b_{t}(L_{t}-D_{t}),\qquad\lambda^b_{t}\ge{}0,\\
    0&=\lambda^D_{t}[U^b_{C,t}-\beta_b{}R_{t}\mathbb{E}_{t}(U^b_{C,t+1})],\qquad\lambda^D_{t}\ge{}0,\\
    0&=\lambda^e_{t}[\mathbb{E}_{t}(Q_{t+1}\xi_{t+1})K_{t}-\mathbb{E}_{t}(B_{t+1})],\qquad\lambda^e_{t}\ge{}0,\\
    0&=\lambda^B_{t}[U^e_{C,t}L_{t}-\beta_e\mathbb{E}_{t}(U^e_{C,t+1}B_{t+1})],\qquad\lambda^B_{t}\ge{}0.
\end{align*}
Inspecting the FOCs for $C^b_{t}$, $C^e_{t}$, and $C^w_{t}$, we see that consumption insurance is generally imperfect.

Consider the steady state. The $\lambda^L_{t}$ constraint implies $L-D=\frac{\beta_b}{1-\beta_b}C^b\ge{}0$, which makes the relaxed leverage constraint redundant, implying $\lambda^b=0$. Since $\beta_b<\beta$, we have $\lambda^D=0$. The FOC for $D_{t}$ then implies $\lambda^L=0$. Guessing that $C^b$ is sufficiently small relative to $D$, since $\beta_e<\beta$, we will have $L>\beta_e{}B$; therefore, $\lambda^B=0$. (A sufficient condition is $\beta_e\le\beta_b$.) The FOC for $L_{t}$ then implies $\lambda^K=\frac{\lambda^e}{(\beta-\beta_e)U^e_{C}}\ge{}0$. If $\lambda^e=0$, we have $\omega_b{}U^b_{C}=\omega_e{}U^e_{C}=\lambda^C+\lambda^Y$; therefore, there is approximately perfect risk sharing between bankers and entrepreneurs. Risk sharing is only approximate because generally $\lambda^L_{t}\neq{}0$ outside of the steady state. If $U^w(C^w,N)=\ln(C^w)-v(N)$ and the steady-state profits of capital good producers are zero, the FOC for $C^w_{t}$ implies $\omega_w{}U^w_{C}=\lambda^C+\lambda^Y$, as in the proof of proposition \ref{prop: FCEA}.

Section \ref{sec: SS OLL with labor taxation} constructs the steady state. The construction boils down to considering two cases: collateral constraint is slack or binding. Each case can be reduced to solving a system of three nonlinear equations. Conditional on solving a nonlinear system, the sequential solution uniquely determines the steady state. Since the problem reduces to a numerical one, we cannot claim that the steady state is unique. However, it is the case under the baseline calibration and other parameterizations considered in the analysis.$\quad\blacksquare$

\subsection{Proposition \ref{prop: CEA}}
The planning problem is
\begin{equation*}
    \max_{\{C^b_{t},C^e_{t},C^w_{t},D_{t},K_{t},L_{t},N_{t}\textcolor{BrickRed}{,\Omega_{1,t}}\}}\mathbb{E}_{0}\left(\sum_{t=0}^\infty\beta^t\sum_{i\in\mathcal{I}}\omega_i{}U^i_{t}\right)
\end{equation*}
subject to
\begin{align*}
    \lambda^b_{t}:\quad{}0&\le(1-\kappa_{t})L_{t}-D_{t},\\
    \lambda^L_{1,t}:\quad{}0&\le{}U^b_{C}(C^b_{t})L_{t}-\beta_b\mathbb{E}_{t}[U^b_{C}(C^b_{t+1})(C^b_{t+1}+L_{t+1}-D_{t+1}+R_{t}D_{t})],\quad\text{equality if }D_{t}=0,\\
    \lambda^L_{2,t}:\quad{}0&=\{U^b_{C}(C^b_{t})L_{t}-\beta_b\mathbb{E}_{t}[U^b_{C}(C^b_{t+1})(C^b_{t+1}+L_{t+1}-D_{t+1}+R_{t}D_{t})]\}[(1-\kappa_{t})L_{t}-D_{t}],\\
    \lambda^C_{t}:\quad{}0&=\textcolor{BrickRed}{\Delta_{t}\left\{\Omega_{1,t}-\frac{\beta\theta\mathbb{E}_{t}[U^w_{C}(C^w_{t+1},N_{t+1})\Pi_{t+1}^{\epsilon}\Omega_{1,t+1}]}{U^w_{C}(C^w_{t},N_{t})}\right\}}-Q(K_{t-1},K_{t},\xi_{t})[K_{t}\\
    &\quad-(1-\delta)\xi_{t}K_{t-1}]-W(C^w_{t},N_{t})N_{t}+D_{t}-R_{t-1}D_{t-1}-C^b_{t}-C^e_{t},\\
    \lambda^e_{t}:\quad{}0&\le{}m_{t}\mathbb{E}_{t}(Q(K_{t},K_{t+1},\xi_{t+1})\xi_{t+1})K_{t}-\mathbb{E}_{t}(C^b_{t+1}+L_{t+1}-D_{t+1}+R_{t}D_{t}),\\
    \lambda^Y_{t}:\quad{}0&=\frac{A_{t}}{\textcolor{BrickRed}{\Delta_{t}}}F(\xi_{t}K_{t-1},N_{t})-\sum_{i\in\mathcal{I}}C^i_{t}-I(K_{t-1},K_{t},\xi_{t}),\\
    \textcolor{BrickRed}{\lambda^\Omega_{t}}:\quad{}0&=\frac{\epsilon-1}{\epsilon}\widetilde{P}_{t}\frac{A_{t}}{\Delta_{t}}F(\xi_{t}K_{t-1},N_{t})-\Omega_{1,t}+\frac{\beta\theta\mathbb{E}_{t}\left[U^w_{C}(C^w_{t+1},N_{t+1})\Pi_{t+1}^{\epsilon-1}\frac{\widetilde{P}_{t}}{\widetilde{P}_{t+1}}\Omega_{1,t+1}\right]}{U^w_{C}(C^w_{t},N_{t})}.
\end{align*}
As before, define $\lambda^L_{t}\equiv\lambda^L_{1,t}+\lambda^L_{2,t}[(1-\kappa_{t})L_{t}-D_{t}]$. The FOCs are
\begin{align*}
    C^b_{t}:\quad{}0&=\omega_b{}U^b_{C,t}-\lambda^Y_{t}-\lambda^C_{t}+\lambda^L_{t}U^b_{CC,t}L_{t}-\frac{\bm{1}_\mathbb{N}(t)}{\beta}[\lambda^L_{t-1}\beta_b(U^b_{CC,t}R^L_{t}L_{t-1}+U^b_{C,t})+\lambda^e_{t-1}],\\
    C^e_{t}:\quad{}0&=\omega_e{}U^e_{C,t}-\lambda^Y_{t}-\lambda^C_{t},\\
    C^w_{t}:\quad{}0&=\omega_w{}U^w_{C,t}-\lambda^Y_{t}-\lambda^C_{t}W_{C,t}N_{t}-[\lambda^L_{t}\beta_b\mathbb{E}_{t}(U^b_{C,t+1})+\beta\mathbb{E}_{t}(\lambda^C_{t+1})+\lambda^e_{t}]R_{1,t}U^w_{CC,t}D_{t}\\
    &\quad-\frac{\bm{1}_\mathbb{N}(t)}{\beta}[\lambda^L_{t-1}\beta_b\mathbb{E}_{t-1}(U^b_{C,t})+\beta\mathbb{E}_{t-1}(\lambda^C_{t})+\lambda^e_{t-1}]R_{2,t-1}U^w_{CC,t}D_{t-1}\\
    &\quad\textcolor{BrickRed}{+\left[(\lambda^C_{t}\Delta_{t}-\lambda^\Omega_{t})\Omega_{1,t}-\left(\lambda^C_{t}P^w_{t}\Delta_{t}-\lambda^\Omega_{t}\frac{\epsilon-1}{\epsilon}\widetilde{P}_{t}\right)Y_{t}\right]\frac{U^w_{CC,t}}{U^w_{C,t}}}\\
    &\quad\textcolor{BrickRed}{-\bm{1}_\mathbb{N}(t)\left(\lambda^C_{t-1}\Delta_{t-1}\Pi_{t}-\lambda^\Omega_{t-1}\frac{\widetilde{P}_{t-1}}{\widetilde{P}_{t}}\right)\theta\Pi_{t}^{\epsilon-1}\Omega_{1,t}\frac{U^w_{CC,t}}{U^w_{C,t-1}}},\\
    D_{t}:\quad{}0&\ge-\lambda^b_{t}-\lambda^L_{2,t}[U^b_{C,t}-\beta_b\mathbb{E}_{t}(U^b_{C,t+1}R^L_{t+1})]L_{t}+\lambda^C_{t}-[\lambda^L_{t}\beta_b\mathbb{E}_{t}(U^b_{C,t+1})+\beta\mathbb{E}_{t}(\lambda^C_{t+1})\\
    &\quad+\lambda^e_{t}]R_{t}+\frac{\bm{1}_\mathbb{N}(t)}{\beta}(\lambda^L_{t-1}\beta_b{}U^b_{C,t}+\lambda^e_{t-1}),\qquad\text{equality if }D_{t}>0,\\
    K_{t}:\quad{}0&=-\lambda^C_{t}\{Q_{2,t}[K_{t}-(1-\delta)\xi_{t}K_{t-1}]+Q_{t}\}+\lambda^e_{t}m_{t}\mathbb{E}_{t}[(Q_{1,t+1}K_{t}+Q_{t+1})\xi_{t+1}]-\lambda^Y_{t}I_{2,t}\\
    &\quad+\beta\mathbb{E}_{t}\biggl[\left(\textcolor{BrickRed}{\lambda^\Omega_{t+1}\frac{\epsilon-1}{\epsilon}\widetilde{P}_{t+1}}+\lambda^Y_{t+1}\right)\frac{A_{t+1}}{\textcolor{BrickRed}{\Delta_{t+1}}}F_{K,t+1}\xi_{t+1}-\lambda^Y_{t+1}I_{1,t+1}+\lambda^C_{t+1}\{Q_{t+1}(1-\delta)\\
    &\quad\times\xi_{t+1}-Q_{1,t+1}[K_{t+1}-(1-\delta)\xi_{t+1}K_{t}]\}\biggr]+\frac{\bm{1}_\mathbb{N}(t)}{\beta}\lambda^e_{t-1}m_{t-1}Q_{2,t}\xi_{t}K_{t-1},\\
    L_{t}:\quad{}0&=\{\lambda^b_{t}+\lambda^L_{2,t}[U^b_{C,t}-\beta_b\mathbb{E}_{t}(U^b_{C,t+1}R^L_{t+1})]L_{t}\}(1-\kappa_{t})+\lambda^L_{t}U^b_{C,t}\\
    &\quad-\frac{\bm{1}_\mathbb{N}(t)}{\beta}(\lambda^L_{t-1}\beta_b{}U^b_{C,t}+\lambda^e_{t-1}),\\
    N_{t}:\quad{}0&=\omega_w{}U^w_{N,t}+\left(\textcolor{BrickRed}{\lambda^\Omega_{t}\frac{\epsilon-1}{\epsilon}\widetilde{P}_{t}}+\lambda^Y_{t}\right)\frac{A_{t}}{\textcolor{BrickRed}{\Delta_{t}}}F_{N,t}-[\lambda^L_{t}\beta_b\mathbb{E}_{t}(U^b_{C,t+1})+\beta\mathbb{E}_{t}(\lambda^C_{t+1})+\lambda^e_{t}]R_{1,t}\\
    &\quad\times{}U^w_{CN,t}D_{t}-\lambda^C_{t}(W_{N,t}N_{t}+W_{t})-\frac{\bm{1}_\mathbb{N}(t)}{\beta}[\lambda^L_{t-1}\beta_b\mathbb{E}_{t-1}(U^b_{C,t})+\beta\mathbb{E}_{t-1}(\lambda^C_{t})+\lambda^e_{t-1}]\\
    &\quad\times{}R_{2,t-1}U^w_{CN,t}D_{t-1}\textcolor{BrickRed}{+\left[(\lambda^C_{t}\Delta_{t}-\lambda^\Omega_{t})\Omega_{1,t}-\left(\lambda^C_{t}P^w_{t}\Delta_{t}-\lambda^\Omega_{t}\frac{\epsilon-1}{\epsilon}\widetilde{P}_{t}\right)Y_{t}\right]\frac{U^w_{CN,t}}{U^w_{C,t}}}\\
    &\quad\textcolor{BrickRed}{-\bm{1}_\mathbb{N}(t)\left(\lambda^C_{t-1}\Delta_{t-1}\Pi_{t}-\lambda^\Omega_{t-1}\frac{\widetilde{P}_{t-1}}{\widetilde{P}_{t}}\right)\theta\Pi_{t}^{\epsilon-1}\Omega_{1,t}\frac{U^w_{CN,t}}{U^w_{C,t-1}}},\\
    \textcolor{BrickRed}{\Omega_{1,t}}:\quad{}0&=\lambda^C_{t}\Delta_{t}-\lambda^\Omega_{t}-\bm{1}_\mathbb{N}(t)\left(\lambda^C_{t-1}\Delta_{t-1}\Pi_{t}-\lambda^\Omega_{t-1}\frac{\widetilde{P}_{t-1}}{\widetilde{P}_{t}}\right)\theta\Pi_{t}^{\epsilon-1}\frac{U^w_{C,t}}{U^w_{C,t-1}}.
\end{align*}
The complementary slackness conditions are
\begin{align*}
    0&=\lambda^b_{t}[(1-\kappa_{t})L_{t}-D_{t}],\qquad\lambda^b_{t}\ge{}0,\\
    0&=\lambda^L_{1,t}[U^b_{C,t}L_{t}-\beta_b\mathbb{E}_{t}(U^b_{C,t+1}B_{t+1})],\qquad{}D_{t}\lambda^L_{1,t}\ge{}0,\\
    0&=\lambda^e_{t}[m_{t}\mathbb{E}_{t}(Q_{t+1}\xi_{t+1})K_{t}-\mathbb{E}_{t}(B_{t+1})],\qquad\lambda^e_{t}\ge{}0.
\end{align*}

\paragraph{Wedges}
Since the FOCs for $C^b_{t}$ and $D_{t}$ are the same as in the FCEA, the deposit wedge $\Psi^D_{t}$ is too. Since the FOCs for $C^e_{t}$ and $L_{t}$ and the $\lambda^L_{1,t}$ constraint are the same as in the FCEA, the loan wedge $\Psi^L_{t}$ is too. The FOCs for $C^e_{t}$, $C^w_{t}$, and $N_{t}$ combined with the definition of $W_{t}$ imply
\begin{equation*}
    W_{t}=\textcolor{BrickRed}{P^w_{t}}A_{t}F_{N,t}+\Psi^N_{t},
\end{equation*}
where
\begin{multline*}
    \Psi^N_{t}=\frac{[(\omega_e{}U^e_{C,t}\textcolor{BrickRed}{+\lambda^\Omega_{t}\frac{\epsilon-1}{\epsilon}\widetilde{P}_{t}-\lambda^C_{t}})\textcolor{BrickRed}{(P^w_{t}\Delta_{t})^{-1}}-\omega_w{}U^w_{C,t}-\lambda^C_{t}]\textcolor{BrickRed}{P^w_{t}}A_{t}F_{N,t}-\lambda^C_{t}W_{N,t}N_{t}}{\omega_w{}U^w_{C,t}+\lambda^C_{t}}\\
    -\frac{U^w_{CN,t}}{U^w_{CC,t}}\frac{\omega_w{}U^w_{C,t}-\omega_e{}U^e_{C,t}+\lambda^C_{t}(1-W_{C,t}N_{t})}{\omega_w{}U^w_{C,t}+\lambda^C_{t}}.
\end{multline*}
The FOCs for $C^e_{t}$ and $K_{t}$ imply
\begin{equation*}
    U^e_{C,t}Q_{t}=\beta_e\mathbb{E}_{t}\{U^e_{C,t+1}[\textcolor{BrickRed}{P^w_{t+1}}A_{t+1}F_{K,t+1}+Q_{t+1}(1-\delta)]\xi_{t+1}\}+\frac{\lambda^e_{t}}{\omega_e}m_{t}\mathbb{E}_{t}(Q_{t+1}\xi_{t+1})+\Psi^K_{t},
\end{equation*}
where
\begin{multline*}
    \omega_e\Psi^K_{t}=(\beta-\beta_e)\mathbb{E}_{t}(\omega_e{}U^e_{C,t+1}\textcolor{BrickRed}{R^K_{t+1}})Q_{t}+\beta\mathbb{E}_{t}\left\{\lambda^Y_{t+1}\left[Q_{t+1}\Phi\left(\frac{I_{t+1}}{K_{t}}\right)-\frac{I_{t+1}}{K_{t}}\right]\right\}\\
    -\lambda^C_{t}Q_{2,t}[K_{t}-(1-\delta)\xi_{t}K_{t-1}]-\beta\mathbb{E}_{t}\{\lambda^C_{t+1}Q_{1,t+1}[K_{t+1}-(1-\delta)\xi_{t+1}K_{t}]\}\\
    +\lambda^e_{t}m_{t}\mathbb{E}_{t}(Q_{1,t+1}\xi_{t+1})K_{t}+\frac{\bm{1}_\mathbb{N}(t)}{\beta}\lambda^e_{t-1}m_{t-1}Q_{2,t}\xi_{t}K_{t-1}\\
    \textcolor{BrickRed}{+\beta\mathbb{E}_{t}\left\{\left[\omega_e{}U^e_{C,t+1}(1-P^w_{t+1}\Delta_{t+1})+\lambda^\Omega_{t+1}\frac{\epsilon-1}{\epsilon}\widetilde{P}_{t+1}-\lambda^C_{t+1}\right]\frac{A_{t+1}}{\Delta_{t+1}}F_{K,t+1}\xi_{t+1}\right\}}.
\end{multline*}

\paragraph{Risk sharing and steady state}
Risk sharing properties follow from inspecting the FOCs for $C^b_{t}$, $C^e_{t}$, and $C^w_{t}$. In particular, the latter now has the term that reflects the market power of retailers. If $\lambda^e=\lambda^L=0$ and workers have separable preferences, the FOC for $C^w$ in the steady state is
\begin{equation*}
    0=\omega_w{}u'(C^w)-\lambda^Y+\lambda^C\frac{u''(C^w)}{u'(C^w)}\left(C^w\textcolor{BrickRed}{-\frac{Y}{\epsilon}}\right).
\end{equation*}
Since $\epsilon<\infty$, even with logarithmic preferences, the worker's steady-state Pareto-weighted marginal utility of consumption is less than that of bankers and entrepreneurs.

As shown in section \ref{sec: SS CEA}, the steady state construction parallels the FCEA, reducing to two cases---whether $\lambda^L=0$ or $\lambda^L>0$. In both cases, the quantity of deposits is indeterminate, but conditional on choosing an admissible value of $D$, there typically exists a unique steady state. The proof that the optimal steady state has $D=0$, provided that $\lambda^C>0$, is identical to the FCEA in proposition \ref{prop: FCEA}.

\paragraph{Decentralization}
After replacing $A_{t}$ with $P^w_{t}A_{t}$ in the entrepreneur's problem, the proof is identical to the proof of proposition \ref{prop: FCEA decentralization}.$\quad\blacksquare$

\subsection{Proposition \ref{prop: optimal leverage limits and monetary policy}}

\subsubsection{Case 1}
The relaxed planning problem is
\begin{equation*}
    \max_{\{C^b_{t},C^e_{t},C^w_{t},D_{t},K_{t},L_{t},N_{t}\textcolor{BrickRed}{,\Omega_{1,t},\Delta_{t},\Pi_{t}}\}}\mathbb{E}_{0}\left(\sum_{t=0}^\infty\beta^t\sum_{i\in\mathcal{I}}\omega_i{}U^i_{t}\right)
\end{equation*}
subject to
\begin{align*}
    \lambda^b_{t}:\quad{}0&\le{}L_{t}-D_{t},\\
    \lambda^L_{t}:\quad{}0&\le{}U^b_{C}(C^b_{t})L_{t}-\beta_b\mathbb{E}_{t}[U^b_{C}(C^b_{t+1})(C^b_{t+1}+L_{t+1}-D_{t+1}+R_{t}D_{t})],\qquad\text{equality if }D_{t}=0,\\
    \lambda^C_{t}:\quad{}0&=\textcolor{BrickRed}{\Delta_{t}\left\{\Omega_{1,t}-\frac{\beta\theta\mathbb{E}_{t}[U^w_{C}(C^w_{t+1},N_{t+1})\Pi_{t+1}^{\epsilon}\Omega_{1,t+1}]}{U^w_{C}(C^w_{t},N_{t})}\right\}}-Q(K_{t-1},K_{t},\xi_{t})[K_{t}\\
    &\quad-(1-\delta)\xi_{t}K_{t-1}]-W(C^w_{t},N_{t})N_{t}+D_{t}-R_{t-1}D_{t-1}-C^b_{t}-C^e_{t},\\
    \lambda^e_{t}:\quad{}0&\le{}\mathbb{E}_{t}(Q(K_{t},K_{t+1},\xi_{t+1})\xi_{t+1})K_{t}-\mathbb{E}_{t}(C^b_{t+1}+L_{t+1}-D_{t+1}+R_{t}D_{t}),\\
    \lambda^Y_{t}:\quad{}0&=\frac{A_{t}}{\textcolor{BrickRed}{\Delta_{t}}}F(\xi_{t}K_{t-1},N_{t})-\sum_{i\in\mathcal{I}}C^i_{t}-I(K_{t-1},K_{t},\xi_{t}),\\
    \textcolor{BrickRed}{\lambda^\Omega_{t}}:\quad{}0&=\frac{\epsilon-1}{\epsilon}\widetilde{P}(\Pi_{t})\frac{A_{t}}{\Delta_{t}}F(\xi_{t}K_{t-1},N_{t})-\Omega_{1,t}+\frac{\beta\theta\mathbb{E}_{t}\left[U^w_{C}(C^w_{t+1},N_{t+1})\Pi_{t+1}^{\epsilon-1}\frac{\widetilde{P}(\Pi_{t})}{\widetilde{P}(\Pi_{t+1})}\Omega_{1,t+1}\right]}{U^w_{C}(C^w_{t},N_{t})},\\
    \textcolor{BrickRed}{\lambda^\Delta_{t}}:\quad{}0&=\theta\Pi_{t}^{\epsilon}\Delta_{t-1}+(1-\theta)(\widetilde{P}(\Pi_{t}))^{-\epsilon}-\Delta_{t},\\
    \textcolor{BrickRed}{\lambda^R_{t}}:\quad{}0&\le{}R_{t}\mathbb{E}_{t}(\Pi_{t+1})-\underline{R},
\end{align*}
with $R_{t}=R(U^w_{C}(C^w_{t},N_{t}),\mathbb{E}_{t}[U^w_{C}(C^w_{t+1},N_{t+1})])$. The FOCs are
\begin{align*}
    C^b_{t}:\quad{}0&=\omega_b{}U^b_{C,t}-\lambda^Y_{t}-\lambda^C_{t}+\lambda^L_{t}U^b_{CC,t}L_{t}-\frac{\bm{1}_\mathbb{N}(t)}{\beta}[\lambda^L_{t-1}\beta_b(U^b_{CC,t}R^L_{t}L_{t-1}+U^b_{C,t})+\lambda^e_{t-1}],\\
    C^e_{t}:\quad{}0&=\omega_e{}U^e_{C,t}-\lambda^Y_{t}-\lambda^C_{t},\\
    C^w_{t}:\quad{}0&=\omega_w{}U^w_{C,t}-\lambda^Y_{t}-\lambda^C_{t}W_{C,t}N_{t}-[\lambda^L_{t}\beta_b\mathbb{E}_{t}(U^b_{C,t+1})+\beta\mathbb{E}_{t}(\lambda^C_{t+1})+\lambda^e_{t}]R_{1,t}U^w_{CC,t}D_{t}\\
    &\quad-\frac{\bm{1}_\mathbb{N}(t)}{\beta}[\lambda^L_{t-1}\beta_b\mathbb{E}_{t-1}(U^b_{C,t})+\beta\mathbb{E}_{t-1}(\lambda^C_{t})+\lambda^e_{t-1}]R_{2,t-1}U^w_{CC,t}D_{t-1}\textcolor{BrickRed}{+\biggl[(\lambda^C_{t}\Delta_{t}-\lambda^\Omega_{t})}\\
    &\quad\textcolor{BrickRed}{\times\Omega_{1,t}-\left(\lambda^C_{t}P^w_{t}\Delta_{t}-\lambda^\Omega_{t}\frac{\epsilon-1}{\epsilon}\widetilde{P}_{t}\right)Y_{t}\biggr]\frac{U^w_{CC,t}}{U^w_{C,t}}-\bm{1}_\mathbb{N}(t)\left(\lambda^C_{t-1}\Delta_{t-1}\Pi_{t}-\lambda^\Omega_{t-1}\frac{\widetilde{P}_{t-1}}{\widetilde{P}_{t}}\right)}\\
    &\quad\textcolor{BrickRed}{\times\theta\Pi_{t}^{\epsilon-1}\Omega_{1,t}\frac{U^w_{CC,t}}{U^w_{C,t-1}}+\left[\lambda^R_{t}R_{1,t}\mathbb{E}_{t}(\Pi_{t+1})+\frac{\bm{1}_\mathbb{N}(t)}{\beta}\lambda^R_{t-1}R_{2,t-1}\mathbb{E}_{t-1}(\Pi_{t})\right]U^w_{CC,t}},\\
    D_{t}:\quad{}0&\ge-\lambda^b_{t}+\lambda^C_{t}-[\lambda^L_{t}\beta_b\mathbb{E}_{t}(U^b_{C,t+1})+\beta\mathbb{E}_{t}(\lambda^C_{t+1})+\lambda^e_{t}]R_{t}+\frac{\bm{1}_\mathbb{N}(t)}{\beta}(\lambda^L_{t-1}\beta_b{}U^b_{C,t}+\lambda^e_{t-1}),\\
    &\quad\text{equality if }D_{t}>0,\\
    K_{t}:\quad{}0&=-\lambda^C_{t}\{Q_{2,t}[K_{t}-(1-\delta)\xi_{t}K_{t-1}]+Q_{t}\}+\lambda^e_{t}\mathbb{E}_{t}[(Q_{1,t+1}K_{t}+Q_{t+1})\xi_{t+1}]-\lambda^Y_{t}I_{2,t}\\
    &\quad+\beta\mathbb{E}_{t}\biggl[\left(\textcolor{BrickRed}{\lambda^\Omega_{t+1}\frac{\epsilon-1}{\epsilon}\widetilde{P}_{t+1}}+\lambda^Y_{t+1}\right)\frac{A_{t+1}}{\textcolor{BrickRed}{\Delta_{t+1}}}F_{K,t+1}\xi_{t+1}-\lambda^Y_{t+1}I_{1,t+1}+\lambda^C_{t+1}\\
    &\quad\times\{Q_{t+1}(1-\delta)\xi_{t+1}-Q_{1,t+1}[K_{t+1}-(1-\delta)\xi_{t+1}K_{t}]\}\biggr]+\frac{\bm{1}_\mathbb{N}(t)}{\beta}\lambda^e_{t-1}Q_{2,t}\xi_{t}K_{t-1},\\
    L_{t}:\quad{}0&=\lambda^b_{t}+\lambda^L_{t}U^b_{C,t}-\frac{\bm{1}_\mathbb{N}(t)}{\beta}(\lambda^L_{t-1}\beta_b{}U^b_{C,t}+\lambda^e_{t-1}),\\
    N_{t}:\quad{}0&=\omega_w{}U^w_{N,t}+\left(\textcolor{BrickRed}{\lambda^\Omega_{t}\frac{\epsilon-1}{\epsilon}\widetilde{P}_{t}}+\lambda^Y_{t}\right)\frac{A_{t}}{\textcolor{BrickRed}{\Delta_{t}}}F_{N,t}-[\lambda^L_{t}\beta_b\mathbb{E}_{t}(U^b_{C,t+1})+\beta\mathbb{E}_{t}(\lambda^C_{t+1})+\lambda^e_{t}]R_{1,t}\\
    &\quad\times{}U^w_{CN,t}D_{t}-\lambda^C_{t}(W_{N,t}N_{t}+W_{t})-\frac{\bm{1}_\mathbb{N}(t)}{\beta}[\lambda^L_{t-1}\beta_b\mathbb{E}_{t-1}(U^b_{C,t})+\beta\mathbb{E}_{t-1}(\lambda^C_{t})+\lambda^e_{t-1}]\\
    &\quad\times{}R_{2,t-1}U^w_{CN,t}D_{t-1}\textcolor{BrickRed}{+\left[(\lambda^C_{t}\Delta_{t}-\lambda^\Omega_{t})\Omega_{1,t}-\left(\lambda^C_{t}P^w_{t}\Delta_{t}-\lambda^\Omega_{t}\frac{\epsilon-1}{\epsilon}\widetilde{P}_{t}\right)Y_{t}\right]\frac{U^w_{CN,t}}{U^w_{C,t}}}\\
    &\quad\textcolor{BrickRed}{-\bm{1}_\mathbb{N}(t)\left(\lambda^C_{t-1}\Delta_{t-1}\Pi_{t}-\lambda^\Omega_{t-1}\frac{\widetilde{P}_{t-1}}{\widetilde{P}_{t}}\right)\times\theta\Pi_{t}^{\epsilon-1}\Omega_{1,t}\frac{U^w_{CN,t}}{U^w_{C,t-1}}}\\
    &\quad\textcolor{BrickRed}{+\left[\lambda^R_{t}R_{1,t}\mathbb{E}_{t}(\Pi_{t+1})+\frac{\bm{1}_\mathbb{N}(t)}{\beta}\lambda^R_{t-1}R_{2,t-1}\mathbb{E}_{t-1}(\Pi_{t})\right]U^w_{CN,t}},\\
    \textcolor{BrickRed}{\Omega_{1,t}}:\quad{}0&=\lambda^C_{t}\Delta_{t}-\lambda^\Omega_{t}-\bm{1}_\mathbb{N}(t)\left(\lambda^C_{t-1}\Delta_{t-1}\Pi_{t}-\lambda^\Omega_{t-1}\frac{\widetilde{P}_{t-1}}{\widetilde{P}_{t}}\right)\theta\Pi_{t}^{\epsilon-1}\frac{U^w_{C,t}}{U^w_{C,t-1}},\\
    \textcolor{BrickRed}{\Delta_{t}}:\quad{}0&=\left(\lambda^C_{t}P^w_{t}\Delta_{t}-\lambda^\Omega_{t}\frac{\epsilon-1}{\epsilon}\widetilde{P}_{t}-\lambda^Y_{t}\right)\frac{Y_{t}}{\Delta_{t}}-\lambda^\Delta_{t}+\beta\theta\mathbb{E}_{t}(\lambda^\Delta_{t+1}\Pi_{t+1}^\epsilon),
\end{align*}
and
\begin{multline*}
    \textcolor{BrickRed}{\Pi_{t}}:\quad{}0=\lambda^\Omega_{t}\widetilde{P}'(\Pi_{t})\left[\frac{\epsilon-1}{\epsilon}Y_{t}+\frac{\beta\theta\mathbb{E}_{t}\left(U^w_{C,t+1}\Pi_{t+1}^{\epsilon-1}\frac{\Omega_{1,t+1}}{\widetilde{P}_{t+1}}\right)}{U^w_{C,t}}\right]\\
    +\lambda^\Delta_{t}\epsilon\left[\theta\Pi_{t}^{\epsilon-1}\Delta_{t-1}-(1-\theta)\frac{\widetilde{P}'(\Pi_{t})}{\widetilde{P}_{t}^{\epsilon+1}}\right]\\
    -\bm{1}_\mathbb{N}(t)\theta\Pi_{t}^{\epsilon-1}\Omega_{1,t}\frac{U^w_{C,t}}{U^w_{C,t-1}}\left[\lambda^C_{t-1}\Delta_{t-1}\epsilon-\lambda^\Omega_{t-1}\frac{\widetilde{P}_{t-1}}{\widetilde{P}_{t}}\left(\frac{\epsilon-1}{\Pi_{t}}-\frac{\widetilde{P}'(\Pi_{t})}{\widetilde{P}_{t}}\right)\right]+\frac{\bm{1}_\mathbb{N}(t)}{\beta}\lambda^R_{t-1}R_{t-1}.
\end{multline*}
The complementary slackness conditions are
\begin{align*}
    0&=\lambda^b_{t}(L_{t}-D_{t}),\qquad\lambda^b_{t}\ge{}0,\\
    0&=\lambda^L_{t}[U^b_{C,t}L_{t}-\beta_b\mathbb{E}_{t}(U^b_{C,t+1}B_{t+1})],\qquad{}D_{t}\lambda^L_{t}\ge{}0,\\
    0&=\lambda^e_{t}[\mathbb{E}_{t}(Q_{t+1}\xi_{t+1})K_{t}-\mathbb{E}_{t}(B_{t+1})],\qquad\lambda^e_{t}\ge{}0,\\
    0&=\lambda^R_{t}[R_{t}\mathbb{E}_{t}(\Pi_{t+1})-\underline{R}],\qquad\lambda^R_{t}\ge{}0.
\end{align*}

The risk-sharing and steady-state properties follow from the proof of proposition \ref{prop: CEA} after setting $\lambda^L_{2,t}=0$, $\kappa_{t}=0$, and $m_{t}=1$. The short-run inflation behavior is represented by the FOC for $\Pi_{t}$. Section \ref{sec: SS OLLMP with unrestricted taxation} shows that in the steady state, the FOC for $\Pi$ is equivalent to
\begin{equation*}
    \lambda^R=\frac{\Pi-1}{\Pi}\frac{\beta\theta\Pi^{\epsilon-1}}{1-\beta\theta\Pi^\epsilon}\frac{(\epsilon-1)\lambda^C+\epsilon\lambda^Y}{1-\theta\Pi^{\epsilon-1}}\beta{}Y.
\end{equation*}
Moreover,
\begin{equation*}
    (\epsilon-1)\lambda^C+\epsilon\lambda^Y=\frac{\epsilon\omega_e{}U^e_{C}[\frac{v''(N)}{u'(C^w)}N+W]+\omega_w{}v'(N)}{\frac{v''(N)}{u'(C^w)}N+W+\frac{1}{\epsilon}\frac{A}{\Delta}F_{N}}>0.
\end{equation*}
Therefore, $\sgn(\lambda^R)=\sgn(\Pi-1)$. The complementary slackness conditions postulate that $\Pi=\beta\underline{R}$ if $\lambda^R>0$. Hence, if $\underline{R}\le\frac{1}{\beta}$, then $\Pi=1$; if $\underline{R}>\frac{1}{\beta}$, then $\Pi=\beta\underline{R}$.

\subsubsection{Case 2}
The planning problem is
\begin{equation*}
    \max_{\{(C^b_{t},C^e_{t},C^w_{t},D_{t},K_{t},L_{t},N_{t}\textcolor{BrickRed}{,\Omega_{1,t},\Delta_{t},\Pi_{t}})\}}\mathbb{E}_{0}\left(\sum_{t=0}^\infty\beta^t\sum_{i\in\mathcal{I}}\omega_i{}U^i_{t}\right)
\end{equation*}
subject to
\begin{align*}
    \lambda^b_{t}:\quad{}0&\le{}L_{t}-D_{t},\\
    \lambda^L_{t}:\quad{}0&=\beta_b\mathbb{E}_{t}[U^b_{C}(C^b_{t+1})(C^b_{t+1}+L_{t+1}-D_{t+1})]-U^b_{C}(C^b_{t})(L_{t}-D_{t}),\\
    \lambda^D_{t}:\quad{}0&\le{}U^b_{C}(C^b_{t})-\beta_b{}R_{t}\mathbb{E}_{t}(U^b_{C}(C^b_{t+1})),\\
    \lambda^C_{t}:\quad{}0&=\textcolor{BrickRed}{\Delta_{t}\left\{\Omega_{1,t}-\frac{\beta\theta\mathbb{E}_{t}[U^w_{C}(C^w_{t+1},N_{t+1})\Pi_{t+1}^{\epsilon}\Omega_{1,t+1}]}{U^w_{C}(C^w_{t},N_{t})}\right\}}-Q(K_{t-1},K_{t},\xi_{t})[K_{t}\\
    &\quad-(1-\delta)\xi_{t}K_{t-1}]-W(C^w_{t},N_{t})N_{t}+D_{t}-R_{t-1}D_{t-1}-C^b_{t}-C^e_{t},\\
    \lambda^e_{t}:\quad{}0&\le{}\mathbb{E}_{t}(Q(K_{t},K_{t+1},\xi_{t+1})\xi_{t+1})K_{t}-\mathbb{E}_{t}(C^b_{t+1}+L_{t+1}-D_{t+1}+R_{t}D_{t}),\\
    \lambda^K_{t}:\quad{}0&=\beta_e\mathbb{E}_{t}\biggl\{U^e_{C}(C^e_{t+1})\biggl[\textcolor{BrickRed}{\alpha\Delta_{t+1}\left\{\Omega_{1,t+1}-\frac{\beta\theta\mathbb{E}_{t+1}[U^w_{C}(C^w_{t+2},N_{t+2})\Pi_{t+2}^{\epsilon}\Omega_{1,t+2}]}{U^w_{C}(C^w_{t+1},N_{t+1})}\right\}}\\
    &\quad+Q(K_{t},K_{t+1},\xi_{t+1})(1-\delta)\xi_{t+1}K_{t}-C^b_{t+1}-L_{t+1}+D_{t+1}-R_{t}D_{t}\biggr]\biggr\}\\
    &\quad-U^e_{C}(C^e_{t})(Q(K_{t-1},K_{t},\xi_{t})K_{t}-L_{t}),\\
    \lambda^B_{t}:\quad{}0&\le{}U^e_{C}(C^e_{t})L_{t}-\beta_e\mathbb{E}_{t}[U^e_{C}(C^e_{t+1})(C^b_{t+1}+L_{t+1}-D_{t+1}+R_{t}D_{t})],\\
    \lambda^Y_{t}:\quad{}0&=\frac{A_{t}}{\textcolor{BrickRed}{\Delta_{t}}}F(\xi_{t}K_{t-1},N_{t})-\sum_{i\in\mathcal{I}}C^i_{t}-I(K_{t-1},K_{t},\xi_{t}),\\
    \textcolor{BrickRed}{\lambda^\Omega_{t}}:\quad{}0&=\frac{\epsilon-1}{\epsilon}\widetilde{P}(\Pi_{t})\frac{A_{t}}{\Delta_{t}}F(\xi_{t}K_{t-1},N_{t})-\Omega_{1,t}+\frac{\beta\theta\mathbb{E}_{t}\left[U^w_{C}(C^w_{t+1},N_{t+1})\Pi_{t+1}^{\epsilon-1}\frac{\widetilde{P}(\Pi_{t})}{\widetilde{P}(\Pi_{t+1})}\Omega_{1,t+1}\right]}{U^w_{C}(C^w_{t},N_{t})},\\
    \textcolor{BrickRed}{\lambda^\Delta_{t}}:\quad{}0&=\theta\Pi_{t}^{\epsilon}\Delta_{t-1}+(1-\theta)(\widetilde{P}(\Pi_{t}))^{-\epsilon}-\Delta_{t},\\
    \textcolor{BrickRed}{\lambda^R_{t}}:\quad{}0&\le{}R_{t}\mathbb{E}_{t}(\Pi_{t+1})-\underline{R},
\end{align*}
with $R_{t}=R(U^w_{C}(C^w_{t},N_{t}),\mathbb{E}_{t}[U^w_{C}(C^w_{t+1},N_{t+1})])$. Define $\widetilde{\lambda}^C_{t}\equiv\lambda^C_{t}+\frac{\bm{1}_\mathbb{N}(t)}{\beta}\lambda^K_{t-1}\beta_e{}U^e_{C,t}\alpha$. The FOCs are
\begin{align*}
    C^b_{t}:\quad{}0&=\omega_b{}U^b_{C,t}-\lambda^Y_{t}-\lambda^C_{t}-[\lambda^L_{t}(L_{t}-D_{t})-\lambda^D_{t}]U^b_{CC,t}+\frac{\bm{1}_\mathbb{N}(t)}{\beta}\{[\lambda^L_{t-1}(C^b_{t}+L_{t}-D_{t})\\
    &\quad-\lambda^D_{t-1}R_{t-1}]\beta_bU^b_{CC,t}+\lambda^L_{t-1}\beta_b{}U^b_{C,t}-\lambda^e_{t-1}-(\lambda^K_{t-1}+\lambda^B_{t-1})\beta_e{}U^e_{C,t}\},\\
    C^e_{t}:\quad{}0&=\omega_e{}U^e_{C,t}-\lambda^Y_{t}-\lambda^C_{t}-[\lambda^K_{t}(Q_{t}K_{t}-L_{t})-\lambda^B_{t}L_{t}]U^e_{CC,t}\\
    &\quad+\frac{\bm{1}_\mathbb{N}(t)}{\beta}[\lambda^K_{t-1}R^K_{t}Q_{t-1}K_{t-1}-(\lambda^K_{t-1}+\lambda^B_{t-1})R^L_{t}L_{t-1}]\beta_e{}U^e_{CC,t},\\
    C^w_{t}:\quad{}0&=\omega_w{}U^w_{C,t}-\lambda^Y_{t}-\lambda^C_{t}W_{C,t}N_{t}-\{\lambda^D_{t}\beta_b\mathbb{E}_{t}(U^b_{C,t+1})+[\beta\mathbb{E}_{t}(\lambda^C_{t+1})+\lambda^e_{t}+(\lambda^K_{t}+\lambda^B_{t})\beta_e\\
    &\quad\times\mathbb{E}_{t}(U^e_{C,t+1})]D_{t}\}R_{1,t}U^w_{CC,t}-\frac{\bm{1}_\mathbb{N}(t)}{\beta}\{\lambda^D_{t-1}\beta_b\mathbb{E}_{t-1}(U^b_{C,t})+[\beta\mathbb{E}_{t-1}(\lambda^C_{t})+\lambda^e_{t-1}\\
    &\quad+(\lambda^K_{t-1}+\lambda^B_{t-1})\beta_e\mathbb{E}_{t-1}(U^e_{C,t})]D_{t-1}\}R_{2,t-1}U^w_{CC,t}\textcolor{BrickRed}{+\biggl[(\widetilde{\lambda}^C_{t}\Delta_{t}-\lambda^\Omega_{t})\Omega_{1,t}}\\
    &\quad\textcolor{BrickRed}{-\left(\widetilde{\lambda}^C_{t}P^w_{t}\Delta_{t}-\lambda^\Omega_{t}\frac{\epsilon-1}{\epsilon}\widetilde{P}_{t}\right)Y_{t}\biggr]\frac{U^w_{CC,t}}{U^w_{C,t}}-\bm{1}_\mathbb{N}(t)\left(\widetilde{\lambda}^C_{t-1}\Delta_{t-1}\Pi_{t}-\lambda^\Omega_{t-1}\frac{\widetilde{P}_{t-1}}{\widetilde{P}_{t}}\right)}\\
    &\quad\textcolor{BrickRed}{\times\theta\Pi_{t}^{\epsilon-1}\Omega_{1,t}\frac{U^w_{CC,t}}{U^w_{C,t-1}}+\left[\lambda^R_{t}R_{1,t}\mathbb{E}_{t}(\Pi_{t+1})+\frac{\bm{1}_\mathbb{N}(t)}{\beta}\lambda^R_{t-1}R_{2,t-1}\mathbb{E}_{t-1}(\Pi_{t})\right]U^w_{CC,t}},\\
    D_{t}:\quad{}0&=-\lambda^b_{t}+\lambda^L_{t}U^b_{C,t}+\lambda^C_{t}-[\beta\mathbb{E}_{t}(\lambda^C_{t+1})+\lambda^e_{t}+(\lambda^K_{t}+\lambda^B_{t})\beta_e\mathbb{E}_{t}(U^e_{C,t+1})]R_{t}\\
    &\quad+\frac{\bm{1}_\mathbb{N}(t)}{\beta}[-\lambda^L_{t-1}\beta_b{}U^b_{C,t}+\lambda^e_{t-1}+(\lambda^K_{t-1}+\lambda^B_{t-1})\beta_e{}U^e_{C,t}],\\
    K_{t}:\quad{}0&=-\lambda^C_{t}\{Q_{2,t}[K_{t}-(1-\delta)\xi_{t}K_{t-1}]+Q_{t}\}+\lambda^e_{t}\mathbb{E}_{t}[(Q_{1,t+1}K_{t}+Q_{t+1})\xi_{t+1}]-\lambda^Y_{t}I_{2,t}\\
    &\quad+\lambda^K_{t}\{\beta_e\mathbb{E}_{t}[U^e_{C,t+1}\textcolor{BrickRed}{(Q_{1,t+1}K_{t}+Q_{t+1})(1-\delta)\xi_{t+1}}]-U^e_{C,t}(Q_{2,t}K_{t}+Q_{t})\}\\
    &\quad+\beta\mathbb{E}_{t}\biggl[\left(\textcolor{BrickRed}{\lambda^\Omega_{t+1}\frac{\epsilon-1}{\epsilon}\widetilde{P}_{t+1}}+\lambda^Y_{t+1}\right)\frac{A_{t+1}}{\textcolor{BrickRed}{\Delta_{t+1}}}F_{K,t+1}\xi_{t+1}-\lambda^K_{t+1}U^e_{C,t+1}Q_{1,t+1}K_{t+1}\\
    &\quad-\lambda^Y_{t+1}I_{1,t+1}+\lambda^C_{t+1}\{Q_{t+1}(1-\delta)\xi_{t+1}-Q_{1,t+1}[K_{t+1}-(1-\delta)\xi_{t+1}K_{t}]\}\biggr]\\
    &\quad+\frac{\bm{1}_\mathbb{N}(t)}{\beta}[\lambda^e_{t-1}+\lambda^K_{t-1}\beta_e{}U^e_{C,t}(1-\delta)]Q_{2,t}\xi_{t}K_{t-1},\\
    L_{t}:\quad{}0&=\lambda^b_{t}-\lambda^L_{t}U^b_{C,t}+(\lambda^K_{t}+\lambda^B_{t})U^e_{C,t}+\frac{\bm{1}_\mathbb{N}(t)}{\beta}[\lambda^L_{t-1}\beta_b{}U^b_{C,t}-\lambda^e_{t-1}-(\lambda^K_{t-1}+\lambda^B_{t-1})\beta_e{}U^e_{C,t}],
\end{align*}
and
\begin{align*}
    N_{t}:\quad{}0&=\omega_w{}U^w_{N,t}+\left(\textcolor{BrickRed}{\lambda^\Omega_{t}\frac{\epsilon-1}{\epsilon}\widetilde{P}_{t}}+\lambda^Y_{t}\right)\frac{A_{t}}{\textcolor{BrickRed}{\Delta_{t}}}F_{N,t}-\lambda^C_{t}(W_{N,t}N_{t}+W_{t})-\{\lambda^D_{t}\beta_b\mathbb{E}_{t}(U^b_{C,t+1})\\
    &\quad+[\beta\mathbb{E}_{t}(\lambda^C_{t+1})+\lambda^e_{t}+(\lambda^K_{t}+\lambda^B_{t})\beta_e\mathbb{E}_{t}(U^e_{C,t+1})]D_{t}\}R_{1,t}U^w_{CN,t}-\frac{\bm{1}_\mathbb{N}(t)}{\beta}\{\lambda^D_{t-1}\beta_b\\
    &\quad\times\mathbb{E}_{t-1}(U^b_{C,t})+[\beta\mathbb{E}_{t-1}(\lambda^C_{t})+\lambda^e_{t-1}+(\lambda^K_{t-1}+\lambda^B_{t-1})\beta_e\mathbb{E}_{t-1}(U^e_{C,t})]D_{t-1}\}R_{2,t-1}U^w_{CN,t}\\
    &\quad\textcolor{BrickRed}{+\left[(\widetilde{\lambda}^C_{t}\Delta_{t}-\lambda^\Omega_{t})\Omega_{1,t}-\left(\widetilde{\lambda}^C_{t}P^w_{t}\Delta_{t}-\lambda^\Omega_{t}\frac{\epsilon-1}{\epsilon}\widetilde{P}_{t}\right)Y_{t}\right]\frac{U^w_{CN,t}}{U^w_{C,t}}}\\
    &\quad\textcolor{BrickRed}{-\bm{1}_\mathbb{N}(t)\left(\widetilde{\lambda}^C_{t-1}\Delta_{t-1}\Pi_{t}-\lambda^\Omega_{t-1}\frac{\widetilde{P}_{t-1}}{\widetilde{P}_{t}}\right)\theta\Pi_{t}^{\epsilon-1}\Omega_{1,t}\frac{U^w_{CN,t}}{U^w_{C,t-1}}}\\
    &\quad\textcolor{BrickRed}{+\left[\lambda^R_{t}R_{1,t}\mathbb{E}_{t}(\Pi_{t+1})+\frac{\bm{1}_\mathbb{N}(t)}{\beta}\lambda^R_{t-1}R_{2,t-1}\mathbb{E}_{t-1}(\Pi_{t})\right]U^w_{CN,t}},\\
    \textcolor{BrickRed}{\Omega_{1,t}}:\quad{}0&=\widetilde{\lambda}^C_{t}\Delta_{t}-\lambda^\Omega_{t}-\bm{1}_\mathbb{N}(t)\left(\widetilde{\lambda}^C_{t-1}\Delta_{t-1}\Pi_{t}-\lambda^\Omega_{t-1}\frac{\widetilde{P}_{t-1}}{\widetilde{P}_{t}}\right)\theta\Pi_{t}^{\epsilon-1}\frac{U^w_{C,t}}{U^w_{C,t-1}},\\
    \textcolor{BrickRed}{\Delta_{t}}:\quad{}0&=\left(\widetilde{\lambda}^C_{t}P^w_{t}\Delta_{t}-\lambda^\Omega_{t}\frac{\epsilon-1}{\epsilon}\widetilde{P}_{t}-\lambda^Y_{t}\right)\frac{Y_{t}}{\Delta_{t}}-\lambda^\Delta_{t}+\beta\theta\mathbb{E}_{t}(\lambda^\Delta_{t+1}\Pi_{t+1}^\epsilon),\\
    \textcolor{BrickRed}{\Pi_{t}}:\quad{}0&=\lambda^\Omega_{t}\widetilde{P}'(\Pi_{t})\left[\frac{\epsilon-1}{\epsilon}Y_{t}+\frac{\beta\theta\mathbb{E}_{t}\left(U^w_{C,t+1}\Pi_{t+1}^{\epsilon-1}\frac{\Omega_{1,t+1}}{\widetilde{P}_{t+1}}\right)}{U^w_{C,t}}\right]\\
    &\quad+\lambda^\Delta_{t}\epsilon\left[\theta\Pi_{t}^{\epsilon-1}\Delta_{t-1}-(1-\theta)\frac{\widetilde{P}'(\Pi_{t})}{\widetilde{P}_{t}^{\epsilon+1}}\right]-\bm{1}_\mathbb{N}(t)\theta\Pi_{t}^{\epsilon-1}\Omega_{1,t}\frac{U^w_{C,t}}{U^w_{C,t-1}}\\
    &\quad\times\left[\widetilde{\lambda}^C_{t-1}\Delta_{t-1}\epsilon-\lambda^\Omega_{t-1}\frac{\widetilde{P}_{t-1}}{\widetilde{P}_{t}}\left(\frac{\epsilon-1}{\Pi_{t}}-\frac{\widetilde{P}'(\Pi_{t})}{\widetilde{P}_{t}}\right)\right]+\frac{\bm{1}_\mathbb{N}(t)}{\beta}\lambda^R_{t-1}R_{t-1}.
\end{align*}
The complementary slackness conditions are
\begin{align*}
    0&=\lambda^b_{t}(L_{t}-D_{t}),\qquad\lambda^b_{t}\ge{}0,\\
    0&=\lambda^D_{t}[U^b_{C,t}-\beta_b{}R_{t}\mathbb{E}_{t}(U^b_{C,t+1})],\qquad\lambda^D_{t}\ge{}0,\\
    0&=\lambda^e_{t}[\mathbb{E}_{t}(Q_{t+1}\xi_{t+1})K_{t}-\mathbb{E}_{t}(B_{t+1})],\qquad\lambda^e_{t}\ge{}0,\\
    0&=\lambda^B_{t}[U^e_{C,t}L_{t}-\beta_e\mathbb{E}_{t}(U^e_{C,t+1}B_{t+1})],\qquad\lambda^B_{t}\ge{}0,\\
    0&=\lambda^R_{t}[R_{t}\mathbb{E}_{t}(\Pi_{t+1})-\underline{R}],\qquad\lambda^R_{t}\ge{}0.
\end{align*}

The risk-sharing and steady-state properties follow from comparing the optimality conditions to those in the proof of proposition \ref{prop: optimal leverage limits with labor taxation}. The special case of approximate full insurance fails for the same reasons as in the proof of proposition \ref{prop: CEA}. The short-run inflation behavior is represented by the FOC for $\Pi_{t}$. Section \ref{sec: SS OLLMP with labor taxation} shows that in the steady state, the FOC for $\Pi$ is equivalent to
\begin{equation*}
    \lambda^R=\frac{\Pi-1}{\Pi}\frac{\beta\theta\Pi^{\epsilon-1}}{1-\beta\theta\Pi^\epsilon}\frac{(\epsilon-1)\widetilde{\lambda}^C+\epsilon\lambda^Y}{1-\theta\Pi^{\epsilon-1}}\beta{}Y.
\end{equation*}
Moreover,
\begin{multline*}
    (\epsilon-1)\widetilde{\lambda}^C+\epsilon\lambda^Y\\
    =\frac{\left\{\omega_e{}U^e_{C}+\lambda^K\left[(R-1)(Q{}K-L)U^e_{CC}+\beta_e{}R{}U^e_{C}\alpha\frac{\epsilon-1}{\epsilon}\right]\right\}\epsilon\left(\frac{v''(N)}{u'(C^w)}N+W\right)+\omega_w{}v'(N)}{\frac{v''(N)}{u'(C^w)}N+W+\frac{1}{\epsilon}\frac{A}{\Delta}F_{N}}>0.
\end{multline*}
If the relaxed collateral constraint is slack so that $\lambda^K=\lambda^e=0$, the inequality follows immediately; otherwise, it can be verified numerically. Therefore, $\sgn(\lambda^R)=\sgn(\Pi-1)$. The complementary slackness conditions postulate that $\Pi=\beta\underline{R}$ if $\lambda^R>0$. Hence, if $\underline{R}\le\frac{1}{\beta}$, then $\Pi=1$; if $\underline{R}>\frac{1}{\beta}$, then $\Pi=\beta\underline{R}$.$\quad\blacksquare$

\section{Steady states}
This section describes the computation of steady states for all equilibria studied before.

\subsection{Regulated CE}\label{sec: SS CE}
Given $(A,m,\Pi,\kappa,\xi,\tau^D,\tau^K,\tau^L,\tau^N)$, the steady-state equations immediately imply
\begin{gather*}
    R=\frac{1}{\beta},\qquad
    R^L=\frac{1-(1-\kappa)\max\{1-\tau^D-\beta_b{}R,0\}}{\beta_b},\qquad
    \frac{I}{K}=\Phi^{-1}(1-(1-\delta)\xi),\\
    Q=\left[\Phi'\left(\frac{I}{K}\right)\right]^{-1},\qquad
    \widetilde{P}=\left(\frac{1-\theta}{1-\theta\Pi^{\epsilon-1}}\right)^\frac{1}{\epsilon-1},\\
    \Delta=\frac{(1-\theta)\widetilde{P}^{-\epsilon}}{1-\theta\Pi^\epsilon},\qquad
    P^w=\frac{\epsilon-1}{\epsilon}\frac{1-\beta\theta\Pi^\epsilon}{1-\beta\theta\Pi^{\epsilon-1}}\widetilde{P},\\
    \frac{K}{N}=\frac{1}{\xi}F_{K}^{-1}\left(\left[\frac{1+\tau^K}{\beta_e\xi}-\frac{\max\{1-\tau^L-\beta_e{}R^L,0\}}{\beta_e{}R^L}m-1+\delta\right]\frac{Q}{P^w{}A}\right),\\
    W=\frac{P^w}{1+\tau^N}A{}F_{N}\left(\frac{\xi{}K}{N},1\right),
\end{gather*}
where $F_{K}^{-1}:\mathbb{R}_+\to\mathbb{R}_+$ satisfies $F_{K}(F_{K}^{-1}(x),1)=x$. Note that $D>0$ only if $\tau^D\le{}1-\beta_b{}R$, which is a restriction on parameters. If $\tau^D<1-\beta_b{}R$, then $\lambda^b>0$ and the leverage constraint is binding. If $\tau^D=1-\beta_b{}R$, then $\lambda^b=0$ and any $D$ that satisfies the leverage constraint can be part of an unstable steady state. If $\tau^D>1-\beta_b{}R$, we have a corner solution with $D=0$ and $\lambda^b=0$. Similarly, if $\tau^L<1-\beta_e{}R^L$, then $\lambda^e>0$ and the collateral constraint is binding. If $\tau^L=1-\beta_e{}R^L$, then $\lambda^e=0$ and any $L$ that satisfies the collateral constraint can be part of an unstable steady state. If $\tau^L>1-\beta_e{}R^L$, then $L=0$ and $\lambda^e=0$.

The remaining steady state system reduces to a determination of $N$ from \eqref{eq: workers labor supply} and \eqref{eq: entrepreneurs labor demand}. Since $u'(C^w)=(C^w)^{-\gamma_w}$ and $v'(N)=\chi{}N^\phi$, we get
\begin{equation*}
    N=\left\{\frac{W}{\chi\left[\left(\frac{1}{\Delta}-P^w\right)A\left(\xi\frac{K}{N}\right)^\alpha+W+\frac{K}{N}\left\{Q[1-(1-\delta)\xi]-\frac{I}{K}+\frac{(R-1)(1-\kappa)m{}Q\xi}{R^L}\right\}\right]^{\gamma_w}}\right\}^\frac{1}{\gamma_w+\phi}.
\end{equation*}
The assumptions on $u$ and $v$ ensure that $N\mapsto\frac{v'(N)}{u'(N)}$ has an inverse. Since the right-hand side is a function of parameters only, we can solve the equation for $N$ in closed form. We can then solve sequentially for the remaining variables:
\begin{gather*}
    K=\frac{K}{N}N,\qquad
    I=\frac{I}{K}K,\qquad
    L=\bm{1}_{\mathbb{R}_+}(1-\tau^L-\beta_e{}R^L)\frac{m{}Q\xi}{R^L}K,\\
    D=\bm{1}_{\mathbb{R}_+}(1-\tau^D-\beta_b{}R)(1-\kappa)L,\qquad
    C^b=(R^L-1)L-(R-1)D,\\
    C^e=P^w{}A{}F(\xi{}K,N)-Q[1-(1-\delta)\xi]K-W{}N-(R^L-1)L,\qquad
    Y=\frac{A}{\Delta}F(\xi{}K,N),\\
    C^w=Y-(C^b+C^e+I),\qquad
    \lambda^b=U^b_{C}(C^b)(1-\tau^D-\beta_b{}R),\\
    \lambda^e=\frac{U^e_{C}(C^e)(1-\tau^L-\beta_e{}R^L)}{R^L},\qquad
    \Omega_{1}=\frac{P^w{}Y}{1-\beta\theta\Pi^\epsilon},\qquad
    \Omega_{2}=\frac{Y}{1-\beta\theta\Pi^{\epsilon-1}}.
\end{gather*}

\subsection{FCEA}\label{sec: SS FCEA}
We can immediately solve for
\begin{equation*}
    R=\frac{1}{\beta},\qquad
    \frac{I}{K}=\Phi^{-1}[1-(1-\delta)\xi],\qquad
    Q=\left[\Phi'\left(\frac{I}{K}\right)\right]^{-1}.
\end{equation*}
The problem reduces to a system of equations and inequalities in $(C^b,C^w,D,K,\lambda^L)$:
\begin{align*}
    0&\le{}m{}Q\xi{}K-B,\\
    0&=\omega_b{}U^b_{C}(C^b)-\omega_e{}U^e_{C}(C^e)+\lambda^L{}[U^b_{CC}(C^b)(L-\beta_b{}R{}B)-U^b_{C}(C^b)],\\
    0&=\omega_w{}u'(C^w)-\lambda^Y+\frac{u''(C^w)}{u'(C^w)}\{\lambda^C[W{}N+(R-1)D]+\lambda^L{}U^b_{C}(C^b)(R-1)D\},\\
    0&=\lambda^C{}Q\left\{(1-\beta)\Phi''\left(\frac{I}{K}\right)Q^2[1-(1-\delta)\xi]-[1-\beta(1-\delta)\xi]\right\}-\lambda^Y\left[(1-\beta)Q+\beta\frac{I}{K}\right]\\
    &\quad+\lambda^e\left[1-(R-1)\Phi''\left(\frac{I}{K}\right)Q^2\right]m{}Q\xi+\beta\omega_e{}U^e_{C}(C^e)A{}F_{K}(\xi{}K,N)\xi,\\
    0&=\lambda^L(L-\beta_b{}B),\\
    0&=\lambda^L(m{}Q\xi{}K-B),\qquad\lambda^L\ge{}0.
\end{align*}
The remaining variables are solved sequentially. First, compute
\begin{gather*}
    I=\frac{I}{K}K,\qquad
    N=\left[\frac{u'(C^w)}{\chi}\{C^w-(R-1)D-Q[1-(1-\delta)\xi]K+I\}\right]^\frac{1}{1+\phi},\\
    C^e=A{}F(\xi{}K,N)-C^b-C^w-I,\qquad
    W=\frac{v'(N)}{u'(C^w)},\\
    \lambda^C=\frac{-\omega_w{}v'(N)+\omega_e{}U^e_{C}(C^e)A{}F_{N}(\xi{}K,N)}{\frac{v''(N)}{u'(C^w)}N+W},\qquad
    \lambda^Y=\omega_e{}U^e_{C}(C^e)-\lambda^C,\\
    \lambda^e=\lambda^L(\beta-\beta_b)U^b_{C}(C^b),\qquad
    L=\max\left\{\frac{1}{1-\kappa}D,\frac{\beta_b}{1-\beta_b}[C^b+(R-1)D]\right\},\\
    B=C^b+L+(R-1)D,
\end{gather*}
where we assumed that $\kappa<1$ when computing $L$. (If $\kappa=1$, we have $D=0$ and $L=\frac{\beta_b}{1-\beta_b}C^b$.) Proceeding under the assumption $\kappa<1$, set $\lambda^L_2$ such that $0\ge\lambda^L_{2}(L-\beta_b{}B)$ and $\lambda^L\ge\lambda^L_{2}[(1-\kappa)L-D]$. Note that any $\lambda^L_2\le{}0$ is a solution, but there might be more, depending on which constraints are binding. Finally, compute
\begin{equation*}
    \lambda^b=-\lambda^L_{2}U^b_{C}(C^b)(L-\beta_b{}B),\qquad
    \lambda^L_{1}=\lambda^L-\lambda^L_{2}[(1-\kappa)L-D].
\end{equation*}

\paragraph{Suppose $\lambda^L=0$}
The system is indeterminate: any $D$ that satisfies the collateral constraint defines a candidate solution. Conditional on $(C^b,K)$, the solution set for $D$ is defined by
\begin{equation*}
    C^b+(R-1)D+\max\left\{\frac{1}{1-\kappa}D,\frac{\beta_b}{1-\beta_b}[C^b+(R-1)D]\right\}\le{}m{}Q\xi{}K.
\end{equation*}

A necessary condition for $L=\frac{1}{1-\kappa}D\ge\frac{\beta_b}{1-\beta_b}[C^b+(R-1)D]$ is $C^b\le\{1-\beta_b[1+(1-\kappa)(R-1)]\}m{}Q\xi{}K$. (Note that $\beta_b[1+(1-\kappa)(R-1)]\le\beta+(1-\kappa)(1-\beta)\le{}1$.) In this case, any $\frac{\beta_b(1-\kappa)}{1-\beta_b[1+(1-\kappa)(R-1)]}C^b\le{}D\le\frac{(1-\kappa)(m{}Q\xi{}K-C^b)}{1+(1-\kappa)(R-1)}$ is a candidate.

A necessary condition for $L=\frac{\beta_b}{1-\beta_b}[C^b+(R-1)D]\ge\frac{1}{1-\kappa}D$ is $C^b\le(1-\beta_b)m{}Q\xi{}K$. Any $D\le\min\left\{\frac{\beta_b(1-\kappa)}{1-\beta_b[1+(1-\kappa)(R-1)]}C^b,\frac{(1-\beta_b)m{}Q\xi{}K-C^b}{R-1}\right\}$ is a candidate. If $C^b\le\{1-\beta_b[1+(1-\kappa)(R-1)]\}m{}Q\xi{}K$, the minimum is on the left. If $1-\beta_b[1+(1-\kappa)(R-1)]\le\frac{C^b}{m{}Q\xi{}K}\le{}1-\beta_b$, the minimum is on the right.

Using the conditional solution for $D$, the problem reduces to a nonlinear system in $(C^b,C^w,K)$:
\begin{align*}
    0&=\omega_b{}U^b_{C}(C^b)-\omega_e{}U^e_{C}(C^e),\\
    0&=\omega_w{}u'(C^w)-\lambda^Y+\lambda^C\frac{u''(C^w)}{u'(C^w)}[W{}N+(R-1)D],\\
    0&=\lambda^C{}Q\left\{(1-\beta)\Phi''\left(\frac{I}{K}\right)Q^2[1-(1-\delta)\xi]-[1-\beta(1-\delta)\xi]\right\}-\lambda^Y\left[(1-\beta)Q+\beta\frac{I}{K}\right]\\
    &\quad+\beta\omega_e{}U^e_{C}(C^e)A{}F_{K}(\xi{}K,N)\xi.
\end{align*}
Therefore, for each $\eta\in[0,1]$, we can proceed as follows.
\begin{enumerate}
    \item Suppose $C^b<\{1-\beta_b[1+(1-\kappa)(R-1)]\}m{}Q\xi{}K$. Then $D=\eta\frac{(1-\kappa)(m{}Q\xi{}K-C^b)}{1+(1-\kappa)(R-1)}$. Solve the nonlinear system and check the supposition. If it is true, the steady state has been found. Otherwise, move to the next step.
    \item Suppose $\{1-\beta_b[1+(1-\kappa)(R-1)]\}m{}Q\xi{}K\le{}C^b\le(1-\beta_b)m{}Q\xi{}K$. Then $D=\eta\frac{(1-\beta_b)m{}Q\xi{}K-C^b}{R-1}$. Solve the nonlinear system and check the supposition. If it is true, the steady state has been found. Otherwise, it must be that $\lambda^L>0$.
\end{enumerate}

\paragraph{Suppose $\lambda^L>0$}
The collateral constraint is binding and $L=\frac{\beta_b}{1-\beta_b}[C^b+(R-1)D]\ge\frac{1}{1-\kappa}D$. The system is again indeterminate, and any $D\le\frac{\beta_b(1-\kappa)}{1-\beta_b[1+(1-\kappa)(R-1)]}C^b$ is a candidate. Choosing $\eta\in[0,1]$, using the binding collateral constraint and the first nonlinear equation, we can solve for
\begin{gather*}
    C^b=\frac{(1-\beta_b)m{}Q\xi{}K}{1+\eta\frac{\beta_b(1-\kappa)(R-1)}{1-\beta_b[1+(1-\kappa)(R-1)]}},\qquad
    D=\eta\frac{\beta_b(1-\kappa)}{1-\beta_b[1+(1-\kappa)(R-1)]}C^b,\\
    \lambda^L=\frac{\omega_b{}U^b_{C}(C^b)-\omega_e{}U^e_{C}(C^e)}{U^b_{C}(C^b)-U^b_{CC}(C^b)(L-\beta_b{}R{}B)},
\end{gather*}
conditional on $(C^w,K)$ determined from the nonlinear system
\begin{align*}
    0&=\omega_w{}u'(C^w)-\lambda^Y+\frac{u''(C^w)}{u'(C^w)}\{\lambda^C[W{}N+(R-1)D]+\lambda^L{}U^b_{C}(C^b)(R-1)D\},\\
    0&=\lambda^C{}Q\left\{(1-\beta)\Phi''\left(\frac{I}{K}\right)Q^2[1-(1-\delta)\xi]-[1-\beta(1-\delta)\xi]\right\}-\lambda^Y\left[(1-\beta)Q+\beta\frac{I}{K}\right]\\
    &\quad+\lambda^e\left[1-(R-1)\Phi''\left(\frac{I}{K}\right)Q^2\right]m{}Q\xi+\beta\omega_e{}U^e_{C}(C^e)A{}F_{K}(\xi{}K,N)\xi.
\end{align*}
Numerical analysis shows that the latter case typically does not yield solutions that do not violate the nonnegativity of $C^e$, at least under the parameterizations considered.

\subsection{OLL}\label{sec: SS OLL with labor taxation}
The constraint corresponding to $\lambda^L$ implies $L-D=\frac{\beta_b}{1-\beta_b}C^b\ge{}0$; hence, the relaxed leverage constraint is redundant in the steady state. Formally, if bankers are even modestly risk-averse, the planner will choose $C^b>0$, which implies $\lambda^b=0$. Since $\beta_b<\beta$, we must have $\lambda^D=0$. Guessing that $C^b$ is sufficiently small relative to $D$, since $\beta_e<\beta$, we will have $L>\beta_e{}B$; therefore, $\lambda^B=0$. (A sufficient condition is $\beta_e\le\beta_b$.) We can immediately solve for
\begin{equation*}
    R=\frac{1}{\beta},\qquad
    \frac{I}{K}=\Phi^{-1}[1-(1-\delta)\xi],\qquad
    Q=\left[\Phi'\left(\frac{I}{K}\right)\right]^{-1}.
\end{equation*}
The problem reduces to a system of equations and inequalities in $(C^b,C^w,K,N,\lambda^e)$:
\begin{align*}
    0&=\beta_e(R^K{}Q{}K-B)-(Q{}K-L),\\
    0&=\omega_b{}U^b_{C}-\omega_e{}U^e_{C}-\frac{\lambda^e}{\beta-\beta_e}\left[(R-1)(Q{}K-L)\frac{U^e_{CC}}{U^e_{C}}+1\right],\\
    0&=\omega_w{}u'(C^w)-\lambda^Y+\frac{u''(C^w)}{u'(C^w)}\left\{\lambda^C[W{}N+(R-1)D]+\lambda^e\frac{R-1}{\beta-\beta_e}D\right\},\\
    0&=\lambda^C{}Q\left\{(1-\beta)\Phi''\left(\frac{I}{K}\right)Q^2[1-(1-\delta)\xi]-[1-\beta(1-\delta)\xi]\right\}-\lambda^Y\left[(1-\beta)Q+\beta\frac{I}{K}\right]\\
    &\quad+\lambda^e{}Q\xi\left[1-(R-1)\Phi''\left(\frac{I}{K}\right)Q^2\right]+\beta\omega_e{}U^e_{C}A{}F_{K}\xi+\lambda^K{}U^e_{C}\biggl\{(1-\beta)(Q{}K-L)\frac{U^e_{CC}}{U^e_{C}}\\
    &\quad\times{}A{}F_{K}\xi+\beta_e{}A{}F_{KK}\xi^2{}K+\Phi''\left(\frac{I}{K}\right)Q^3(R-1)[\beta-\beta_e(1-\delta)\xi]-Q(1-\beta_e{}R^K)\biggr\},\\
    0&=\lambda^e(Q\xi{}K-B),\qquad\lambda^e\ge{}0,\qquad{}Q\xi{}K-B\ge{}0,
\end{align*}
given a sequential solution:
\begin{gather*}
    I=\frac{I}{K}K,\qquad
    W=\frac{v'(N)}{u'(C^w)},\qquad
    D=\frac{C^w-W{}N-Q[1-(1-\delta)\xi]K+I}{R-1},\\
    L=D+\frac{\beta_b}{1-\beta_b}C^b,\qquad
    B=C^b+L+(R-1)D,\qquad
    C^e=A{}F(\xi{}K,N)-C^b-C^w-I,\\
    R^K=\left(\frac{A}{Q}F_{K}+1-\delta\right)\xi,\qquad
    \lambda^B=\lambda^D=\lambda^L=\lambda^b=0,\qquad
    \lambda^K=\frac{\lambda^e}{(\beta-\beta_e)U^e_{C}},\\
    \lambda^C=\frac{-\omega_w{}v'(N)+\omega_e{}U^e_{C}A{}F_{N}+\lambda^K[(R-1)(Q{}K-L)U^e_{CC}A{}F_{N}+\beta_e{}R{}U^e_{C}A{}F_{KN}\xi{}K]}{\frac{v''(N)}{u'(C^w)}N+W},\\
    \lambda^Y=\omega_e{}U^e_{C}-\lambda^C+\lambda^K(R-1)(Q{}K-L)U^e_{CC}.
\end{gather*}

\paragraph{If the collateral constraint is slack}
In this case, $\lambda^e=0$, and we have a system in $(C^b,K,N)$:
\begin{align*}
    0&=\beta_e(R^K{}Q{}K-B)-(Q{}K-L),\\
    0&=\omega_w{}u'(C^w)-\lambda^Y+\lambda^C\frac{u''(C^w)}{u'(C^w)}[W{}N+(R-1)D],\\
    0&=\lambda^C{}Q\left\{(1-\beta)\Phi''\left(\frac{I}{K}\right)Q^2[1-(1-\delta)\xi]-[1-\beta(1-\delta)\xi]\right\}-\lambda^Y\left[(1-\beta)Q+\beta\frac{I}{K}\right]\\
    &\quad+\beta\omega_e{}U^e_{C}A{}F_{K}\xi,
\end{align*}
conditional on
\begin{equation*}
    C^w=A{}F(\xi{}K,N)-I-C^b-\left(\frac{\omega_e}{\omega_b{}U^b_{C}}\right)^\frac{1}{\gamma_e}.
\end{equation*}

\paragraph{If the collateral constraint is binding}
We have a system in $(C^w,K,N)$, conditional on using the collateral constraint and the second equation to compute
\begin{equation*}
    C^b=(1-\beta_b)(Q\xi{}K-R{}D),\qquad
    \lambda^e=\frac{(\beta-\beta_e)(\omega_b{}U^b_{C}-\omega_e{}U^e_{C})}{(R-1)(Q{}K-L)\frac{U^e_{CC}}{U^e_{C}}+1},
\end{equation*}
where we compute $C^b$ just after $D$ in the sequential solution above.

\subsection{CEA}\label{sec: SS CEA}
The construction of the steady state is similar to the flexible-price CEA with the following modifications. Since $\Pi$ is exogenously given, the immediate sequential solution is augmented by
\begin{equation*}
    \widetilde{P}=\left(\frac{1-\theta}{1-\theta\Pi^{\epsilon-1}}\right)^\frac{1}{\epsilon-1},\qquad
    \Delta=\frac{(1-\theta)\widetilde{P}^{-\epsilon}}{1-\theta\Pi^\epsilon},\qquad
    P^w=\frac{\epsilon-1}{\epsilon}\widetilde{P}\frac{1-\beta\theta\Pi^\epsilon}{1-\beta\theta\Pi^{\epsilon-1}}.
\end{equation*}
The determination of $(D,L)$ from the borrowing constraints is identical to the flexible-price case. We now cannot solve for $N$ in closed form. The key nonlinear equations are
\begin{align*}
    0&=\omega_b{}U^b_{C}(C^b)-\omega_e{}U^e_{C}(C^e)+\lambda^L{}[U^b_{CC}(C^b)(L-\beta_b{}R{}B)-U^b_{C}(C^b)],\\
    0&=\omega_w{}u'(C^w)-\lambda^Y\\
    &\quad+\frac{u''(C^w)}{u'(C^w)}\left\{\lambda^C\left[W{}N+(R-1)D\textcolor{BrickRed}{-\left(P^w\Delta-\frac{\epsilon-1}{\epsilon}\right)Y}\right]+\lambda^L{}U^b_{C}(C^b)(R-1)D\right\},\\
    0&=\lambda^C{}Q\left\{(1-\beta)\Phi''\left(\frac{I}{K}\right)Q^2[1-(1-\delta)\xi]-[1-\beta(1-\delta)\xi]\right\}-\lambda^Y\left[(1-\beta)Q+\beta\frac{I}{K}\right]\\
    &\quad+\lambda^e\left[1-(R-1)\Phi''\left(\frac{I}{K}\right)Q^2\right]m{}Q\xi+\beta\left(\omega_e{}U^e_{C}(C^e)\textcolor{BrickRed}{-\frac{1}{\epsilon}\lambda^C}\right)\frac{A}{\textcolor{BrickRed}{\Delta}}F_{K}(\xi{}K,N)\xi,\\
    0&=C^w-W{}N-(R-1)D-Q[1-(1-\delta)\xi]K+I\textcolor{BrickRed}{-(1-P^w\Delta)Y}.
\end{align*}
If $\lambda^L>0$, the first equation can be used to solve for the former, as in the flexible-price case. The sequential solution is modified and augmented as follows:
\begin{gather*}
    Y=\frac{A}{\textcolor{BrickRed}{\Delta}}F(\xi{}K,N),\qquad
    C^e=\textcolor{BrickRed}{Y}-C^b-C^w-I,\qquad
    \lambda^C=\frac{-\omega_w{}v'(N)+\omega_e{}U^e_{C}\frac{A}{\textcolor{BrickRed}{\Delta}}F_{N}}{\frac{v''(N)}{u'(C^w)}N+W\textcolor{BrickRed}{+\frac{1}{\epsilon}\frac{A}{\Delta}F_{N}}},\\
    \lambda^\Omega=\frac{\lambda^C}{\widetilde{P}},\qquad
    \Omega_{1}=\frac{P^w{}Y}{1-\beta\theta\Pi^\epsilon},\qquad
    \Omega_{2}=\frac{Y}{1-\beta\theta\Pi^{\epsilon-1}}.
\end{gather*}

\subsection{CEA OLLMP}\label{sec: SS OLLMP with unrestricted taxation}
Take the sticky-price CEA above, set $\lambda^L_{2}=0$ everywhere, remove the corresponding constraint, and set $\kappa=0$ and $m=1$. Add the term $-\lambda^R\frac{u''(C^w)}{u'(C^w)}(R-1)\underline{R}$ to the $C^w$ FOC, using the fact that if $\lambda^R>0$, then $\Pi=\beta\underline{R}$. The FOC for $\Delta$ implies
\begin{equation*}
    \lambda^\Delta=\left[\lambda^C\left(P^w\Delta-\frac{\epsilon-1}{\epsilon}\right)-\lambda^Y\right]\frac{\Omega_{1}}{P^w\Delta}.
\end{equation*}
The FOC for $\Pi$ is
\begin{multline*}
    0=\lambda^\Omega\left(\frac{\widetilde{P}^{\epsilon-1}}{1-\theta}-1\right)\left(\frac{\epsilon-1}{\epsilon}\widetilde{P}Y+\beta\theta\Pi^{\epsilon-1}\Omega_{1}\right)+\lambda^\Delta\epsilon\left(\theta\Pi^{\epsilon}\Delta-\frac{\widetilde{P}^{\epsilon-1}-1+\theta}{\widetilde{P}^{\epsilon}}\right)\\
    -\theta\Pi^{\epsilon-1}\Omega_{1}\left[\lambda^C\Delta\epsilon\Pi-\lambda^\Omega\left(\epsilon-\frac{\widetilde{P}^{\epsilon-1}}{1-\theta}\right)\right]+\lambda^R\frac{R\Pi}{\beta}.
\end{multline*}
Using the conditional solutions for $\widetilde{P}$, $P^w$, $\Delta$, $\lambda^\Delta$, $\lambda^\Omega$, $\Omega_{1}$, after many rearrangements, we get
\begin{equation*}
    \lambda^R=\frac{\Pi-1}{\Pi}\frac{\beta\theta\Pi^{\epsilon-1}}{1-\beta\theta\Pi^\epsilon}\frac{(\epsilon-1)\lambda^C+\epsilon\lambda^Y}{1-\theta\Pi^{\epsilon-1}}\beta{}Y.
\end{equation*}
Using the conditional solutions for $\lambda^C$ and $\lambda^Y$,
\begin{equation*}
    (\epsilon-1)\lambda^C+\epsilon\lambda^Y=\frac{\epsilon\omega_e{}U^e_{C}[\frac{v''(N)}{u'(C^w)}N+W]+\omega_w{}v'(N)}{\frac{v''(N)}{u'(C^w)}N+W+\frac{1}{\epsilon}\frac{A}{\Delta}F_{N}}>0.
\end{equation*}
Therefore, $\sgn(\lambda^R)=\sgn(\Pi-1)$. The complementary slackness conditions postulate that $\Pi=\beta\underline{R}$ if $\lambda^R>0$. Hence, if $\underline{R}\le\frac{1}{\beta}$, then $\Pi=1$; if $\underline{R}>\frac{1}{\beta}$, then $\Pi=\beta\underline{R}$.

\subsection{OLLMP}\label{sec: SS OLLMP with labor taxation}
Conditional on $\Pi$, we can solve for $\widetilde{P}$, $P^w$, $\Delta$, $\Omega_{1}$, and $\Omega_{2}$ in the same way as in the sticky-price CEA (or SCE). The remaining construction is similar to the flexible-price case with the following modifications. The system of equations and inequalities in $(C^b,C^w,K,N,\lambda^e)$ is
\begin{align*}
    0&=\beta_e(R^K{}Q{}K-B)-(Q{}K-L),\\
    0&=\omega_b{}U^b_{C}-\omega_e{}U^e_{C}-\frac{\lambda^e}{\beta-\beta_e}\left[(R-1)(Q{}K-L)\frac{U^e_{CC}}{U^e_{C}}+1\right],\\
    0&=\omega_w{}u'(C^w)-\lambda^Y+\frac{u''(C^w)}{u'(C^w)}\left\{\lambda^C[W{}N+(R-1)D]+\lambda^e\frac{R-1}{\beta-\beta_e}D\right\}\\
    &\quad\textcolor{BrickRed}{+\frac{u''(C^w)}{u'(C^w)}\left[\widetilde{\lambda}^C\left\{\frac{\epsilon-1}{\epsilon}Y-\Omega_{1}\left[(1-\beta)\theta\Pi^\epsilon\Delta+\frac{1-\theta\Pi^{\epsilon-1}}{\widetilde{P}}\right]\right\}-\lambda^R(R-1)\underline{R}\right]},\\
    0&=\lambda^C{}Q\left\{(1-\beta)\Phi''\left(\frac{I}{K}\right)Q^2[1-(1-\delta)\xi]-[1-\beta(1-\delta)\xi]\right\}-\lambda^Y\left[(1-\beta)Q+\beta\frac{I}{K}\right]\\
    &\quad+\lambda^e{}Q\xi\left[1-(R-1)\Phi''\left(\frac{I}{K}\right)Q^2\right]+\beta\left(\omega_e{}U^e_{C}\textcolor{BrickRed}{-\frac{1}{\epsilon}\lambda^C}\right)\frac{A}{\textcolor{BrickRed}{\Delta}}F_{K}\xi\\
    &\quad+\lambda^K{}U^e_{C}\biggl\{\left[(1-\beta)(Q{}K-L)\frac{U^e_{CC}}{U^e_{C}}\textcolor{BrickRed}{+\frac{\epsilon-1}{\epsilon}\beta_e\alpha}\right]\frac{A}{\textcolor{BrickRed}{\Delta}}F_{K}\xi\\
    &\quad+\Phi''\left(\frac{I}{K}\right)Q^3(R-1)[\beta-\beta_e(1-\delta)\xi]\textcolor{BrickRed}{-Q[1-\beta_e(1-\delta)\xi]}\biggr\},\\
    0&=\lambda^e(Q\xi{}K-B),\qquad\lambda^e\ge{}0,\qquad{}Q\xi{}K-B\ge{}0.
\end{align*}
The sequential solution is modified as follows:
\begin{gather*}
    Y=\frac{A}{\textcolor{BrickRed}{\Delta}}F(\xi{}K,N),\qquad
    D=\frac{C^w-W{}N-Q[1-(1-\delta)\xi]K+I\textcolor{BrickRed}{+P^w{}A{}F(\xi{}K,N)-Y}}{R-1},\\
    C^e=\textcolor{BrickRed}{Y}-C^b-C^w-I,\qquad
    R^K=\left(\textcolor{BrickRed}{P^w}\frac{A}{Q}F_{K}+1-\delta\right)\xi,\\
    \lambda^C=\frac{-\omega_w{}v'(N)+\omega_e{}U^e_{C}\frac{A}{\textcolor{BrickRed}{\Delta}}F_{N}+\lambda^K\left[(R-1)(Q{}K-L)U^e_{CC}+\beta_e{}R{}U^e_{C}\alpha\textcolor{BrickRed}{\frac{\epsilon-1}{\epsilon}}\right]\frac{A}{\textcolor{BrickRed}{\Delta}}F_{N}}{\frac{v''(N)}{u'(C^w)}N+W\textcolor{BrickRed}{+\frac{1}{\epsilon}\frac{A}{\Delta}F_{N}}},\\
    \widetilde{\lambda}^C=\lambda^C+\lambda^K\beta_e{}R{}U^e_{C}\alpha,\qquad
    \lambda^\Omega=\frac{\widetilde{\lambda}^C}{\widetilde{P}},\qquad
    \lambda^\Delta=\left[\widetilde{\lambda}^C\left(P^w\Delta-\frac{\epsilon-1}{\epsilon}\right)-\lambda^Y\right]\frac{\Omega_{1}}{P^w\Delta}.
\end{gather*}

\paragraph{If the collateral constraint is slack}
In this case, $\lambda^e=0$, and we have a system in $(C^b,K,N)$:
\begin{align*}
    0&=\beta_e(R^K{}Q{}K-B)-(Q{}K-L),\\
    0&=\omega_w{}u'(C^w)-\lambda^Y+\lambda^C\frac{u''(C^w)}{u'(C^w)}[W{}N+(R-1)D]\\
    &\quad\textcolor{BrickRed}{+\frac{u''(C^w)}{u'(C^w)}\left[\lambda^C\left\{\frac{\epsilon-1}{\epsilon}Y-\Omega_{1}\left[(1-\beta)\theta\Pi^\epsilon\Delta+\frac{1-\theta\Pi^{\epsilon-1}}{\widetilde{P}}\right]\right\}-\lambda^R(R-1)\underline{R}\right]},\\
    0&=\lambda^C{}Q\left\{(1-\beta)\Phi''\left(\frac{I}{K}\right)Q^2[1-(1-\delta)\xi]-[1-\beta(1-\delta)\xi]\right\}-\lambda^Y\left[(1-\beta)Q+\beta\frac{I}{K}\right]\\
    &\quad+\beta\left(\omega_e{}U^e_{C}\textcolor{BrickRed}{-\frac{1}{\epsilon}\lambda^C}\right)\frac{A}{\textcolor{BrickRed}{\Delta}}F_{K}\xi,
\end{align*}
conditional on
\begin{equation*}
    C^w=\textcolor{BrickRed}{Y}-C^b-I-\left(\frac{\omega_e}{\omega_b{}U^b_{C}}\right)^\frac{1}{\gamma_e}.
\end{equation*}

\paragraph{If the collateral constraint is binding}
The algorithm is the same as in the flexible-price case.

\paragraph{Inflation}
Identically to the computation of the ``best CEA'', using the conditional solutions for $\widetilde{P}$, $P^w$, $\Delta$, $\lambda^\Delta$, $\lambda^\Omega$, $\Omega_{1}$, after many rearrangements, we get
\begin{equation*}
    \lambda^R=\frac{\Pi-1}{\Pi}\frac{\beta\theta\Pi^{\epsilon-1}}{1-\beta\theta\Pi^\epsilon}\frac{(\epsilon-1)\widetilde{\lambda}^C+\epsilon\lambda^Y}{1-\theta\Pi^{\epsilon-1}}\beta{}Y.
\end{equation*}
Using the conditional solutions for $\lambda^C$, $\widetilde{\lambda}^C$, and $\lambda^Y$,
\begin{multline*}
    (\epsilon-1)\widetilde{\lambda}^C+\epsilon\lambda^Y\\
    =\frac{\left\{\omega_e{}U^e_{C}+\lambda^K\left[(R-1)(Q{}K-L)U^e_{CC}+\beta_e{}R{}U^e_{C}\alpha\frac{\epsilon-1}{\epsilon}\right]\right\}\epsilon\left(\frac{v''(N)}{u'(C^w)}N+W\right)+\omega_w{}v'(N)}{\frac{v''(N)}{u'(C^w)}N+W+\frac{1}{\epsilon}\frac{A}{\Delta}F_{N}}>0.
\end{multline*}
If the relaxed collateral constraint is slack so that $\lambda^K=\lambda^e=0$, the inequality follows immediately; otherwise, it can be verified numerically. Therefore, $\sgn(\lambda^R)=\sgn(\Pi-1)$. The complementary slackness conditions postulate that $\Pi=\beta\underline{R}$ if $\lambda^R>0$. Hence, if $\underline{R}\le\frac{1}{\beta}$, then $\Pi=1$; if $\underline{R}>\frac{1}{\beta}$, then $\Pi=\beta\underline{R}$.

\section{Method of simulated moments estimation}\label{sec: MSM estimation}
The first MSM step estimates the autocorrelations and standard deviations of exogenous stochastic processes $(\rho_a,\rho_\xi,\sigma_a,\sigma_\xi)$ based on the FCE. The first-step moments include the first-order autocorrelation of output, the standard deviations of output and investment, and the contemporaneous correlation between output and investment, where all variables are logged and detrended using the HP filter with $\lambda=1600$. The empirical moments are based on the National Income and Product Accounts (NIPA) data. Consumption corresponds to personal consumption expenditures on nondurable goods and services net of imports of nondurable goods and services, investment is mapped to gross private domestic investment and personal consumption expenditures on durable goods net of imports of durable goods, and output is a sum of consumption and investment, as in the model. Conditional on the first step, the second step determines the Taylor rule parameters $(\rho_R,\eta_\pi,\eta_y)$ by targeting the first-order autocorrelations and standard deviations of inflation and output, the contemporaneous correlation between inflation and output, and the ZLB frequency, where the target for the latter is consistent with \citet{fernandez-villaverde15} and \citet{guerrieri15}. The inflation series is constructed from the GDP deflator in the NIPA data and is logged and detrended as the other variables. Targeting the first-order autocorrelation and standard deviation of output in both steps makes the flexible-price and sticky-price economies produce relatively comparable business cycle fluctuations.

The MSM estimation at each step proceeds as follows. Let $\bm{\theta}\in\Theta$ denote the parameter vector to be estimated, where $\Theta\equiv[0,0.99]^2\times[0.0001,0.02]^2$ in the first MSM step, and $\Theta\equiv[0,0.99]\times[1.01,5]\times[0,5]$ in the second step. Let $m^s(\bm{x}_s\mid\bm{\theta})$ be the estimate of a vector of simulated moments at $\bm{\theta}$, let $m^d(\bm{x})$ be the vector of empirical moments, and define $h(\bm{x},\bm{x}_s,\theta)\equiv{}m^s(\bm{x}_s\mid\bm{\theta})-m^d(\bm{x})$. The estimate of $\bm{\theta}$ is obtained as $\widehat{\bm{\theta}}=\arg\min_{\bm{\theta}\in\Theta}h(\bm{x},\bm{x}_s,\theta)'\bm{W}h(\bm{x},\bm{x}_s,\theta)$, where $\bm{W}$ is a positive semidefinite weighting matrix. I set $\bm{W}=\diag(|m^d(\bm{x})|)^{-2}$ to ensure that all moments are in comparable units.

Table \ref{tab: moments} reports the moments in the data, FCE, and CE.
\begin{table}[ht!]
    \caption{MSM estimation moments: data and model}\label{tab: moments}
    \centering
    \begin{tabular*}{\textwidth}{l@{\extracolsep{\fill}}cccccccc}
        \toprule
        & $\rho(\widehat{Y}_{t},\widehat{Y}_{t-1})$ & $\sigma(\widehat{Y}_{t})$ & $\sigma(\widehat{I}_{t})$ & $\rho(\widehat{I}_{t},\widehat{Y}_{t})$ & $\rho(\widehat{\Pi}_{t},\widehat{\Pi}_{t-1})$ & $\sigma(\widehat{\Pi}_{t})$ & $\rho(\widehat{\Pi}_{t},\widehat{Y}_{t})$ & $\Pr(R^N_{t}=\underline{R})$ \\
        \midrule
        Data & 0.886 & 0.013 & 0.045 & 0.921 & 0.294 & 0.002 & 0.178 & 0.075\\
        FCE & 0.696 & 0.013 & 0.043 & 0.795 & -- & -- & -- & -- \\
        CE & 0.3 & 0.019 & 0.073 & 0.901 & 0.349 & 0.001 & 0.187 & 0.095 \\
        \bottomrule
    \end{tabular*}
    \begin{tabular}{@{}p{\textwidth}@{}}
        {\small Notes: $\widehat{X}_{t}$ denotes the cyclical component of $\ln(X_{t})$ extracted using the HP filter with $\lambda=1600$. $\rho(x,y)$ denotes $\corr(x,y)$ and $\sigma(x)$ denotes $\sd(x)$.}
    \end{tabular}
\end{table}
The first-step estimation matches the standard deviations of investment and output quite well, and there is a strong correlation between investment and output as in the data. The greatest discrepancy occurs in the autocorrelation of output, which is around 0.7 in the model and 0.9 in the data. Although the distance can be reduced to about 0.1 by increasing the value of $\psi$, it would decrease the variation in the asset price, which governs one of the key externalities in the model. In the data, output is nonstationary, but there is no stochastic trend in the model, contributing to a lower autocorrelation of the filtered time series. In the CE, this discrepancy is widened, although one could bring it closer to the FCE level by decreasing the price rigidity parameter $\theta$. There is more output and investment volatility, but the ratio of standard deviations is consistent with the data, and the correlation is matched closely, not being targeted. The moments associated with inflation are all close to the targets. Overall, the model accounts for the business cycle dynamics reasonably well, considering that it does not contain fit-enhancing features such as habit formation, price indexation, or quadratic investment adjustment costs.

\end{document}